      [1999/12/01 v1.4c Il Nuovo Cimento]
\documentclass[varenna]{cimento}
\newcommand{\Msun}{{\ifmmode{{\rm M}_{\odot}}\else{${\rm M}_{\odot}$}\fi}}

\def\simpropto{\lower.2ex\hbox{$\; \buildrel \propto \over \sim \;$}}
\def\ltsim{\lower.5ex\hbox{$\; \buildrel < \over \sim \;$}}
\def\gtsim{\lower.5ex\hbox{$\; \buildrel > \over \sim \;$}}

\usepackage[numbers]{natbib}
\usepackage{amssymb,epsfig,epstopdf,graphicx}
\usepackage[utf8x]{inputenc}

\title{Galaxy Formation}

\author{Joseph~Silk}

\institute{
Institut d'Astrophysique de Paris, UMR 7095, CNRS, UPMC Univ. Paris VI, 98 bis boulevard Arago, 75014 Paris, France\\
Department of Physics and Astronomy, The Johns Hopkins University,
3400 N. Charles Street, Baltimore, Maryland 21218, USA\\
 Beecroft Institute of Particle Astrophysics and Cosmology, Department of Physics,
University of Oxford, Denys Wilkinson Building, 1 Keble Road, Oxford OX1 3RH, UK}

\author{Arianna~Di Cintio}

\institute{Departamento de F\'isica Te\'orica, Universidad Aut\'onoma de Madrid - Madrid, Spain\\
Physics Department, Universit\`{a} di Roma Sapienza - Rome, Italy}

\author{Irina~Dvorkin}

\institute{School of Physics and Astronomy, Tel Aviv University - Tel Aviv,
Israel}

\shortauthor{J.~Silk, A.~Di Cintio, I.~Dvorkin}

\begin{document}

\maketitle

\begin{abstract}
Galaxy formation is at the forefront of observation and theory in cosmology. An
improved understanding is essential for improving our knowledge both of the
cosmological parameters, of the contents of the universe, and of our origins. In
these lectures intended for graduate students, galaxy formation theory is reviewed and confronted with recent
observational issues. In Lecture 1,  the following topics are presented:
star formation considerations, including IMF, star formation efficiency and star
formation rate,
the origin of the galaxy luminosity function, and feedback in dwarf galaxies. In
Lecture 2, we describe formation of disks and massive spheroids, including the
growth of supermassive black holes, negative feedback in spheroids, the AGN-star
formation connection, star formation rates at high redshift and the baryon fraction
in galaxies.
\end{abstract}

\section{Introduction}
Galaxy formation is a core theme of cosmology. Galaxies provide the beacons with which we measure the expansion and acceleration of the universe. If they are biased tracers of the underlying matter distribution then progress in cosmology becomes difficult. And there is no question that they are biased tracers: for example we only observe about half of the baryons in the universe in galaxies, the conversion of baryons into stars peaks at about 3 \%, without feedback star formation rates are unacceptably early and high, and we have at best a fragmentary understanding of feedback.
And our understanding of these issues, and others, is based on relatively local observations, with no fundamental theory nor any real grasp of how the relevant physical processes may vary in extreme conditions such as those encountered in the early universe or near supermassive black holes. Nor have we mentioned the initial mass function of stars or the complex  interplay between the fuel for forming stars, the gas reservoirs, and galaxy formation.

However we cannot abandon the search for a theory of galaxy formation, if only because an improved understanding of this subject is central to essentially all major telescope projects, under construction or being planned for the future. To be fair, there has been one enormous advance since the pioneering study of Lifschitz that may be said to have triggered the entire field of how galaxies formed from primordial density perturbations. This has been the detection of cosmic microwave background temperature fluctuations in their damped acoustic oscillations, now beautifully confirmed over many harmonics, which provide the evidence for the seed fluctuations. The physical effect was predicted
soon after the discovery of the cosmic microwave background radiation, and  evaluated quantitatively for the current cold dark matter-dominated cosmology by solving the
Boltzmann equation for the photon coupling with matter. Armed with initial conditions for the primordial density fluctuations, one can now successfully account for the large-scale structure of the universe. However it is the smaller scale astrophysics that is still poorly understood. This review will begin with star formation, and then discuss  various aspects of galaxy formation that are beginning to be probed at low and high redshift.

\section{Initial mass function and star formation}

The stellar initial mass function (IMF) is defined as the distribution of
stellar masses that form in a given region in one starburst event. Apart
from being interesting in its own right, the IMF has important consequences for
the observable properties of galaxies. The relative abundance of stars in
different mass ranges affects various measurable quantities: most of the
\emph{stellar mass} is contributed by low-mass stars, the \emph{luminosity} is
mainly due to massive stars, while intermediate-mass and massive stars are
responsible for enriching the interstellar medium (ISM) with \emph{metals}.
Understanding the physics of star formation and the form of the IMF is therefore
crucial for correct interpretation of observations of galaxies. In particular,
we would like to have an understanding of the masses of forming stars as a
function of the physical conditions in the star formation region, the efficiency
of star formation, and its rate.

Historically, the first model of star formation was offered by Newton, who put
forward the idea of gravitational collapse.

\emph{But if the matter was evenly disposed throughout an infinite space, it
could never convene into one mass, but some of it would convene into one mass,
and some into another, so as to make an infinite number of great masses,
scattered at great distances from one another throughout all that infinite
space. And thus might the sun and fixed stars be formed, supposing the matter
were of a lucid nature. But how the matter should divide itself into two sorts,
and that part of it, which is fit to compose a shining body, should fall down
into one mass and make a sun, and the rest, which is fit to compose an opaque
body, should coalesce, not into one great body, like the shining matter, but
into many little ones; or if the sun at first were an opaque body like the
planets, or the planets lucid bodies like the sun, how he alone should be
changed into a shining body, whilst all they continue opaque, or all they be
changed into opaque ones, whilst he remains unchanged, I do not think explicable
by mere natural causes, but am forced to ascribe it to the counsel and
contrivance of a voluntary Agent.} [Sir Isaac Newton, Letters to Dr. Bentley,
1692]

The next  breakthrough was
due to James Jeans, who put Newton's conjectures on a more
quantitative ground and formulated the criterion of gravitational instability.

\emph{We have found that, as Newton first conjectured, a chaotic mass of gas of
approximately uniform density and of very great extent would be dynamically
unstable: nuclei would tend to form in it, around which the whole of the matter
would ultimately condense.} [James Jeans, \emph{Astronomy and Cosmogony}, 1929]

To see how this instability arises, imagine a static infinite medium of density
$\rho$ and temperature $T$ with a density fluctuation on a scale
$\lambda=2\pi/k$:
\begin{equation}
 \delta\equiv\frac{\delta \rho}{\rho}\propto e^{i(\omega t-kx)}\:.
\end{equation}
The free-fall time is $t_{ff}\simeq 1/\sqrt{G\rho}$ and the dispersion relation
is given by:
\begin{equation}
 \omega^2=k^2 v_s^2-4\pi G\rho
\end{equation}
where $v_s$ is the speed of sound in the medium. Exponential growth of the
perturbation, and hence instability, will occur for wavelengths that satisfy:
\begin{equation}
 k<\frac{\sqrt{4\pi G\rho}}{v_s}\equiv k_J\:.
\end{equation}
In other words, perturbations on scales larger than the Jeans scale, defined as
follows:
\begin{equation}
 R_J=\frac{\pi}{k_J}
\end{equation}
will become unstable and collapse. Note the dependence of the Jeans scale on
the density and temperature of the medium. The physical meaning of this
criterion is that sound waves must cross the overdense region to communicate
pressure changes before collapse occurs. Indeed, a simple way to derive the
Jeans scale is to compare the sound crossing time $t_{sc}=R/v_s$ to the
free-fall time of a sphere of radius $R$. The resulting scale is $R_J\approx
t_{ff}v_s$. This simple argument provides us with the typical scale of
collapsed objects in a given medium, and serves as the basis for more
sophisticated models.

The division of collapsed objects into those that shine (stars) and
those that do not (planets) which, as Newton believed, could only be
explained by the intervention of a \emph{voluntary Agent} was described in
a very elegant way by Eddington:

\emph{We can imagine a physicist on a cloud-bound planet who has never heard
tell of the stars calculating the ratio of radiation pressure to gas pressure
for a series of globes of gas of various sizes, starting, say, with a globe of
mass of 10 gm., then 100 gm., 1000 gm., and so on, so that his $n$-th globe
contains $10^n$ gm... Regarded as a tussle between matter and aether (gas
pressure and radiation pressure) the contest is overwhelmingly one-sided except
between Nos. 33-35, where we may expect something interesting to happen.
What happens is the stars.} [Sir Arthur Eddington, \emph{The internal
constitution
of stars}, 1926]

Eddington was the first to suggest that the source of the stellar energy is
nuclear fusion of hydrogen into helium. This idea was later confirmed and the
theory of stellar nucleosynthesis is now well established. However, a complete
theory that describes the process of \emph{star formation} is still missing.
We cannot predict the initial mass function of stars which we shall see is crucial for understanding issues such as feedback.
Below we briefly review some of the key aspects of this problem as well as
attempts at solving it.

\subsection{Star formation: general considerations}

Stars form in the dense cores of giant molecular clouds (GMC) - dense and clumpy
concentrations of
cold gas and dust. According to the Jeans
criterion, initial collapse occurs whenever gravity overcomes pressure. Put
differently, the important scales in star formation are those on which gravity
operates against electromagnetic forces, and thus the natural dimensionless
constant that quantifies star formation processes is given by:

\begin{equation}
 \alpha_g=\frac{Gm_p^2}{e^2}\approx 8\cdot 10^{-37}\: .
\end{equation}

For example, we can derive the maximal mass of a white dwarf star in terms of
$\alpha_g$. For a star with $N$ baryons, the gravitational
energy per baryon is $E_G\sim -GNm_p^2/R$, and the kinetic energy of
relativistic degenerate gas is
$E_K\sim p_F c\sim \hbar c N^{1/3}/R$ where $p_F$ is the Fermi
momentum. Consequently, the total energy is:

\begin{equation}
 E=\frac{\hbar c N^{1/3}}{R}-\frac{GNm_p^2}{R}\;.
\end{equation}
For the system to be stable, the maximal number of baryons $N$ is obtained by
setting $E=0$ in the expression above. The result is the Chandrasekhar mass:
\begin{equation}
 M_{Chandra}\approx m_p \left(\frac{\hbar c}{Gm_p^2} \right)^{3/2}=m_p
(\alpha\cdot \alpha_g)^{-3/2} =1.8M_{\odot}
\end{equation}
where $\alpha=e^2/(\hbar c)$ is the
fine structure constant. This simple derivation result is close to the more precise value, derived via the equations  of
stellar structure for degenerate matter, of $1.4M_{\odot}$.

Another example is the upper limit on the mass of a hydrogen burning star, which
can be
derived if we compare the radiation pressure $P_{rad}=1/3aT^4$ to the thermal
pressure of the star, under the condition that the temperature be high enough so
as to allow two protons to overcome their mutual Coulomb
repulsion. The resulting upper limit is $M_{Edd}\approx
m_p\alpha_g^{-5/3}\alpha^{2/3}\approx 50M_{\odot}$, which is comparable to the masses of
the most massive stars observed (at the Eddington limit, again derived from the equations of stellar structure):  $\sim 100\rm M_\odot $.

\begin{figure}[h!]
\centering
\epsfig{file=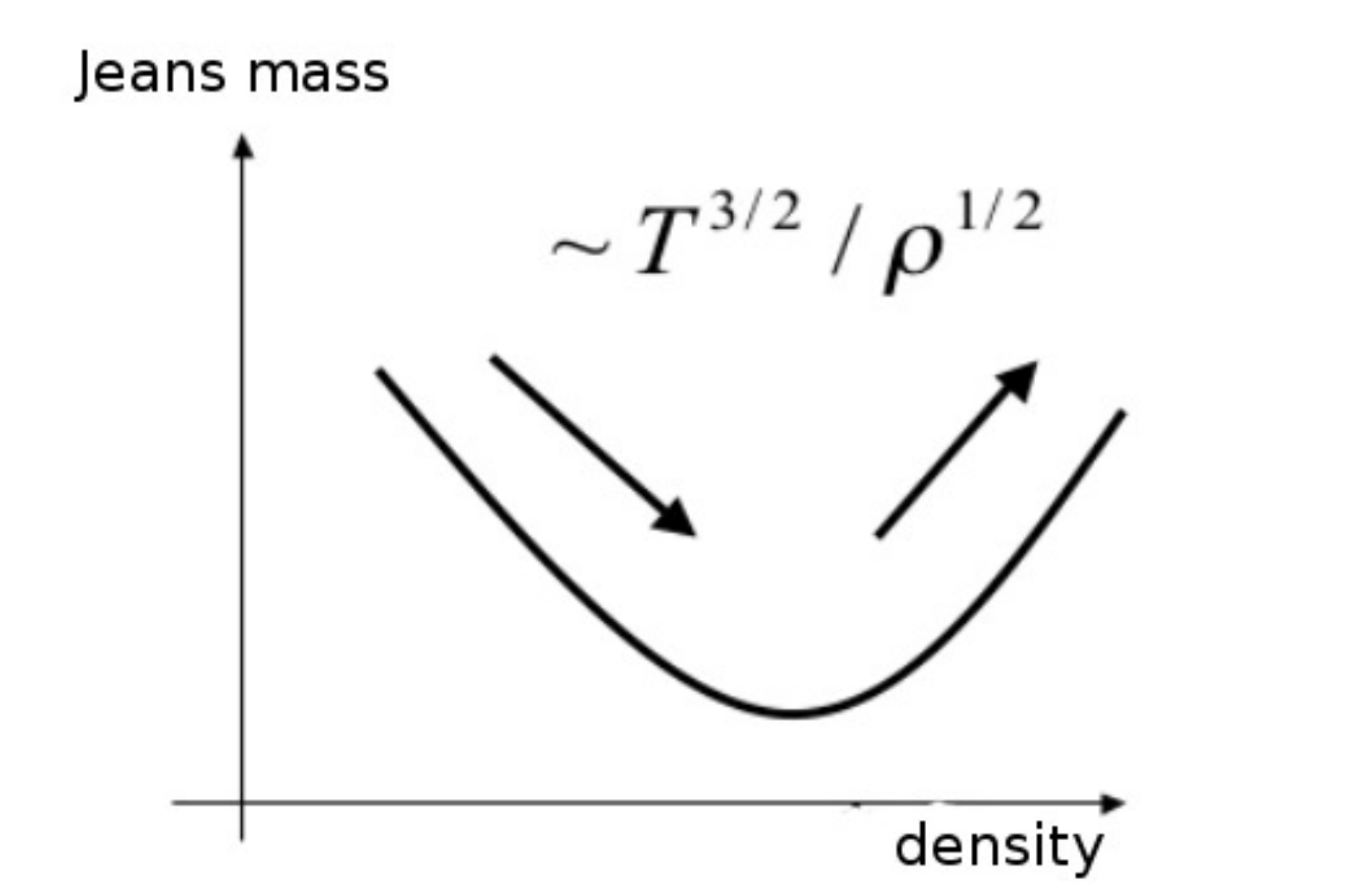, height=4cm}
\caption{Jeans mass vs. density. The minimal Jeans mass sets the limit of
fragmentation.}
\label{fig:jmass}
\end{figure}

The formation of a star, or indeed a star cluster, begins with the collapse of an
overdense region whose mass is larger than the Jeans mass, defined in terms of
the Jeans scale $R_J$:
\begin{equation}
 M_J=\frac{4\pi}{3}\rho \left(\frac{R_J}{2} \right)^3\propto
\frac{T^{3/2}}{\rho^{1/2}}\:.
\end{equation}

Overdensities can arise as a result of turbulent motions in the cloud.
At the first stage of the collapse, the gas is optically thin and isothermal,
whereas the density increases and $M_J\propto \rho^{-1/2}$. As a result, the
Jeans mass decreases and
smaller clumps inside the originally collapsing region begin to collapse
separately. \emph{fragmentation} is halted when the gas
becomes optically thick and adiabatic, so that $M_J\propto \rho^{1/2}$, as
illustrated in fig. \ref{fig:jmass}.
This process determines the opacity-limited minimum fragmentation scale for low
mass
stars, and is given by:
\begin{equation}
 M_{min}\approx m_p \alpha_g^{-3/2}\alpha^{-1}\left(\frac{m_e}{m_p}
\right)^{1/4} \approx 0.01M_{\odot}\:.
\end{equation}
Of course, this number, which is a robust scale and confirmed in simulations, is far smaller than the observed current epoch stellar mass range, for which the characteristic stellar mass is $\sim 0.5 \rm M_\odot.$
Fragmentation also leads to the formation of star clusters, where many stars
with different masses form through the initial collapse of a large cloud.

In reality, however, the process of star formation is more complex, and the
initial collapse of an overdense clump is followed by \emph{accretion} of cold
gas at a typical rate of $v_s^3/G$, where $v_s$ is the speed of sound. This assumes spherical symmetry, but accretion along filaments, which is closer to what is actually observed, yields similar rates.
The gas
surrounding the protostellar object
typically has too much angular momentum to fall directly onto the
protostar, and as a result an accretion disk forms around the central object.
The final mass of the star is fixed only when accretion is halted by some
feedback process.

\begin{figure}[h!]
\centering
\epsfig{file=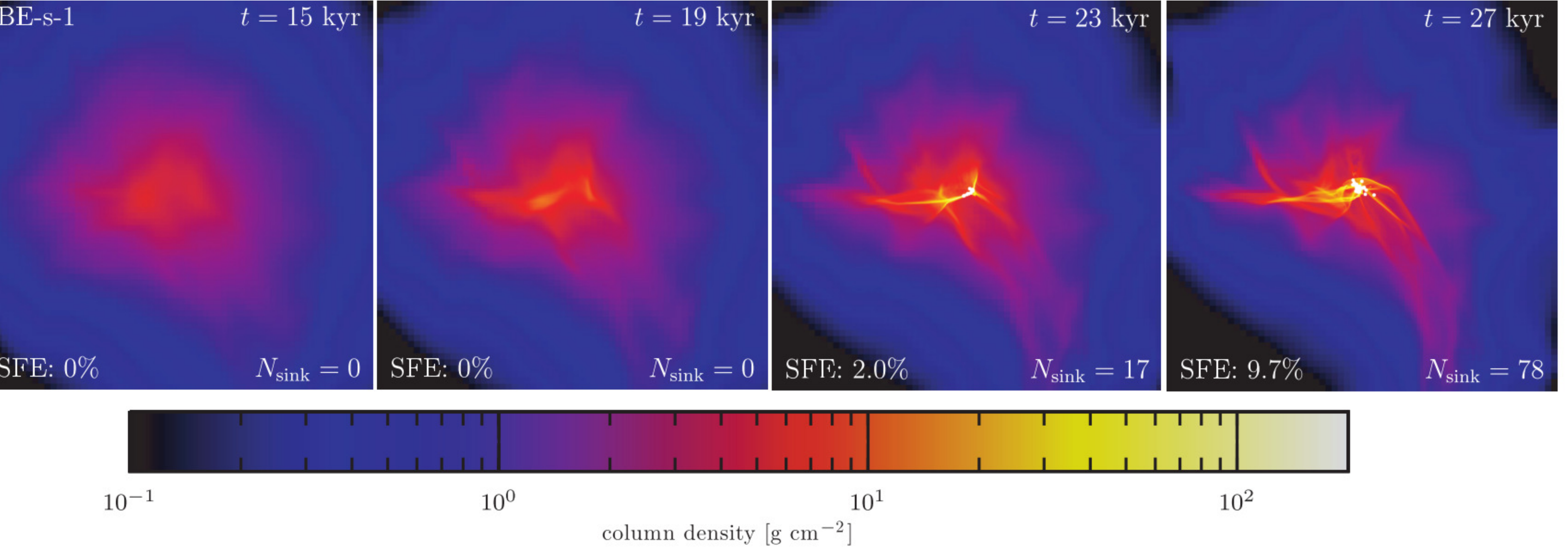, height=5cm}
\caption{Column density in a star forming cloud with an initial $\rho\propto
r^{-1.5}$ density profile. Each box is $0.13$ pc on a side. Figure from
\citep{2011MNRAS.413.2741G}.}
\label{fig:girichidis}
\end{figure}

This scenario is confirmed by numerical simulations
\citep{1998ApJ...501L.205K, 2011MNRAS.413.2741G, 2013arXiv1304.4600K}. For
example, fig. \ref{fig:girichidis} shows the formation of a star cluster from an
initially
spherical molecular cloud with a mass of $100 M_{\odot}$. In general, all
simulations find that filamentary structures form in the initially smooth cloud
before fragmenting further into stars. However, the details of this process, and
in particular
the relative numbers of stars of different masses depend on the initial density
profile of the cloud and the assumed turbulent field that seeded the
overdensities.

An additional mechanism that can influence star formation is outflowing
\emph{supersonic winds} from a protostar that can accelerate the surrounding
molecular
gas to high velocities and oppose accretion \citep{1980ApJ...238..158N}.
These winds can affect the
density structure of the cloud \citep{2007ApJ...668.1028B} and thereby alter the
mass spectrum of formed
stars and the star formation efficiency
\citep{2000ApJ...545..364M, 2007ApJ...662..395N}. Such outflows have been
detected in numerous protostars,
and it was found that in some cases they have strong impact on the surrounding
medium even at very large distances from the source. For example, the outflows
in the Perseus complex have enough power
to maintain turbulence and hence halt star formation in their immediate
surroundings, but not in the entire complex \citep{2010ApJ...715.1170A}.

\begin{figure}[h!]
\centering
\epsfig{file=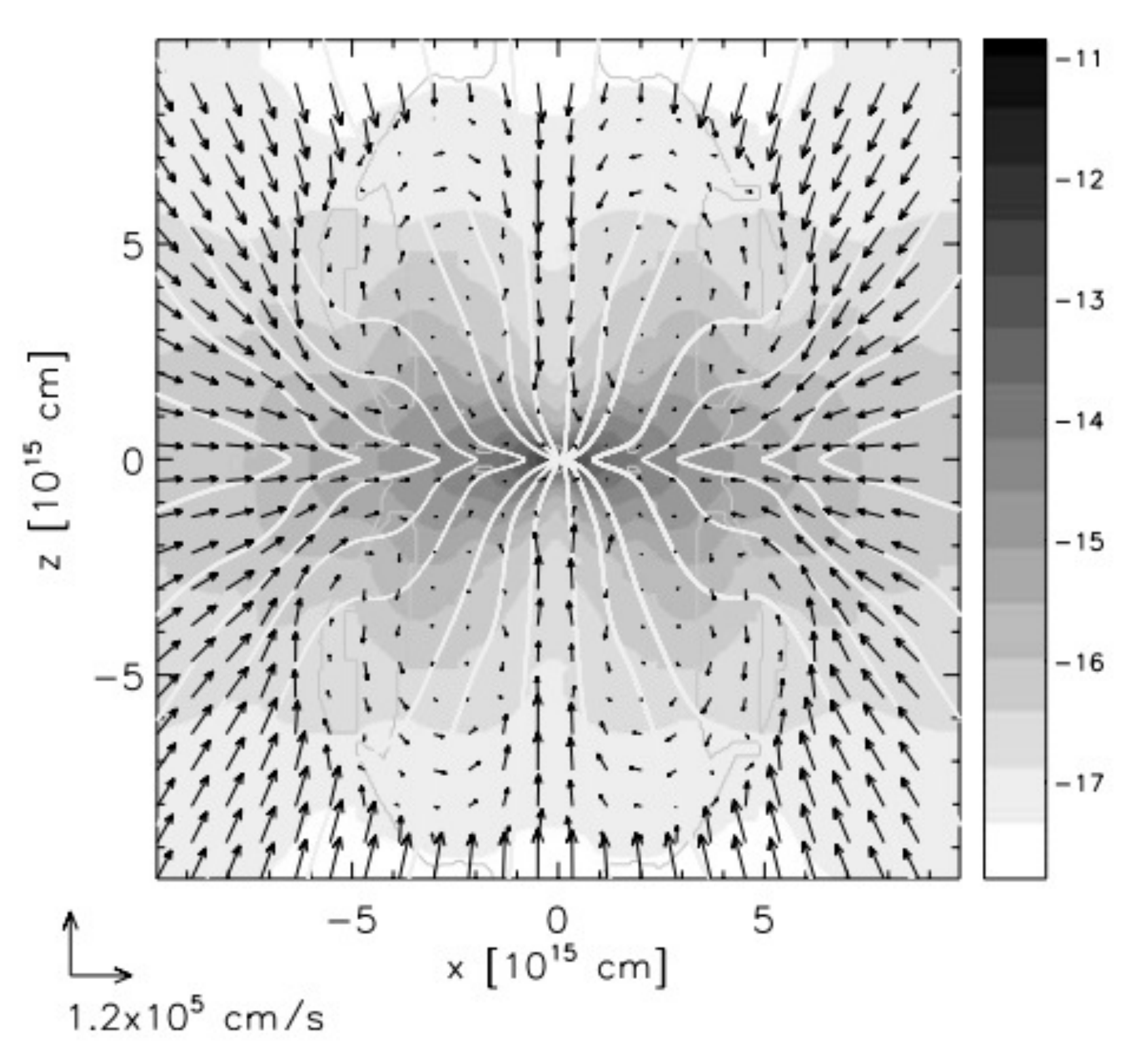, height=6cm}
\caption{Onset of a large-scale outflow in a numerical simulation of collapsing
magnetized cloud core. The magnetic pressure drives a bubble that is surrounded
by shock fronts and reverses the gas flow. Grey scale shows the density in
$gr/cm^3$ (logarithmic). The vector field represents the velocity flow, magnetic
flux is in white. Figure from \citep{2006ApJ...641..949B}.}
\label{fig:banerjee}
\end{figure}

\emph{Magnetic fields} can play an important role in star formation if the
mass of
the overdense cloud is of the order of the supercritical mass, such that the
magnetic energy density is comparable to the gravitational energy density:

\begin{equation}
 \frac{B^2}{8\pi}\sim \frac{GM}{R}\rho\sim G\Sigma^2
\end{equation}
where $\Sigma$ is the surface mass density. The classical picture is that when the cloud first collapses, its
mass must be above the critical mass. Previously,  the
magnetic field is energetically dominant and halts further collapse. However,
magnetic fields are frozen only into ionized gas and dust, whereas neutral
matter can continue to accrete onto the cloud. This process is known as
gravity-driven \emph{ambipolar diffusion}. When the mass of the cloud exceeds
the critical
mass, magnetic pressure becomes irrelevant and the cloud
collapses.

Clouds with supercritical values of the magnetic field have been observed in
several systems \citep{2008ApJ...680..457T}, however the question remains
whether they play a dominant role in star formation
\citep{2012ARA&A..50...29C}. In particular, the relative importance of magnetic
fields and \emph{turbulence-driven feedback} remains unclear
\cite{2012A&ARv..20...55H}.

Molecular clouds are best probed by studying thermal dust emission in the infrared. The Herschel telescope has imaged nearby molecular clouds between 70$\mu$m and 500$\mu$m
as part of  a survey that is optimized for studying the formation conditions of
solar-type stars. The results cover a wide range of environments, from warm to
cold cloud complexes, where stars form in clusters or in small numbers or even
not at all. Two main results have emerged from these studies. The mass function
of prestellar cores has been measured \citep{2010A&A...518L.102A}. It is found
to match the initial stellar mass function but with a displacement of a factor
of about 3 in mass, discussed below. The stellar cores are found to form in long filaments ($\gtsim 1\rm pc$) of near-universal width $\sim 0.1\rm pc, $ and are fed by what
appears to be magnetically-regulated accretion of the surrounding cloud material
along striations that are nearly perpendicular to the filaments
\citep{2013A&A...550A..38P}. The process is inefficient: on GMC scales, only
about 2\% of the gas is forming stars, whereas in the densest cores, the star
formation efficiency exceeds 30\% \citep{2013arXiv1309.7762A}. The peak in the
protostellar core mass function at $\sim 0.6\rm M_\odot$ corresponds to the local
Bonnor-Ebert mass (the equivalent of the Jeans mass discussed above in the case
of net external pressure).

Magnetic fields also cause strong bipolar outflows \citep{1996ARA&A..34..111B}.
These outflows occur because the infalling matter bends magnetic field lines
inwards, and the rotating accretion disk leads to field configurations that
eject the gas. Numerical simulations that include magnetohydrodynamics
\citep{2006ApJ...641..949B} indicate
that this process results in large-scale outflows, as can be seen in fig.
\ref{fig:banerjee}. Thus, magnetically-driven feedback reduces
star formation efficiency.

\begin{figure}[h!]
\centering
\epsfig{file=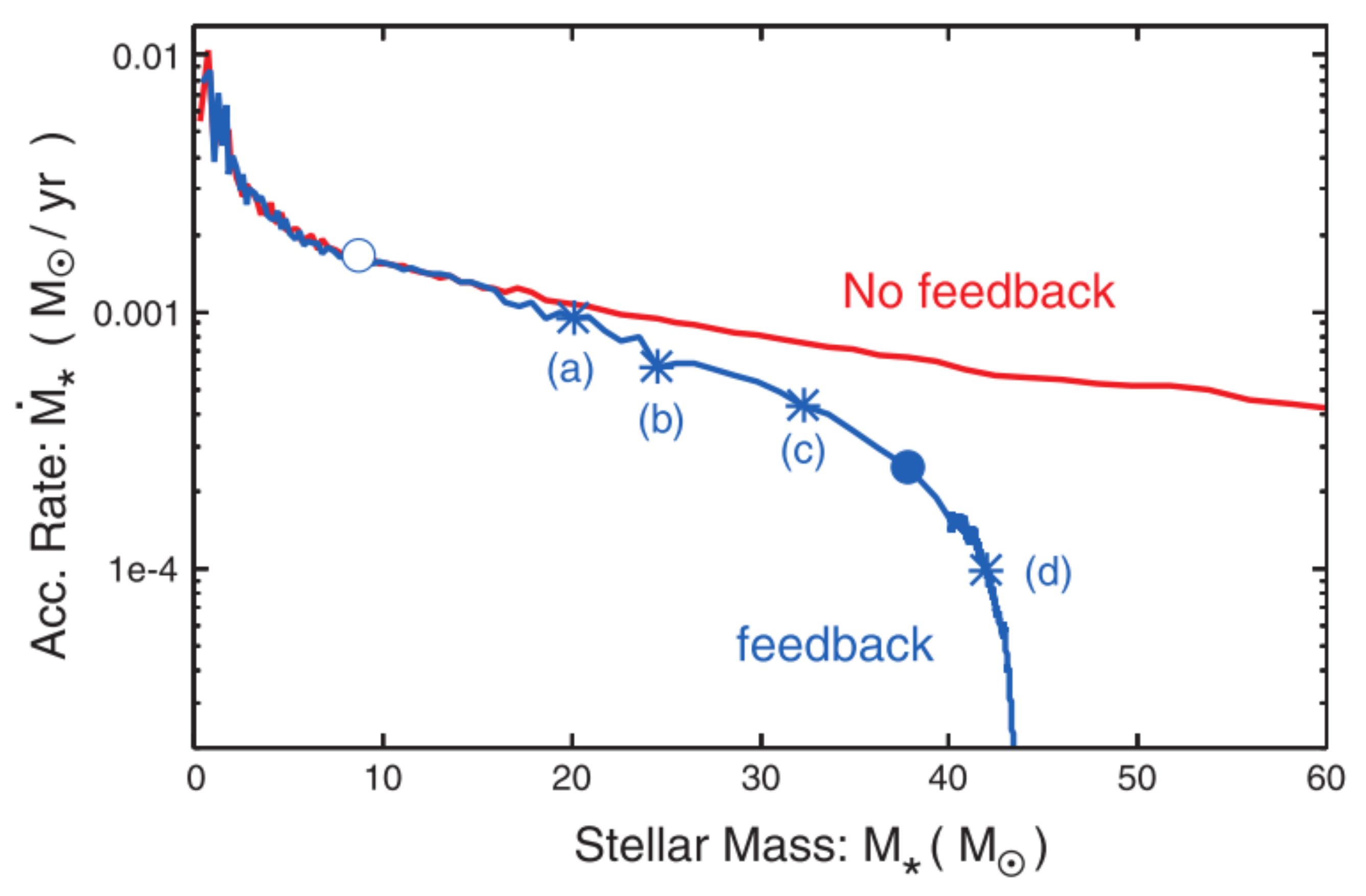, height=6cm}
\caption{Evolution of the accretion rate onto the primordial protostar with
(blue line) and without (red line) UV feedback. Figure from
\citep{2011Sci...334.1250H}.}
\label{fig:hosokawa}
\end{figure}

\subsection{The first stars}

Since the first stars form in a metal-poor environment, $H_2$ cooling is
dominant, the gas temperatures are relatively high and so is the accretion rate
onto the protostar: $\dot{M}\sim v_s^3/G\sim 10^{-3}M_{\odot}/\rm yr$. Thus, the
first stars are expected to be extremely massive, reaching $\sim 100M_{\odot}$
and more \citep{2005SSRv..117..445G}.

A viable feedback mechanism that sets the limit of the most
massive first stars is accretion disk evaporation by the \emph{ionizing radiation} of
the protostar. Numerical simulations \citep{2011Sci...334.1250H} show that
radiation feedback is indeed important in metal-poor stars. The evolution of the
accretion rate onto a primordial star with and without this effect is shown in
fig. \ref{fig:hosokawa}. The characteristic first star mass is reduced to $\sim 40\rm M_\odot.$

Another possibility to limit the mass of the primordial stars is via
gravitational fragmentation. Numerical simulations that followed cloud collapse
from cosmological initial conditions demonstrate the possibility of the
formation of metal-free binary systems through fragmentation
\citep{2009Sci...325..601T}.

Cooling controls the lower limit of fragmentation. The cooling in primordial stars occurs via trace amounts of $H_2$ formation in the parent clouds. This leads to a high accretion rate and high masses,
as modulated by feedback.
Once enrichment occurs, cooling is dominated either by dust or by atomic cooling. Either pathway, uncertain because we have no robust theory of dust formation in the early universe, allows low mass star formation by fragmentation.

Chemical abundance tracers in the most extreme metal-poor stars testify to the existence of a prior generation of metal-free stars.
Numerous metal-poor stars have been detected to date
\citep{2013ApJ...762...25N}. The abundance and properties of these objects put
interesting constraints on the theories of star formation. These expectations can be reconciled with
observations, as  the most extremely metal-poor solar-mass stars reveal abundance anomalies. The current record holders are two stars at $[Fe/H]\sim -5.2$
which in common with some $\sim$30\% of the extremely metal-poor stars at $[Fe/H]\sim -3$ are carbon and oxygen-rich \citep{2013ApJ...762...28N}.
These are the oldest stars ever observed. This means that a previous generation of short-lived, hence massive and essentially zero metallicity, stars polluted the environments where the observed extremely low metallicity low mass stars formed.

\subsection{Initial stellar mass function}

The initial stellar mass function (IMF) describes all the stars formed in a
given region following a starburst. In practice, the IMF is obtained by
counting all the stars in a given region and correcting for the most massive
stars that by the time of observation had already been transformed into compact
objects. This correction procedure was first introduced by Salpeter
\citep{1955ApJ...121..161S}, who
obtained a power-law relation in the mass range $\sim 0.4-50M_{\odot}$:
\begin{equation}
 \frac{dn}{d M}=AM^{-\alpha}
\end{equation}
where $\alpha=2.35$ and $A$ is a normalization constant. Much recent work has
incorporated more sophisticated stellar lifetime modeling and encompassed a
wider mass range, resulting in slightly different shapes. For
example
Kroupa \citep{2001MNRAS.322..231K} found a broken power-law with $\alpha=0.3$
for $M<0.08M_{\odot}$,
$\alpha=1.3$ for $0.08M_{\odot}<M< 0.5M_{\odot}$ and $\alpha=2.3$ for
$M>0.5M_{\odot}$, while Chabrier \citep{2003PASP..115..763C} introduced a
log-normal distribution of stellar masses.

\begin{figure}[h!]
\centering
\epsfig{file=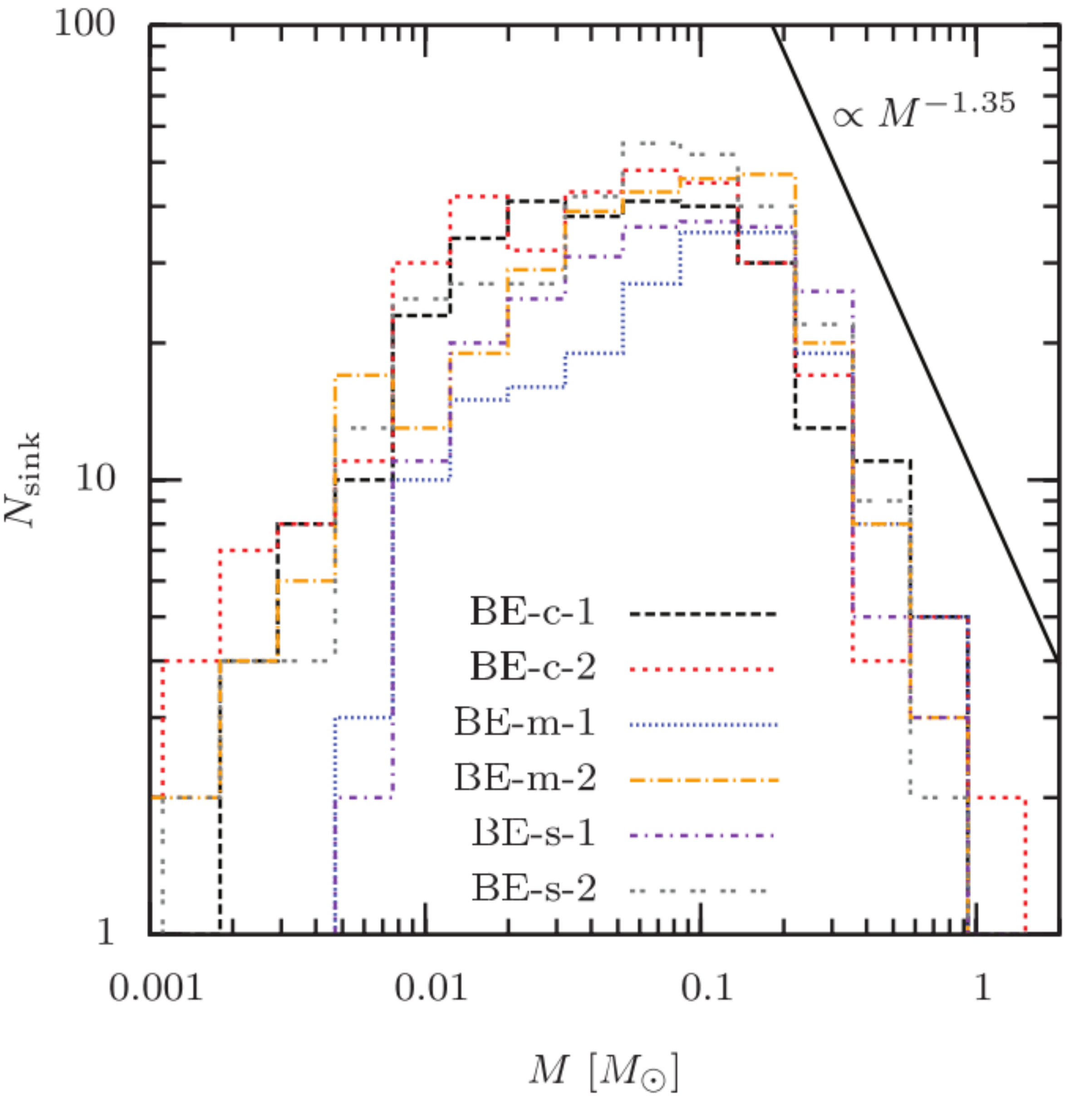, height=7cm}
\caption{IMF (per unit logarithmic mass) for the simulations from fig.
\ref{fig:girichidis}, for different turbulent fields. The Salpeter slope is
shown for comparison. Figure from \citep{2011MNRAS.413.2741G}.}
\label{fig:imf_giri}
\end{figure}

Clearly, a complete theory of star formation should be able to reproduce the
observed IMF, and some general features can be
explained in relatively simple terms \citep{2004ASPC..323...37S,
2007prpl.conf..149B}. The initial slope is set by the perturbation spectrum and
is subsequently modified by the different feedback processes.
The broad peak of the IMF is probably determined by the physics of
gravitational fragmentation and is due to the dispersion in the thermodynamical
properties of the gas. The formation of low-mass stars poses a problem, as they
would need to avoid excessive accretion, for example by being ejected from their birth
site. High-mass stars form through accretion, and the mass dependence of
this process determines the high-mass slope of the IMF.

A complete treatment of the problem is possible only with numerical simulations
that can take into account all the relevant physical processes (for an example, see fig.
\ref{fig:imf_giri}). Recent
advances in the field and some outstanding problems are
discussed in the comprehensive review by McKee and
Ostriker \citep{2007ARA&A..45..565M}.

\begin{figure}[h!]
\centering
\epsfig{file=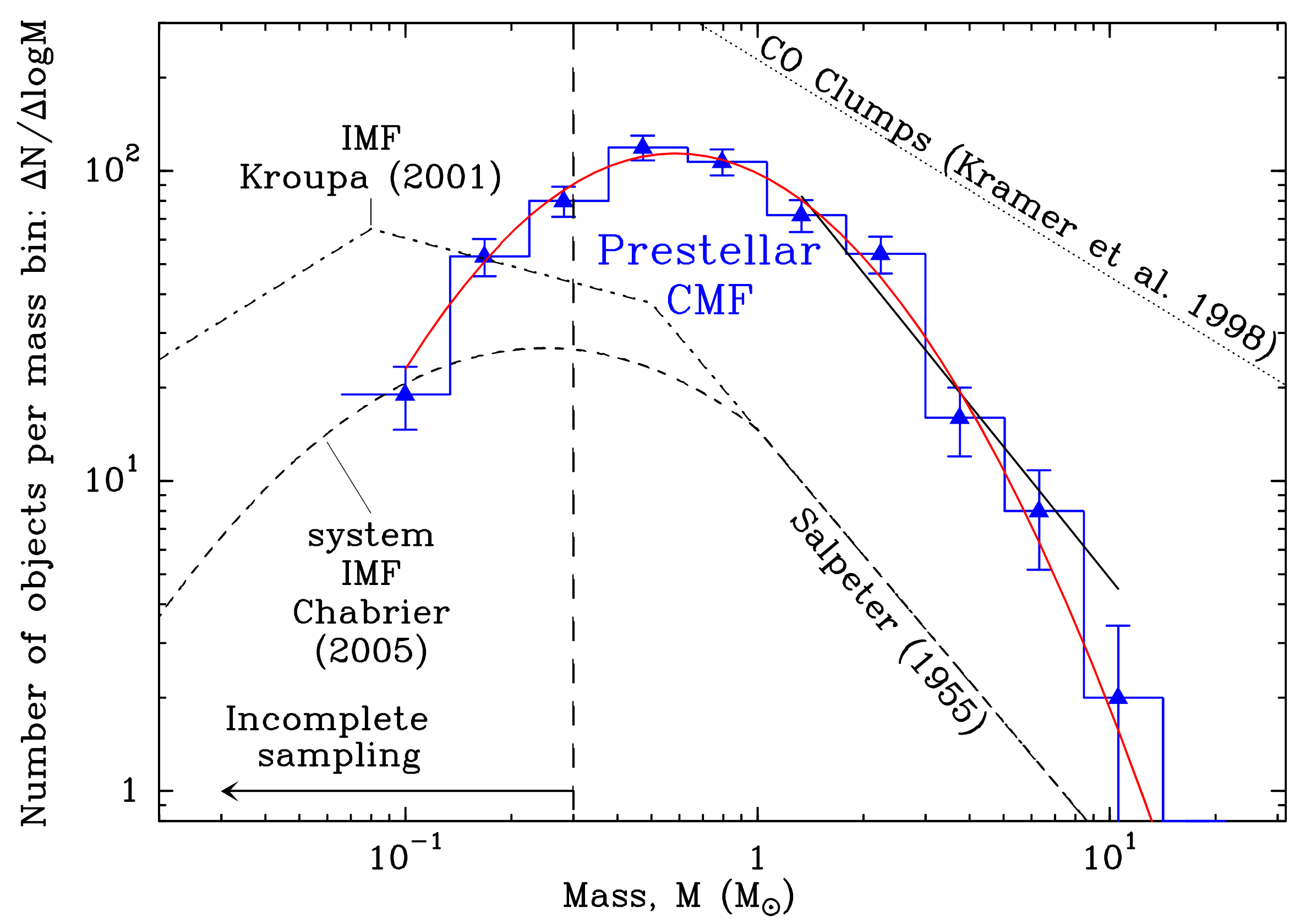, height=6cm}
\caption{Prestellar core mass function in molecular clouds compared to the initial stellar mass function. Figure from reference \cite{2013arXiv1309.7762A}.}
\label{fig:CMF}
\end{figure}

A recent development has been the compilation of the mass function of prestellar
cores, as discussed above. These are found to be highly embedded in dense
molecular clouds, and are
identified by Herschel telescope imaging surveys of nearby star-forming
complexes and molecular clouds. The prestellar core IMF, shown in fig.
\ref{fig:CMF}, has the same shape as
the initial stellar IMF, apart from a displacement to larger masses by about a
factor of 3 in mass. Both can be fit by a lognormal function. Comparing the
peaks of the two functions one infers that the efficiency of star formation is
universally, over all masses, a factor of $\sim 3$. The efficiency factor is
most likely due to protostellar outflows, which themselves are usually
considered responsible for generating the turbulence observed in star-forming
molecular clouds both from simulations, cf.  \citep{2010ApJ...709...27W} and
observations, cf. \citep{2013ApJ...774...22P}.

\section{From linear theory to galaxy formation}

The process that governs galaxy formation is the collapse of
gravitationally unstable regions, and in some aspects resembles the process of
star formation discussed above. The most likely origin of the initial
perturbations is inflation, an epoch of exponential
expansion of the Universe, during which quantum fluctuations in the field
driving the inflation (the inflaton) were
stretched to macroscopic scales. The exact nature of the inflaton is not yet
known, but the basic predictions of the theory are confirmed by observations, in
particular the flatness of the Universe \citep{2013arXiv1303.5076P}.

There are three
types of metric perturbations: scalar, vector and tensor.
\emph{Vector} and  \emph{tensor}
perturbations decay (in the matter-dominated era) in an expanding Universe, and the large-scale structure is
seeded by the \emph{scalar} perturbations, since these are coupled to the
stress-energy tensor of the matter-radiation field.

Scalar perturbations can be classified into perturbations of the spatial
curvature (commonly referred to as adiabatic) and perturbations in the entropy.
Adiabatic perturbations arise if the relative number
densities of all species remain constant: $\delta(n_{\gamma}/n_b)=0$,
$\delta(n_{\gamma}/n_m)=0$ and so on. In this case there is no energy transfer
between the various components, the energy conservation equation is satisfied by
each component separately and we obtain the familiar results: $\rho_m a^3=const.$,
$\rho_b a^3=const.$ and
$\rho_\gamma a^4=const.$ Therefore:
\begin{equation}
 \frac{\delta \rho_m}{\rho_m}=\frac{\delta
\rho_b}{\rho_b}=\frac{3}{4}\frac{\delta
\rho_{\gamma}}{\rho_{\gamma}}
\end{equation}

Entropy perturbations can be defined as deviations from adiabaticity, for
example for dark matter and radiation:
\begin{equation}
 \delta S=\frac{\delta \rho_m}{\rho_m}-\frac{3}{4}\frac{\delta
\rho_{\gamma}}{\rho_{\gamma}}
\end{equation}
and similarly for other pairs of species.

These two types of fluctuations are orthogonal, in the sense that all other
types can be described as a linear combination of adiabatic and entropy
modes which evolve independently.

The most natural, single-field inflationary models typically produce adiabatic
fluctuations, and these predictions agree well with observations
\citep{2013arXiv1303.5082P}. However, a sub-dominant isocurvature component
might exist and lead to observable effects, such as blue tilted scale-dependent spectra or to non-gaussian density
perturbations \citep{2012JPhCS.405a2003L}.

These initial conditions were set by the end of the inflationary period,
when the Universe was $\sim 10^{-35}$ second old. The primordial curvature
perturbations were then at a level of $\sim 10^{-5}$ and provided
the seeds for future formation of structure in the universe.

\subsection{Linear theory}

In order to understand the evolution of the density fluctuations in the
expanding Universe we need to solve the perturbed Friedmann equations. However,
when the scale of the perturbation is small compared
to the horizon ($\ell\ll ct_0$) and the flow is non-relativistic ($v\ll c$),
a Newtonian derivation can be used. Furthermore, when the
perturbations are still small, as happens in the early stages of
structure formation, the equations that govern their evolution can
be linearized.

We begin with the equation of mass conservation:
\begin{equation}
 \frac{\partial \rho}{\partial t}+\nabla \cdot (\rho u)=0\:,
\label{eq:continuity}
\end{equation}
the Euler equation:
\begin{equation}
 \frac{du}{dt}=-\nabla \phi-\frac{1}{\rho}\nabla P\:,
\label{eq:euler}
\end{equation}
the Poisson equation:
\begin{equation}
 \nabla^2 \Phi=4\pi G\rho
\label{eq:poisson}
\end{equation}
and an equation of state: $P=P(\rho)$. If the flow is homogeneous, the solution
to the above equations is:
\begin{equation}
 \bar{\rho}\propto a^{-3}\:, u=\frac{\dot{a}}{a}r\:, \bar{\Phi}=\frac{2\pi
G}{3}\bar{\rho} r^2\: .
\end{equation}

Now we perturb this solution as follows:
\begin{equation}
 \rho=\bar{\rho}+\delta\rho\:, v=u-Hr\:, \Phi=\bar{\Phi}+\phi
\end{equation}
and define the comoving coordinate
$x=r/a(t)$ and the proper time $d\tau=dt/a$. We obtain the following equations,
where all the time and spatial derivatives are with respect to $\tau$ and $x$:
\begin{equation}
 \dot{\delta}+\nabla\cdot \left((1+\delta)v \right)=0\:,
\label{eq:delta}
\end{equation}
\begin{equation}
 \frac{dv}{dt}=-\frac{\dot{a}}{a}v-\frac{1}{\rho}\nabla P-\nabla \phi \:,
\label{eq:vel}
\end{equation}
and
\begin{equation}
 \nabla^2\phi=4\pi G \bar{\rho}a^2\delta\:.
\label{eq:phi}
\end{equation}

If the perturbations are small, we can linearize the equations, so that the
first equation becomes $\dot{\delta}+\nabla\cdot v=0$ and solve for $\delta$.
Note that from the condition of adiabaticity (no heat exchange between fluid
elements) it follows that:
\begin{equation}
 \frac{1}{\rho}\nabla P=\frac{1}{\rho}c_s^2\nabla \rho
=\frac{c_s^2}{(1+\delta)}\nabla\delta
\end{equation}
where $c_s=\sqrt{dP/d\rho}$ is the speed of sound and the last equality is the
result of linearization.

The velocity field can be
decomposed as follows: $\vec{v}=\vec{v}_{||}+\vec{v}_{\bot}$ where
$\vec{\nabla}\times \vec{v}_{||}=0$ and $\vec{\nabla}\cdot \vec{v}_{\bot}=0$.
The rotational component, equivalent to a vorticity, satisfies:
\begin{equation}
 \vec{\nabla}\times \vec{v}_{\bot}\propto \frac{1}{a}
\end{equation}
and therefore decays in the expanding Universe. The irrotational component
satisfies:
\begin{equation}
 \vec{\nabla}\cdot\vec{v}_{||}=-\dot{\delta} \:.
\end{equation}
Peculiar velocities associated with this component arise due to density
fluctuations.

Finally, combining eqs. (\ref{eq:delta})-(\ref{eq:phi}) and moving back to
physical units, we obtain the following equation for the
overdensity field:
\begin{equation}
 \frac{d^2\delta}{dt^2}+2\frac{\dot{a}}{a}\frac{d\delta}{dt}=4\pi
G\bar{\rho}\delta +
\frac{1}{a^2}c_s^2\nabla^2\delta
\label{eq:overdensity}
\end{equation}
where the second term on the left is the damping term due to the expansion of
the Universe, the first term on the right is the gravitational driving term and
the second term on the right represents the pressure support.

Now let us decompose $\delta$ into plain wave modes $\delta(x,t)=\Sigma
\delta_k e^{i\textbf{x}\cdot \textbf{k}}$ where $\lambda=2\pi a/k$ is the
physical wavelength. The perturbation is unstable if its scale exceeds the
Jeans scale, defined here as:
\begin{equation}
 \lambda_{J}=c_s\sqrt{\frac{\pi}{G\rho_0}}\:.
\end{equation}

In the case of weakly interacting dark matter the matter pressure vanishes and
eq. (\ref{eq:overdensity}) reduces to:
\begin{equation}
 \frac{d^2\delta}{dt^2}+2\frac{\dot{a}}{a}\frac{d\delta}{dt}=4\pi G
\bar{\rho}\delta\:.
\label{eq:overdensity_dm}
\end{equation}

During the epoch of matter domination, $\rho_m>\rho_{\gamma}$ and all the wavelengths
are unstable. A lower bound arises due to the free streaming of the
dark matter particles, and corresponds to the cosmologically irrelevant scale of
$\sim 10^{-6}$ Earth mass for a $100$
GeV dark matter particle.
%%%%
In fact there are interesting variations in this scale that arise from considerations  of kinetic as opposed to thermal decoupling \cite{2001PhRvD..64b1302C} and from the possible dominance of warm dark matter \cite{2005A&A...438..419B}.   %

In an Einstein-de Sitter background ($\Omega_m=1$), the
scale factor grows as $a\propto t^{2/3}$ while $\rho=\rho_m\propto a^{-3}$ and
there are two solutions to eq. (\ref{eq:overdensity_dm}): $\delta\propto
t^{2/3}$ and $\delta\propto t^{-1}$. At late epochs, the constant energy density
associated with the cosmological constant dominates over matter and the
solution is $\delta\approx const$.

Recall that in the general case the background Universe is governed by the
following equation:
\begin{equation}
 \left(\frac{\dot{a}}{a} \right)^2=\frac{8\pi G}{3}\left(\rho_m+\rho_{\gamma}
\right) \:.
\end{equation}
One solution to the perturbation equations is therefore:
\begin{equation}
 \delta=1+\frac{3}{2}\frac{\rho_m}{\rho_{\gamma}}\:.
\end{equation}
In other words, the overdensity is nearly constant during radiation domination
(in fact it grows logarithmically), and grows as $t^{2/3}$ in the matter
domination epoch.

\begin{figure}[h!]
\centering
\epsfig{file=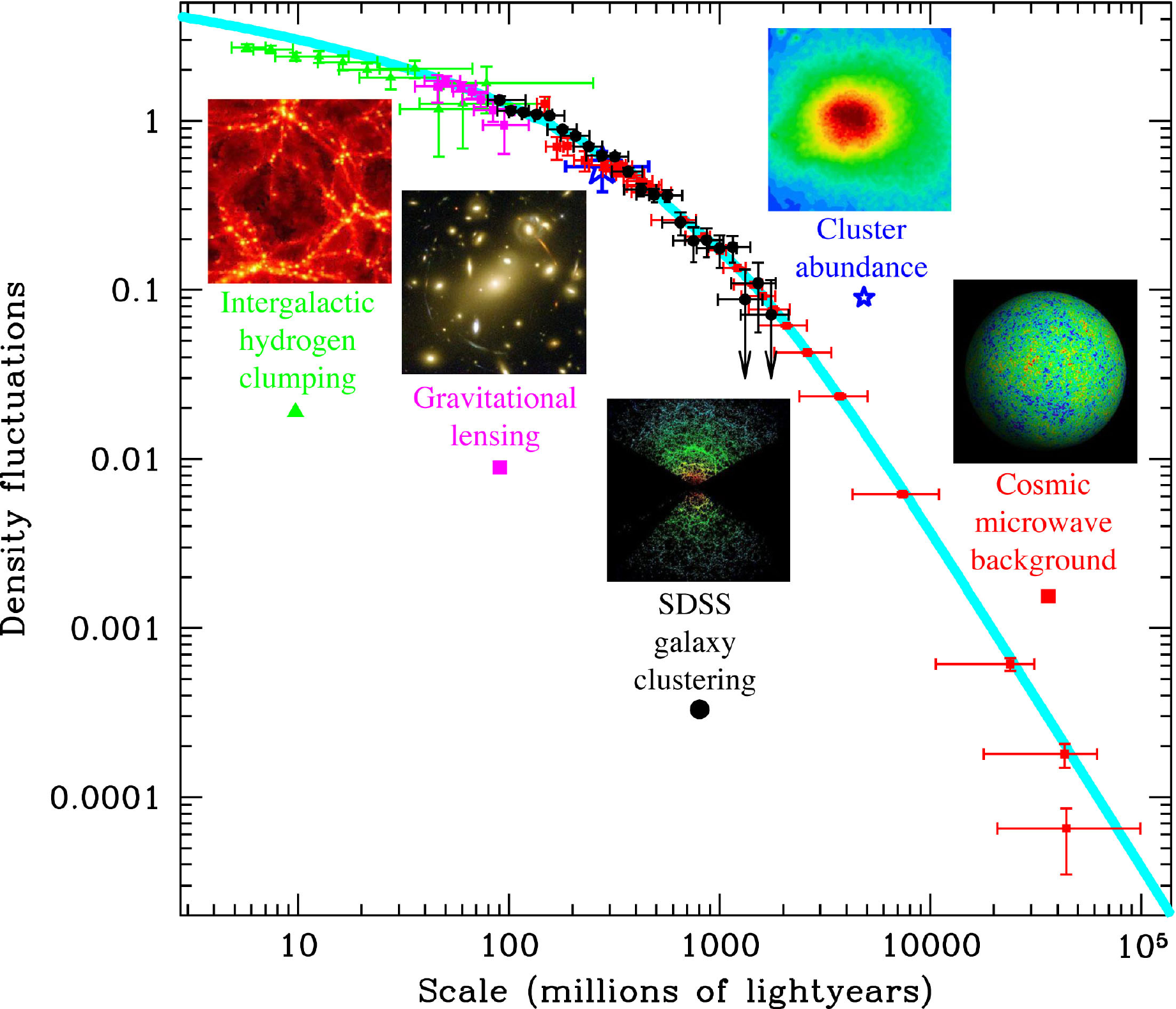, height=9cm}
\caption{Power spectrum measurements (dots) and the theoretical prediction
(blue curve). Image credit: M. Tegmark (SDSS).}
\label{fig:fluctuations}
\end{figure}

These results can be reformulated statistically by Fourier transforming these equations and looking at the
power spectrum of the density fluctuations:
\begin{equation}
  <|\delta^2|>=\int P(k) d^3k
\label{eq:Pk}
\end{equation}
where $P(k)=Ak^n$. For large scales, $n\simeq 1$ (for example, from
measurements of the cosmic microwave background anisotropies
\citep{2013arXiv1303.5076P}) while on small
scales, growth is suppressed during radiation domination, which results in a
characteristic peak in $P(k)$ on a scale corresponding to matter-radiation
equality. From eq. (\ref{eq:Pk}) it follows that $\delta \rho/\rho\propto
k^{(n+3)/2}$ or $\delta\rho/\rho \propto M^{-(n+3)/6}$. Asymptotically for
large $M$, $\delta\rho/\rho \propto M^{-2/3}$. More precisely, the power
spectrum today can be described by $P_{obs}(k)\propto k^nT^2(k)$ where $T(k)$
is a transfer function which represents the modifications to the primordial
power spectrum due to the transition from radiation to matter domination. The
power spectrum observed on different scales is shown schematically  in fig.
\ref{fig:fluctuations}. This figure nicely illustrates the complementarity of diverse observations that either directly (CMB) or indirectly (clusters, galaxies, IGM) sample the linear density fluctuation spectrum.

\subsection{Nonlinear theory}

When the fluctuations are large enough
so that $\delta \simeq 1$, the linear approximation breaks down and a full
solution
for the non-linear equations is needed. In the general case, eqs.
(\ref{eq:continuity})-(\ref{eq:poisson})
cannot be solved analytically and it is necessary to use numerical simulations
to obtain the matter distribution at late times.

The general scheme for structure formation in a cold-dark matter dominated universe was proposed in \citep{1984Natur.311..517B}.  Dissipation was incorporated and a two-stage theory of baryonic cores in CDM halos was first described by \citep{1978MNRAS.183..341W}. We develop these arguments below  via describing the recent simulations  but first give the only analytical results for fully nonlinear dissipationless dark matter structures.

An exact solution exists for the nonlinear evolution of a spherically
symmetric density perturbation \citep{1972ApJ...176....1G} (a pedagogical
treatment is given in \citep{1993sfu..book.....P}).
Consider a mass $M$ enclosed in a spherical volume with radius $R$ and
constant overdensity relative to the background universe $1+\delta=3M/4\pi R^3
\rho_b$, where $\rho_b$ is the background density. By Birkhoff's theorem, the
evolution of this spherical region is determined solely by its mass:
\begin{equation}
 \frac{d^2 R}{dt^2}=-\frac{GM}{R^2}=-\frac{4\pi G}{3}\rho_b(1+\delta)R
\label{eq:sc_equation}
\end{equation}
while the expansion of the background universe in the Einstein-de Sitter case
is given by:
\begin{equation}
 \frac{d^2 a}{dt^2}=-\frac{4\pi G}{3}\rho_b a \:.
\end{equation}

Thus, the overdense spherical region evolves like a Universe with a different
mean density but the same initial expansion rate. Integration of eq.
(\ref{eq:sc_equation}) gives:
\begin{equation}
 \frac{1}{2}\left(\frac{dR}{dt} \right)^2-\frac{GM}{R}=E
\label{eq:energy_sc}
\end{equation}
where $E$ is the energy of the spherical region. In analogy with the background
Universe, if $E<0$ the overdense region behaves like a closed universe and
collapses, whereas for $E>0$, $\dot{R}$ is always positive and the spherical
region continues to expand forever.

The solution to eq. (\ref{eq:energy_sc}) is given in parametric form by:
\begin{equation}
 R=A(1-\cos\theta)
\label{eq:sc1}
\end{equation}
and
\begin{equation}
 t=B(\theta-\sin\theta)\:,
\label{eq:sc2}
\end{equation}
where the constants $A$ and $B$ are related by $A^3=GMB^2$. The parameter
$\theta$ increases with $t$, while $R$ increases to a maximum
value $R_m=2A$ called the \emph{turnaround radius} at $\theta=\pi$ and $t=\pi B$
and then decreases to zero at $t_{coll}=2t_m$.

However, in reality the spherical perturbation does not collapse to a point but, at least in the case of dissipationless collapse,
attains a finite radius, taken here to be the \emph{virial radius} at which the
kinetic and gravitational energies satisfy $|U|=2K$. At the turnaround radius
the kinetic energy vanishes, and the total energy is
$E=U=-3GM^2/5R_m$. At virialization, the total energy is
$E=U+K=U/2=-3GM^2 /10R_{vir}$.
Equating the last two expressions we conclude that the virialization radius
satisfies
$R_{vir}=R_m/2$. Consequently, the nonlinear overdensity at virialization
$\Delta_V\equiv 1+\delta$ is not infinite.

In order to relate to the linear theory developed above, we expand eqs.
(\ref{eq:sc1})-(\ref{eq:sc2}) in powers of small $\theta$ and then eliminate $\theta$:
\begin{equation}
 R_{lin}=R_m\frac{1}{4}\left(6\pi\frac{t}{t_m} \right)^{2/3}\left[1-\frac{1}{20}
\left(6\pi\frac{t}{t_m} \right)^{2/3}\right]\:.
\end{equation}
For $t\rightarrow 0,$ we recover the background scale factor:
\begin{equation}
 R_b=R_m\frac{1}{4}\left(6\pi\frac{t}{t_m} \right)^{2/3}
\end{equation}
which, as expected, evolves like $R_b\propto t^{2/3}$. The linear overdensity
is given by $1+\delta=R_b^3/R_{lin}^3$ which, after using the above expressions
and expanding to lowest order in $t$ results in:
\begin{equation}
 \delta\simeq \frac{3}{20}\left(6\pi\frac{t}{t_m} \right)^{2/3}\:.
\end{equation}
Consequently, the linear overdensity at the time of collapse $t=t_{coll}$ is
$\delta_c=(12\pi)^{2/3}3/20=1.686$. This result should be understood in the
following way: clearly the linear approximation breaks down long before the
spherical region collapses, however if we continue to evolve the linear
overdensity field as a mere mathematical construct, it would have obtained the
value $\delta_c$ when the real spherical overdensity collapses.

In order to find the true overdensity at virialization, we first calculate the
overdensity at the turnaround point: $1+\delta=R_b^3(t_m)/R_m^3=9\pi^2/16$. As
we have seen, the radius at virialization is reduced by a factor of $2$,
therefore the density increases by a factor of $8$, while the background
density decreases by a factor of
$R_b^3(t_{coll})/R_b^3(t_m)=(t_{coll}/t_m)^2=4$. Thus, the overdensity at
virialization is given by:
\begin{equation}
 \Delta_V\equiv 1+\delta=18\pi^2 \simeq 178 \:.
\end{equation}

Another, almost exact, solution exists for one-dimensional collapse and is a
variant of the
Zeldovich approximation \citep{1989RvMP...61..185S}. Imagine a homogeneous
medium perturbed by a sinusoidal wave of
wavelength $L_d$, such that the positions are given by:
\begin{equation}
 Z(t)=a(t)\left[Z_i-\frac{a(t)}{a_p}\frac{L_d}{\pi}\sin\left(\pi\frac{Z_i}{L_d}
\right) \right]\:,
\end{equation}
where $Z_i$ are the initial positions, $a(t)$ is the growth factor and $a_p$
defines the amplitude of the perturbation. Using the mass conservation equation
for the one-dimensional case we obtain the density
distribution:
\begin{equation}
 \rho=\frac{\rho_0}{1-\frac{a(t)}{a_p}\cos\left(\pi\frac{Z_i}{L_d} \right)}\:.
\end{equation}
In this case, a caustic singularity develops for $Z_i=nL_d/2$. This simplified
analysis demonstrates the formation of \emph{caustics}, which is also seen as sheets and filaments in
numerical
simulations of the so-called cosmic web of large-scale structure \cite{1996Natur.380..603B}.

\subsection{Halo mass function}

The spherical collapse model discussed above is the basis for the calculation
of the halo mass function, or the abundance of virialized objects of a given
mass.
Firstly, let us assume that the initial density field can be described by
random Gaussian fluctuations. Consider the (linearly extrapolated)
density field smoothed on a scale $R$, $\delta(R,z)$. The main assumption is
that regions where $\delta(R,z)>\delta_c$, where $\delta_c$ is given by the
spherical collapse model, reside in collapsed obejects of mass $M=4\pi/3
\rho_b(z)R^3$.

We define the peak height:
\begin{equation}
 \nu=\frac{\delta_c}{\sigma(R,z)}
\end{equation}
where $\sigma$ is the rms density fluctuation. Then the
fraction of collapsed objects with mass greater than $M$ is given by:
\begin{equation}
 F(>M)=\frac{2}{\sqrt{2\pi \sigma^2(R,z)}}\int_{\delta_c}^{\infty}\exp\left(
-\frac{\delta^2}{2\sigma^2(R,z)}\right)d\delta=erfc \left(\frac{\nu}{\sqrt{2}}
\right)
\end{equation}
where erfc($x$) is the complementary error function. The mass
function is then given by $\partial F/\partial M$ and the comoving number
density of collapsed objects of mass $M$ is obtained by dividing this
expression by $M/\rho_b$. The result is the Press-Schechter mass function
\citep{1974ApJ...187..425P}:
\begin{equation}
 \frac{dn}{dM}=\sqrt{\frac{2}{\pi}}\frac{\rho_b}{M}\left(-\frac{d\ln
\sigma}{d\ln M} \right)\nu e^{-\nu^2/2}
\end{equation}
This mass function is exponentially suppressed on large scales and varies as
$M^{-2}$ on small scales. Note that, since $\sigma^2\propto M^{-(n+3)/3}$,
small objects collapse first, and larger objects form later through mergers and
accretion.

The Press-Schechter mass function provides a reasonable fit to numerical
simulations, but is not sufficiently accurate
as
it predicts too many low-mass halos and too few high-mass halos. Alternative
mass functions have been derived by direct fit to
simulations, for example the Sheth-Tormen mass function
\citep{1999MNRAS.308..119S}:
\begin{equation}
	\frac{dn(M,z)}{dM}=A_s\sqrt{\frac{2a_s}{\pi}}\left[1+\left(\frac{
\sigma^2}{a_s\delta_c^2}\right)^{p_s} \right]
\frac{\rho_b}{M^2}
\left(-\frac{d\ln \sigma}{d\ln M} \right)\nu e^{-\nu^2/2}
\end{equation}
with $A_s=0.3222,\: a_s=0.707$ and $p_s=0.3$, or the more recent Tinker mass
function \citep{2008ApJ...688..709T}.

\subsection{Comparison with observations}

\begin{figure}[t]
\centering
\epsfig{file=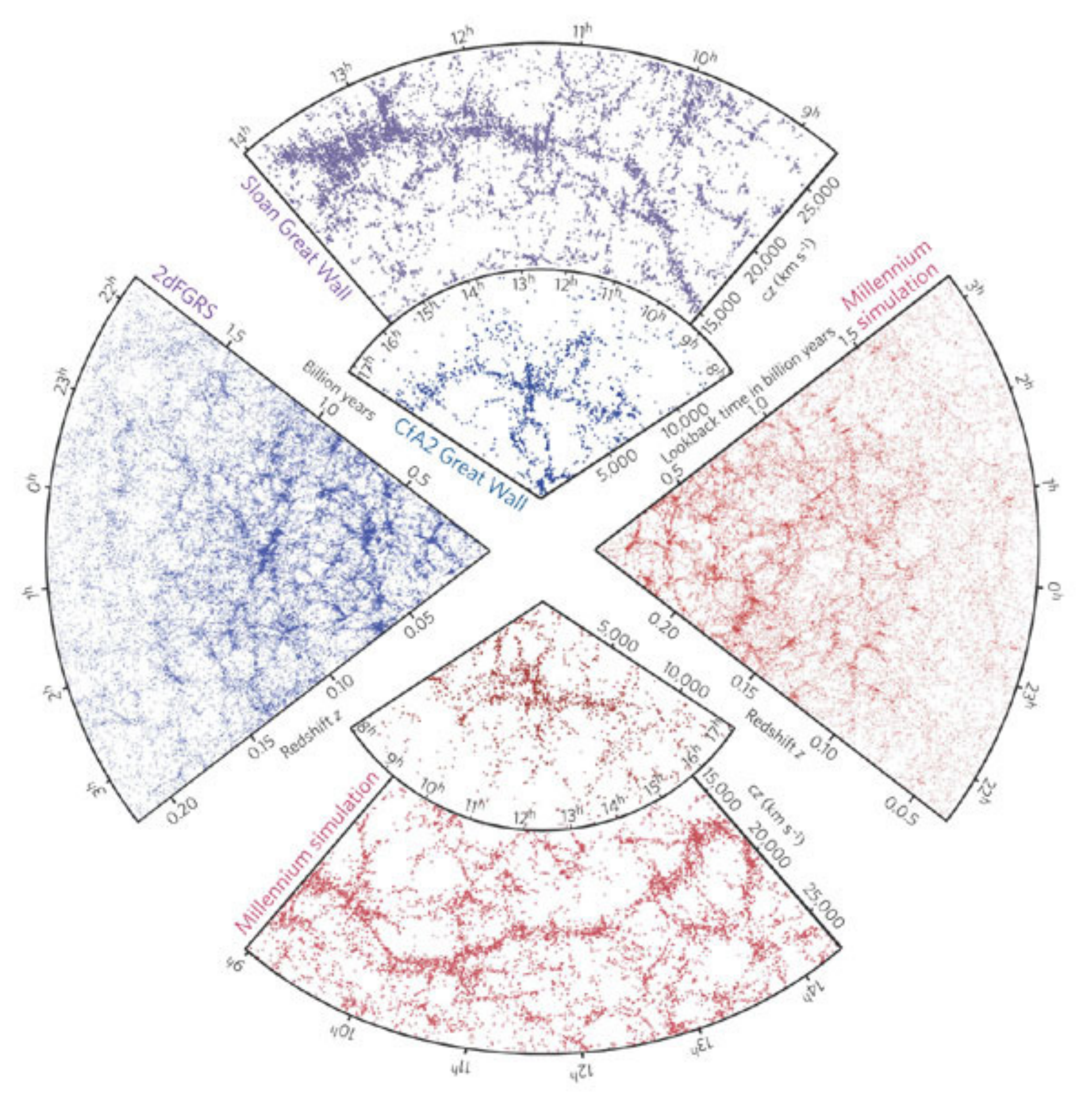, height=9cm}
\caption{Galaxy redshift surveys (blue) and mock surveys from numerical
simulations (red). Figure from \citep{2006Natur.440.1137S}.}
\label{fig:springel}
\end{figure}

The theory of structure formation can be tested using large-scale numerical
simulations, which are initialized at  high redshift and then
evolved according to the hydrodynamic and Poisson equations. The outcome of
these simulations is the density and velocity
distribution that can be compared with observations. For
example, the Millennium
Simulation \citep{2005Natur.435..629S} was run with $10^{10}$ particles from
redshift $z=127$ to the present in a cubic region $500h^{-1}$ Mpc on a side.
As shown in fig. \ref{fig:springel}, the distribution of galaxies seen in
spectroscopic redshift surveys, such as the 2-degree Field Galaxy Redshift
Survey (2dFGRS) and the Sloan Digital Sky Survey (SDSS), looks remarkably like
mock galaxy catalogues extracted from the Millennium Simulation.
The similarity of the observed and the simulated Universe, verified by
quantitative measures of galaxy clustering, is a powerful confirmation of the
validity of the theory of galaxy formation.

It is important to keep in mind that the basic predictions of the theory refer
to the dark matter component, which is not directly observable. In
fig. \ref{fig:springel}, semi-analytic models were used to
estimate the evolution of the baryonic component within the dark matter halos
of the Millennium Simulation. This approach relies on our knowledge of the
complex baryonic physics, but it turns out that numerical simulations fail to
reproduce some of the observed properties of galaxies. This tension between
theory and observation on small scales, as well as possible routes to its
resolution, are discussed below.

\begin{figure}[h!]
\centering
\epsfig{file=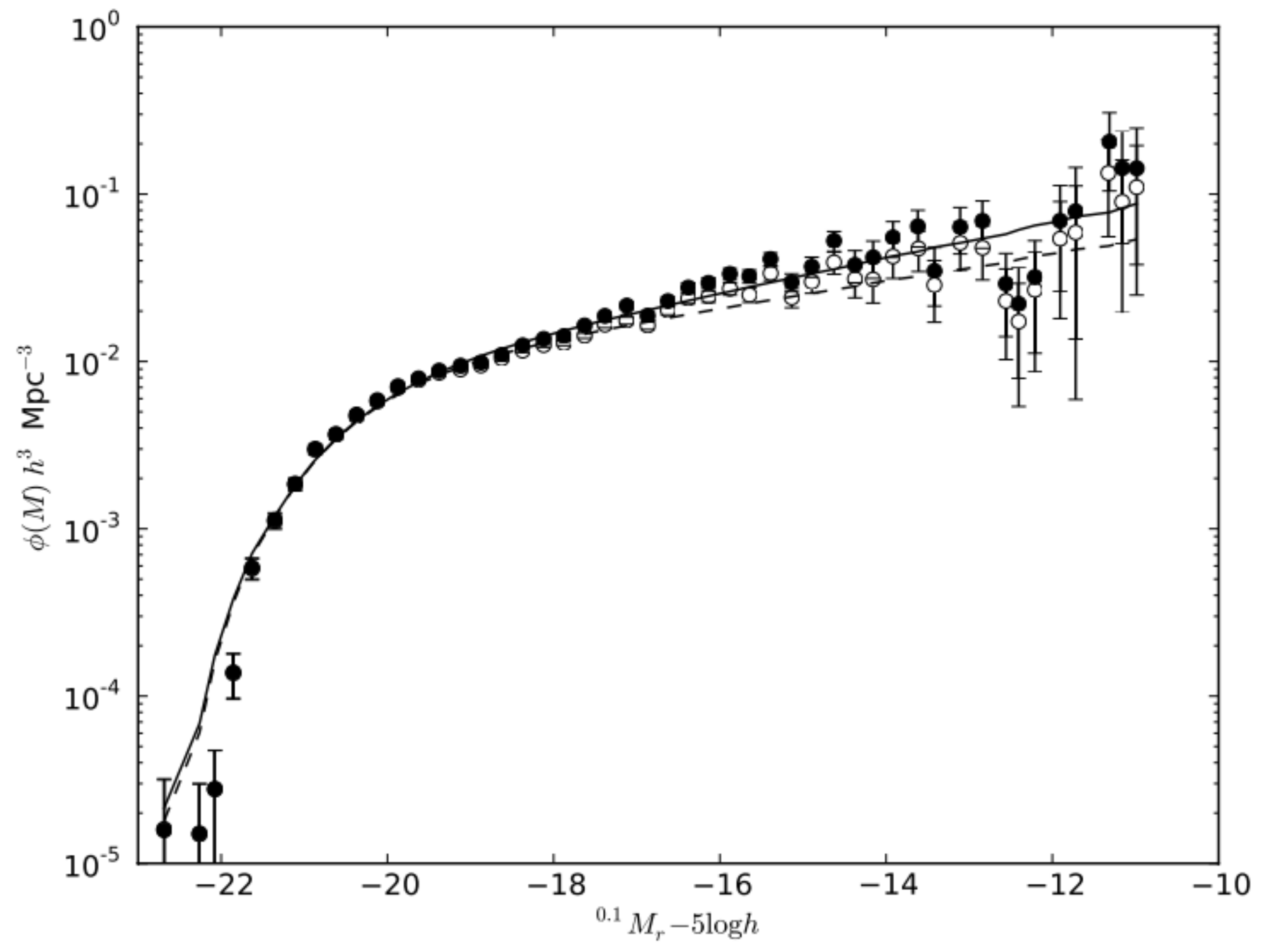, height=6cm}
\caption{Luminosity function measured for low-redshift galaxies ($z<0.1$) from
the Galaxy and Mass Assembly (GAMA) survey. Solid line shows the fit to the
Schechter function. Figure from \citep{2012MNRAS.420.1239L}.}
\label{fig:loveday}
\end{figure}

\section{From mass to light: reconciling theory with observations}

\subsection{Galaxy luminosity function}

The galaxy luminosity or stellar mass function, an example of which is shown in
fig. \ref{fig:loveday}, is usually
described by the Schechter function \citep{1976ApJ...203..297S}:
\begin{equation}
 \frac{dn}{dL}=\phi_{*}\left(\frac{L}{L_*} \right)^{\alpha}e^{-L/L_*}
\end{equation}
where $L_*$ is the cut-off luminosity. Empirically, $L_*$ is found to
correspond to a mass of about $10^{12}M_{\odot}$ at the present epoch. A typical
galactic mass is indeed expected from the
following considerations \citep{1977ApJ...211..638S}.

\begin{figure}[t]
\centering
\epsfig{file=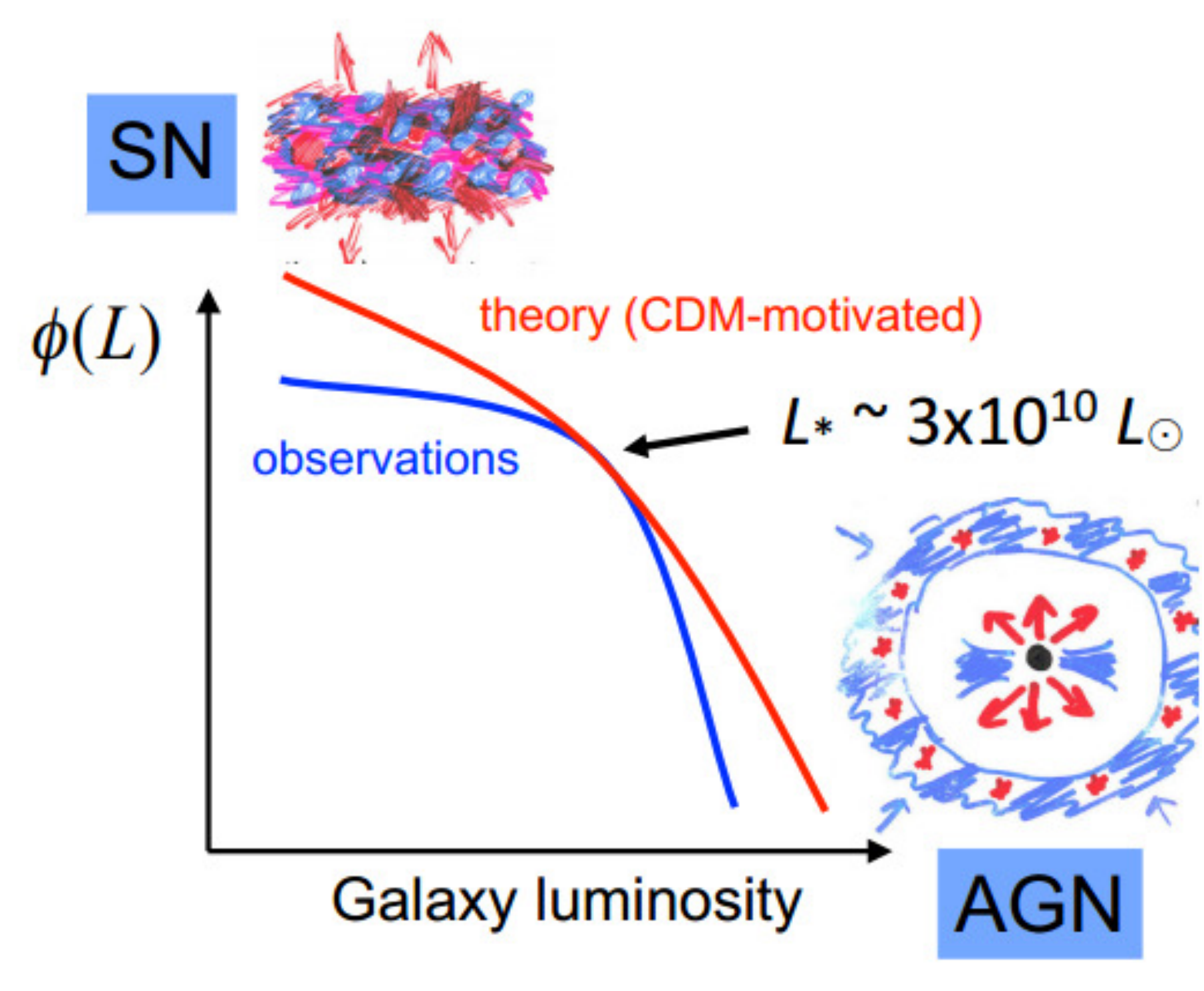, height=6cm}
\caption{The expected (red line) and observed (blue line) galaxy luminosity
function. The discrepancies in the low- and high-mass ends is probably due to
SN and AGN feedback, respectively. Figure from \citep{2012RAA....12..917S}.}
\label{fig:silk}
\end{figure}

Inside the galaxy, the gas generally resides in smaller clumps which collide at
the virial velocity. In order for the gas in these clumps to form stars it is
necessary that the cooling time of the shocked gas $t_{cool}\sim nk_B
T/\Lambda(T)$ be shorter than the dynamical time $t_{dyn}\sim
1/\sqrt{G\rho}$. We define the cooling function $\Lambda(T)\propto T^\beta$ for
the relevant virial temperature range of the halo ($10^5$-$10^7$ K), where
$\beta \simeq -1/2$ for a low metallicity plasma \citep{2007ApJS..168..213G}.
Then the ratio
$t_{cool}/t_{dyn}$ is proportional to $T^{3/2}/\rho^{1/2}$ which in turn is
proportional to halo mass. Then the typical mass of stars formed inside
the halo is given in terms of the familiar constants $\alpha$ and $\alpha_g$:
\begin{equation}
 M_{*}\simeq
m_p\frac{\alpha^3}{\alpha_g^2}\frac{m_p}{m_e}\frac{t_{cool}}{t_{dyn}}
T^{1+2\beta}\simeq 10^{12}M_{\odot}\: .
\end{equation}

In order to compare the observed luminosity function of galaxies with
theoretical predictions, one has to assume a certain mass-to-light ratio (or,
equivalently, star formation efficiency). As
shown schematically in fig. \ref{fig:silk}, a single value can be calculated so
that the observed and the theoretically predicted curves overlap at $M_{*}$.
However, their shapes are different, since stellar mass does not necessarily
follow halo mass. Assuming a universal mass-to-light ratio leads to too many
small galaxies, too many big galaxies in the nearby Universe, too few red
massive
galaxies at high redshift and too many baryons in galactic halos. There are
additional problems, such as overconcentration and excessive cuspiness in
simulated dark matter halos.

The resolution of all these problems must be related to the dynamics of baryons
within the dark matter halos, and more specifically the feedback mechanisms
that
would lower star formation efficiency on various scales. Possible sources of
feedback include supernovae, photoionization, massive stellar winds, tidal disruption, input from active galactic nuclei
and cosmic reionization. Below we will discuss some of the issues related to
feedback, focusing primarily on dwarf galaxies. A more complete treatment is
given in several recent reviews \citep{2010PhR...495...33B,
2012RAA....12..917S}.

\subsection{Supernova-driven winds}

One of the possible feedback mechanisms that may suppress star formation is
galactic winds driven by the star formation process itself
\citep{1974MNRAS.169..229L,1986ApJ...303...39D}.  After an initial
population of stars has formed, a certain fraction of those stars (depending on the
IMF) explode as supernovae, releasing large amounts of energy into the
surrounding medium. If the outflow is accelerated to a velocity that is higher
than the escape velocity of the galaxy, it is ejected into the IGM, suppressing
the star formation rate. Such outflows have been detected in many systems and
are believed to be the primary mechanism by which metals are
deposited into the IGM \citep{2005ARA&A..43..769V}.

Whether or not this process can significantly affect
star formation efficiency depends on the assumed IMF, the acceleration of the
outflow, the manner in which this excess energy is deposited into the IGM and
the depth of the potential well of the halo. Multiphase simulations have
been able to reproduce supernova-driven winds and trace the metal enrichment
processes, but their impact on the star formation efficiency in low-mass
galaxies remains unclear \citep{2003MNRAS.339..289S,
2011MNRAS.414.3671P}.

\subsection{Entropy barrier}

Very low-mass halos ($\lesssim 10^5M_{\odot}$) are not able to accrete
the
gas in the first place, because its specific entropy is too
high \citep{1986MNRAS.218P..25R}. This imposes a relatively sharp lower limit on
the mass of observable dwarfs. Cosmic
reionization reinforces this entropy barrier by heating the IGM and suppressing
gas infall onto low-mass galaxies \citep{2000ApJ...542..535G,
2008MNRAS.390..920O}.

\begin{figure}[h!]
\centering
\epsfig{file=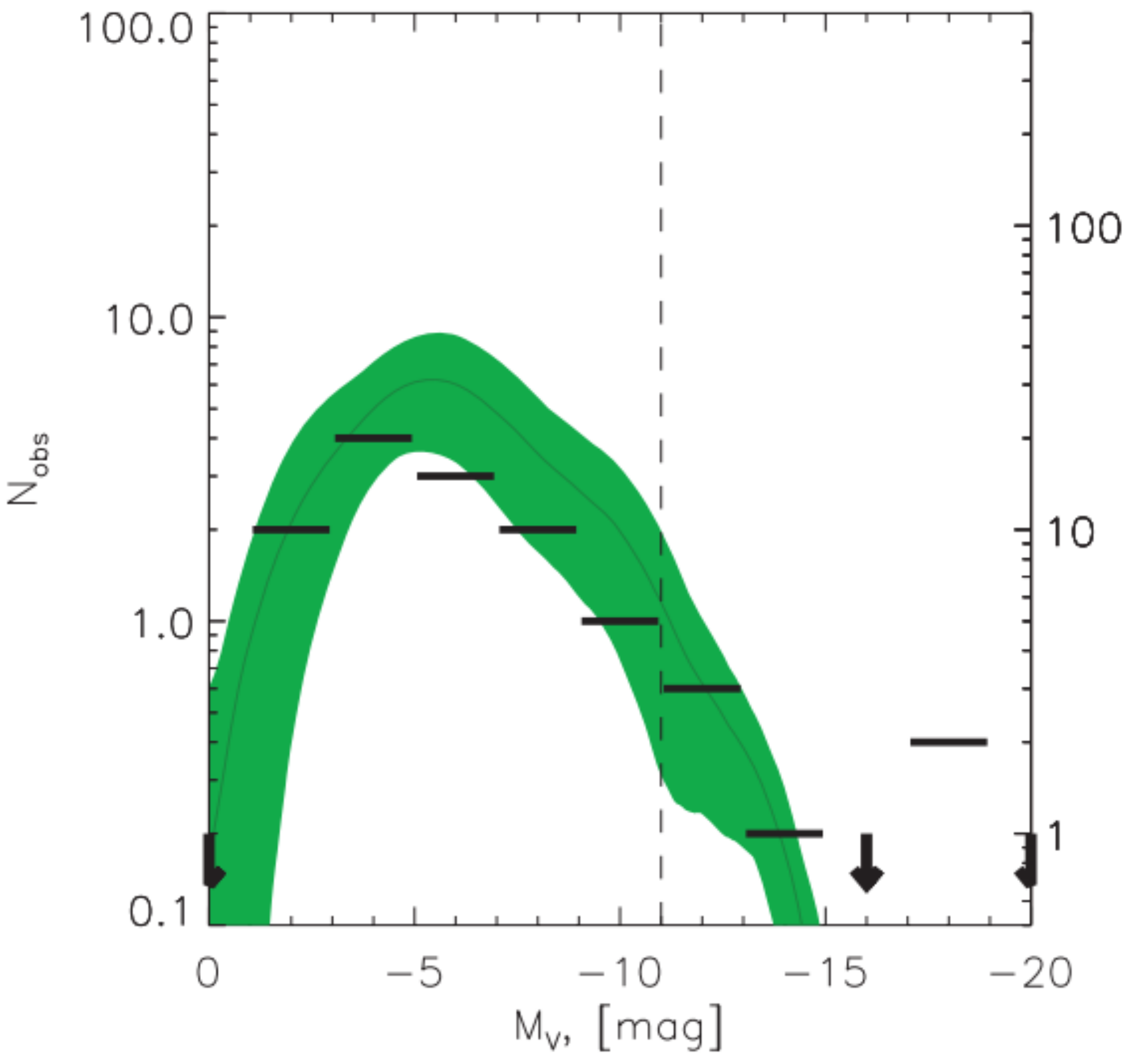, height=8cm}
\caption{MW satellite magnitude distribution (bars) and predictions from a
semi-analytic model with mass-dependent suppression of star formation. Figure
from \citep{2009ApJ...696.2179K}.}
\label{fig:koposov}
\end{figure}

Interestingly, this mechanism is a possible solution to the \emph{missing
satellite problem}, which amounts to the discrepancy between the theoretically
expected
and observed number of \emph{subhalos} in a Milky Way (MW)-like galaxy
\citep{1999ApJ...522...82K, 1999ApJ...524L..19M}. As we
saw earlier, the
spectrum of the density fluctuations continues to small scales, and the mass
function of DM halos is expected to rise steeply towards low masses. The
majority of these halos fall into more massive structures, and some are
gravitationally disrupted. However, numerical simulations predict that a
significant fraction of these subhalos survive. The problem is that the number
of subhalos produced in numerical
simulations greatly exceeds the number of observed dwarf satellites in the MW
and the Local Group.

One way to resolve this problem would be to modify the spectrum of the
primordial density fluctuations, so as to reduce the power on small scales, for
example by introducing warm dark matter. However, other possibilities
exist within the framework of CDM, such as mass-dependent suppression
of star formation. In the latter scenario, which however lacks any convincing physical mechanism, the expected satellites exist but
are not observed. Semi-analytic models of galaxy formation that account for an
entropy barrier created after reionization are able to reproduce the
observed present epoch  luminosity function \citep{2009ApJ...696.2179K}, as shown in
fig. \ref{fig:koposov}. However these models, if tuned to the faint dwarfs, underpredict the numbers of massive dwarfs.

\subsection{Tidal disruption}

\begin{figure}[t]
\centering
\epsfig{file=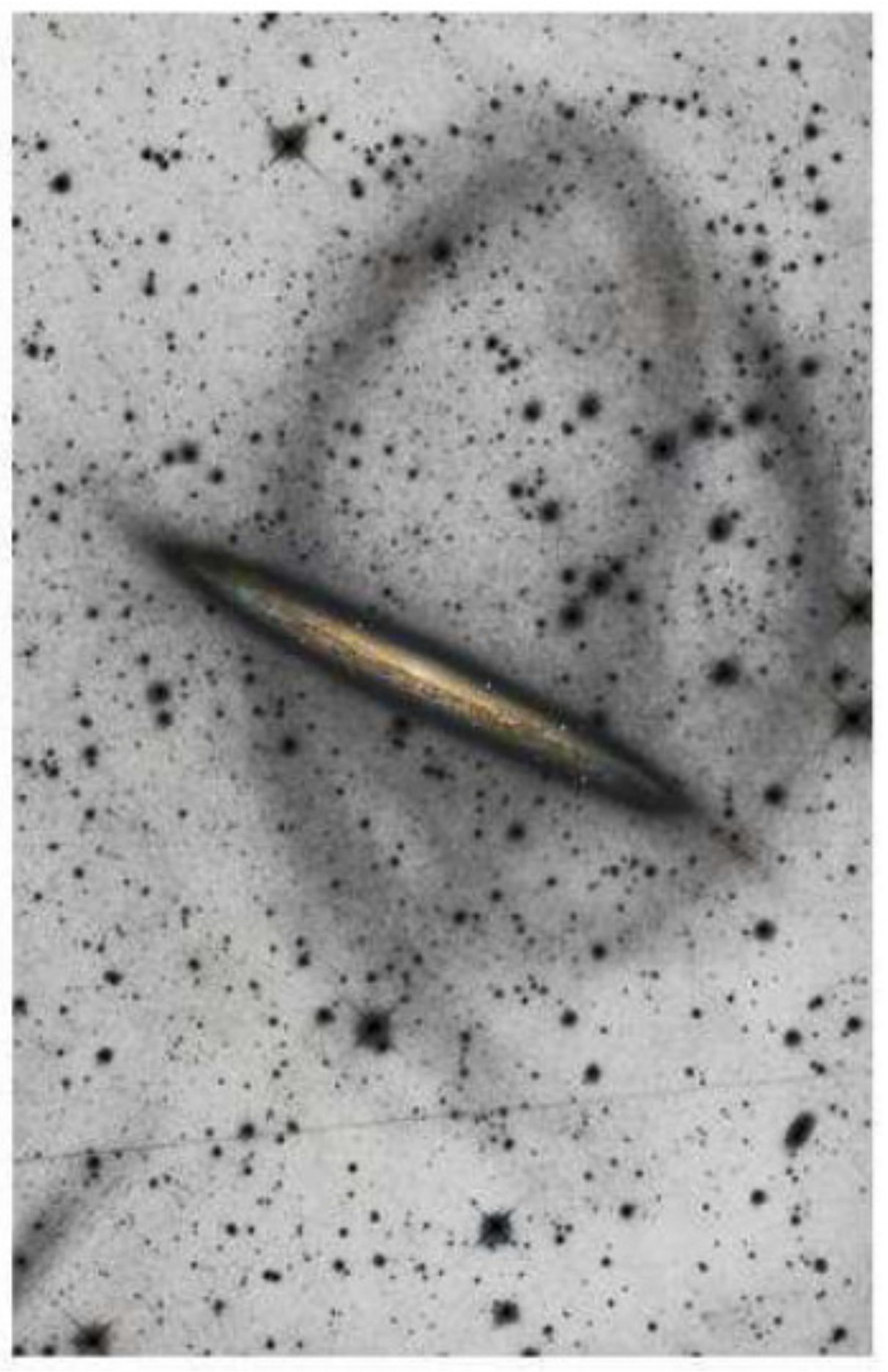, height=8cm}
\caption{Deep image of the stellar tidal stream around NGC 5907. Figure from
\citep{2010AJ....140..962M}.}
\label{fig:martinez-delgado}
\end{figure}

Some of the satellites may be tidally disrupted due to close encounters with more
massive subhalos or during infall into the main halo, if their orbits intersect
the disk or bulge. In this case, distinct stellar structures in the outskirts of
the more massive systems are expected, and these are indeed observed in deep
images of nearby galaxies, as shown in fig. \ref{fig:martinez-delgado}. The
origin of these \emph{tidal tails} is
confirmed by numerical simulations \citep{2010MNRAS.406..744C}. An independent
confirmation of the origin of these structures is the fascinating discovery of
\emph{gaps} in the tidal tails around the outermost MW globular star clusters
\citep{2009ApJ...693.1118G}, which indicate the presence of dark satellites.
This
effect is also predicted by numerical simulations \citep{2011ApJ...731...58Y}.

These \emph{galactic fossils} can shed light on the hierarchical evolution of
the host halo. The
\emph{galactic archeologist} can, in principle, reconstruct the
history of the MW by observing the surviving dwarf galaxies and the debris of
those that perished \citep{2013arXiv1307.0041B}. Both populations should be
explained by a viable theory of galaxy formation, since the same processes are
responsible for the formation of present-day as well as high-redshift dwarfs.

\subsection{The core/cusp problem}

Another problem in the standard theory of galaxy formation is related to the
density profiles of dark matter halos. Numerical N-body simulations predict a
universal profile which diverges toward the halo center as $\rho\sim 1/r$
\citep{1996ApJ...462..563N}. However, observations reveal flat central
profiles in dwarf and low surface brightness disk galaxies
\citep{2011AJ....142...24O}.

\begin{figure}[h!]
\centering
\epsfig{file=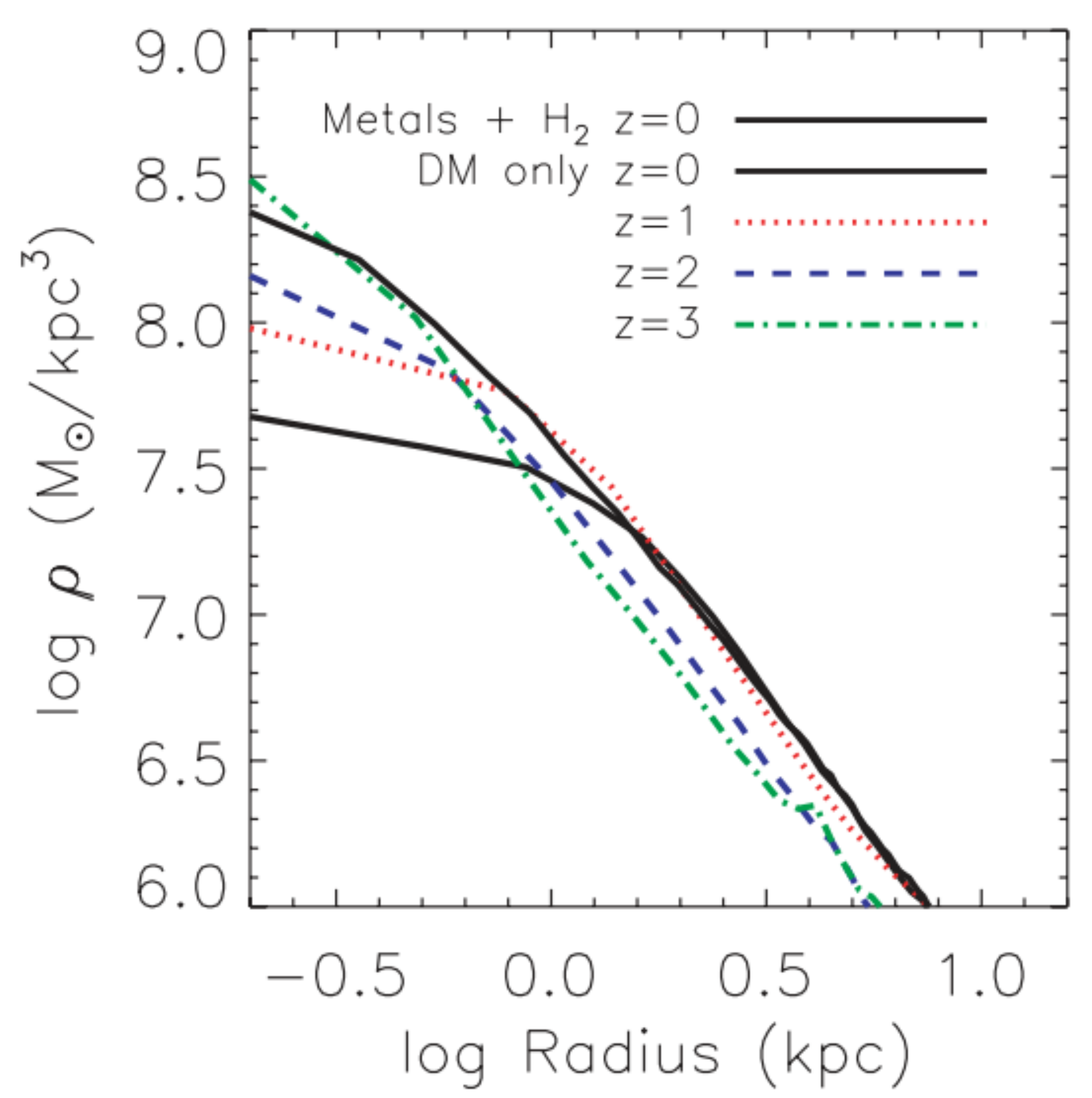, height=5cm,width=7cm}
\caption{The DM density profile of a simulated dwarf galaxy at different
redshifts. In the DM only simulation, the galaxy retains its cuspy profile,
whereas in the run with repeated outflows a shallow inner profile is formed by
$z=0$. Figure from \citep{2012MNRAS.422.1231G}.}
\label{fig:governato}
\end{figure}

A possible solution of the core/cusp problem may lie in supernova-driven
gas outflows \citep{2012MNRAS.422.1231G}. In this scenario, fast and repeated gas
outflows following bursts of star formation transfer energy to the dark matter component and flatten the density profile of the halo \citep{Pontzen12}. As shown in fig.\ref{fig:governato}, galaxies form with a steep profile which significantly flattens over cosmic time. Cusps are retained only in very small galaxies, where the star formation rates are too low to significantly modify the density profile \citep{Penarrubia12}.
Therefore, this model predicts the existence of observable small cuspy galaxies.

The situation is less clear for more massive galaxies, where the measurements of dark matter contributions to the rotation curve (and therefore to the density profile) within an effective radius or two are notoriously difficult due to uncertainties in stellar mass to light ratios, since baryons constitute a significant fraction of the total mass in the inner regions,  as well as  uncertainties in  orbits of the trace stellar probes such as planetary nebulae.

\begin{figure}[h!]
\centering
\epsfig{file=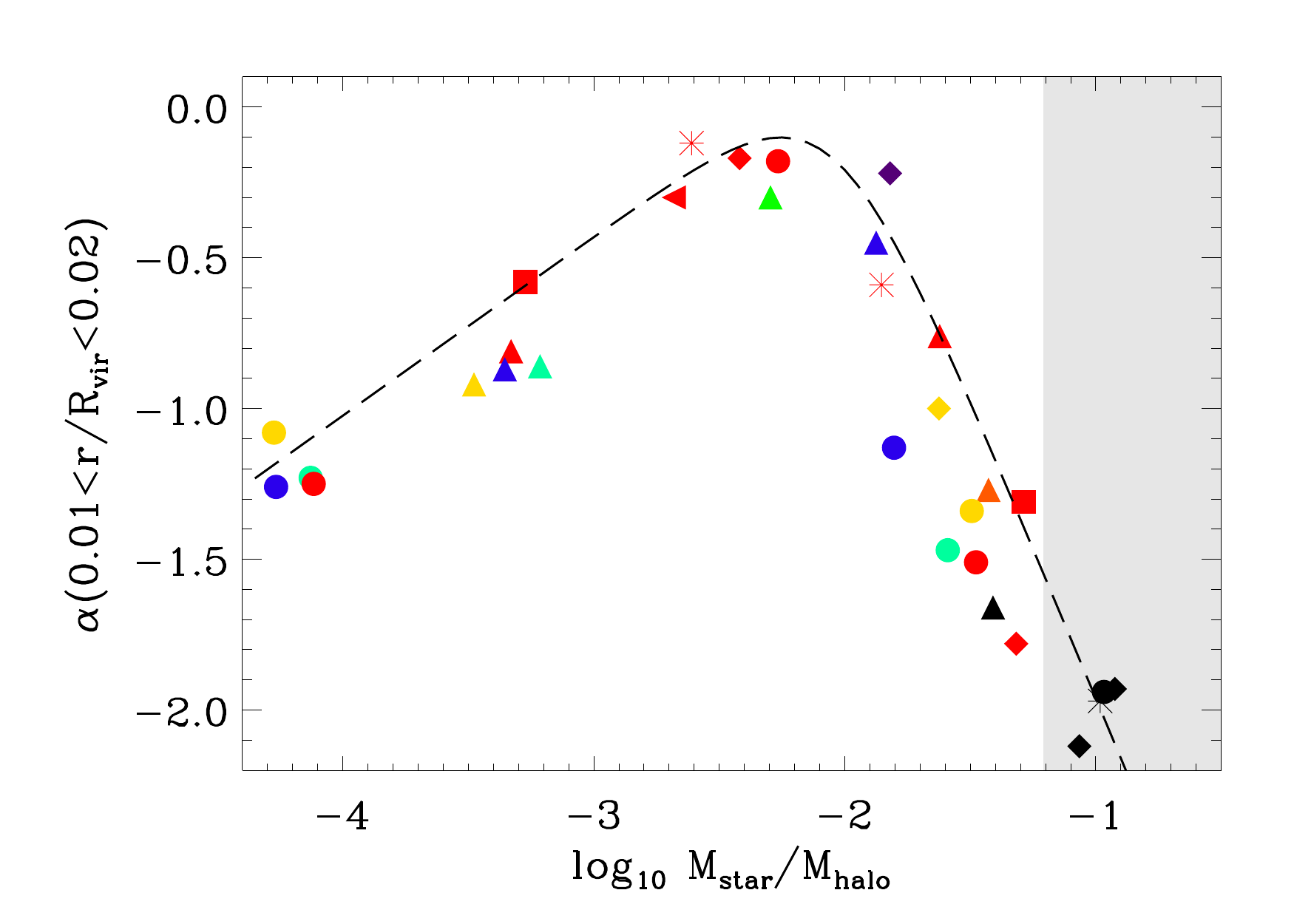, height=6cm}
\caption{The relation between dark matter density profile slope $\alpha$, measured
between $0.01<r/R_{vir}<0.02$, and the stellar-to-halo-mass ratio of each simulated galaxy. Figure from \citep{2013MNRAS.tmp.2583D}.}
\label{fig:dicintio}
\end{figure}

It has indeed been shown \citep{2013MNRAS.tmp.2583D} that at the high halo mass end, above $M_{halo}\sim10^{11.2}\Msun$, the deepened potential well induced by the greater efficiency of star formation is able to resist the expansion process, and such massive galaxies become increasingly cuspy. This paper used a suite of hydrodynamical simulations from the MaGICC project \citep{stinson13} that well reproduce many observed galaxy relationships over a wide stellar mass range \citep{brook12b}, and make  predictions for the relation between the inner slope of DM density profiles and galaxy mass: they found, in particular, that the effect of baryons on the dark matter halo depends on the integrated efficiency of star formation $M_{\star}/M_{halo}$ in the way shown in fig.\ref{fig:dicintio}. At the low mass end, their prediction is in agreement with previous studies, and at the high mass end the profiles steepen again due to the increased stellar mass at the galaxy center that oppose the expansion process. The flattest slope is found at $M_{halo}\sim10^{10.8}\Msun$, or $V_{rot}\sim\rm 50\,km/s$ in agreement with the most reliable observational measurements of cored profiles in disc galaxies, found in low surface brightness (LSB) galaxies with $V_{rot}<100\rm \,km/s$ \citep{oh11b}.

\subsection{Cosmic reionization by dwarf galaxies}

\begin{figure}[hb!]
\centering
\epsfig{file=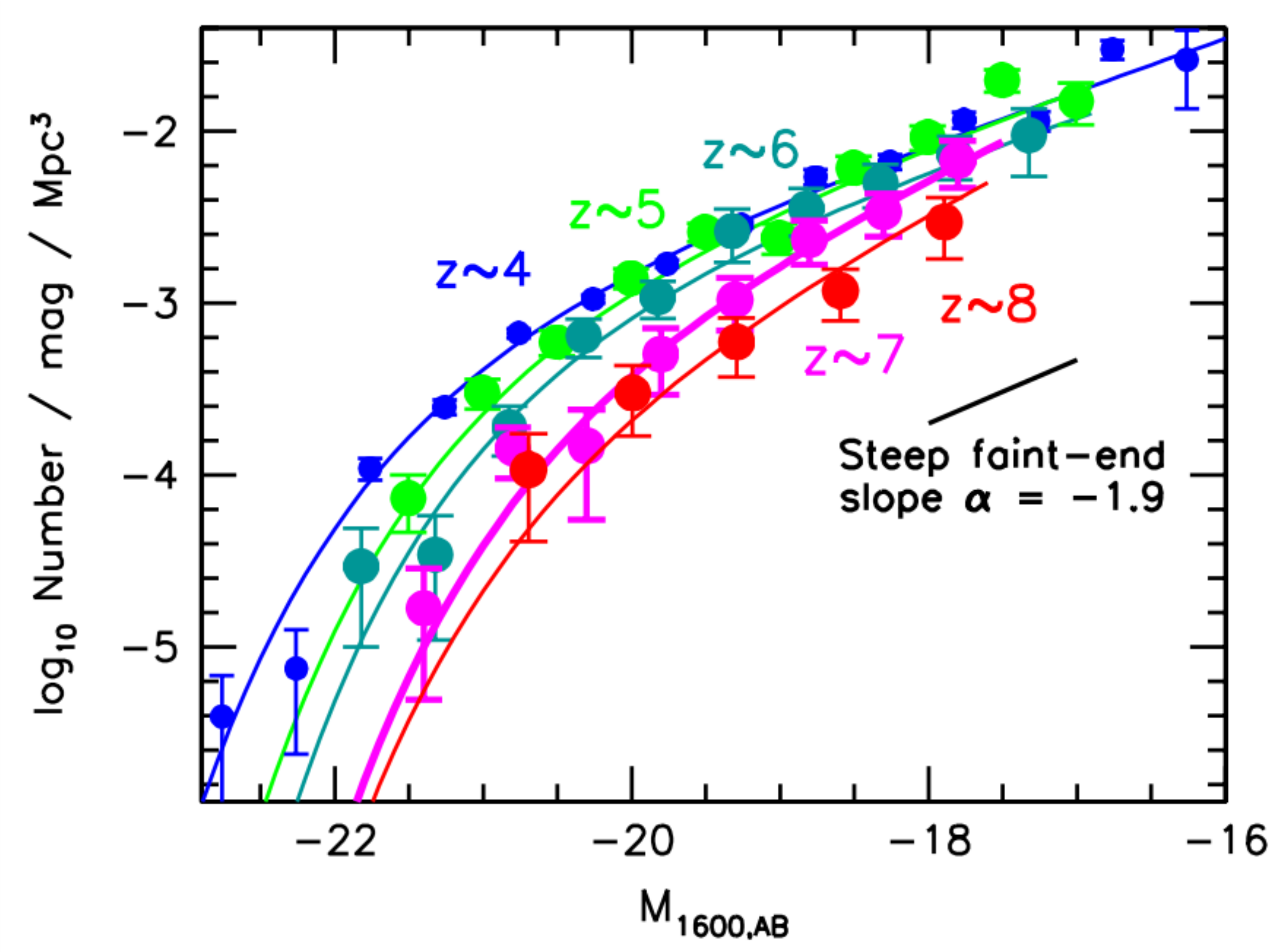, height=5cm,width=8cm}
\caption{UV luminosity functions from Hubble Ultra Deep Field observations. Figure from \citep{2012ApJ...752L...5B}.}
\label{fig:bouwens}
\end{figure}

The very steep slope of the faint end of the galaxy luminosity function, as
measured in very deep surveys \citep{2012ApJ...752L...5B},  raises the possibility
that the universe was reionized by dwarfs (see fig. \ref{fig:bouwens}).
Star-forming galaxies are the most
obvious source of ionizing photons, and it seems plausible that
dwarfs, which outnumber large luminous galaxies, could produce
enough UV photons to reionize the Universe
\citep{2012ApJ...752L...5B,2013ApJ...773...75O}. Under reasonable assumptions
about the escape fraction of photons and the clumping factor of the
intergalactic medium, and the requirement that a low level of star formation
extends out to redshift $\sim 12$, the optical depth to reionization by early
low-luminosity galaxies as seen to $z\sim 8$  is marginally  consistent with the constraints of cosmic microwave
background measurements \citep{2013ApJ...768...71R}.

However observations suggest that the UV rest-frame stellar luminosity density
may drop sharply beyond $z\sim 8$ \citep{2013arXiv1309.2280O}, which would
reduce the possible contribution of ionizing photons at the epoch when they are
most needed.

%%%%%%%%%%%%%%%%%%%%%%%%%%%%%%%%%

\section{Computational methods in galaxy formation}
The cosmological growth of density fluctuations, which will end up forming galaxies, is a highly non-linear process which is impossible to treat analytically. Apart from the more empirical approach, called "halo-occupation-distribution", two main techniques have been developed in the past decades to solve this issue.
On the one side, semi-analytical models (SAMs) try to address the non-linear growth of structures using approximate, analytic techniques. On the other side, hydrodynamical simulations directly address a wide range of dynamical scales and solve numerically the combined non-linear N-body and hydrodynamic equations describing the formation of a galaxy.

The main advantage of the semi-analytic approach is that it is computationally inexpensive compared to N-body simulations and it is therefore easy to address the relative importance of the different physical processes involved, simply by turning them off in the model and looking at the outcome: this allows a rapid exploration of parameter space with respect to N-body simulations \citep{2012NewA...17..175B}. The weakness of SAMs is that the majority of the physics is controlled by hand, and that the codes are so complex to include a very high number of non-cosmological parameters, leading to a great degree of uncertainty.

With N-body and hydrodynamical simulations, instead, the basic equations of gravity and hydrodynamics are solved numerically in a much more consistent way. Pure N-body, dark matter only simulations have been found extremely effective at reproducing the large scale structure of the Universe, and powerful tools have been developed to deal with them, such has halo finders \citep{Knollmann09, 2004MNRAS.351..399G} and dark matter halo merger-tree algorithms \citep{2011MNRAS.412..665D}; at small scales, however, baryonic physics must be taken into account.
The main disadvantage of hydrodynamical simulations, apart from being computationally heavy, is that the vast range of dynamical scales between megaparsecs and astronomical units cannot yet be addressed numerically in a coherent fashion.  The best scales achievable to date are at the level of parsecs. All the relevant physics happening at scales below the resolution of the simulation are put in by hand using a semi-analytic approach: this is the case of the complex processes of star formation and feedback from supernovae and AGN, whose treatment is at the "sub-grid" level.

Hydrodynamical simulations are essentially divided into two main branches, grid-based and particle-based. While grid-based codes cannot resolve at the same time large, cosmological and small, galactic scales, the particle-based methods, and in particular the smoothed particle hydrodynamics ones (SPH), do not suffer from such a problem (GADGET \citep{2005MNRAS.364.1105S}, GASOLINE \citep{wadsley04}). SPH methods, however, fail to resolve shocks and Kelvin-Helmholtz instabilities. An improvement on grid-based codes is the Adaptive Mesh Refinement (AMR) scheme, in which cells can be adaptively refined according to a density criterion: schemes with deformable cells are becoming more widely used nowadays (RAMSES \citep{Teyssier02}, AREPO \citep{2010MNRAS.401..791S}).

\begin{figure}[h!]
\centering
\epsfig{file=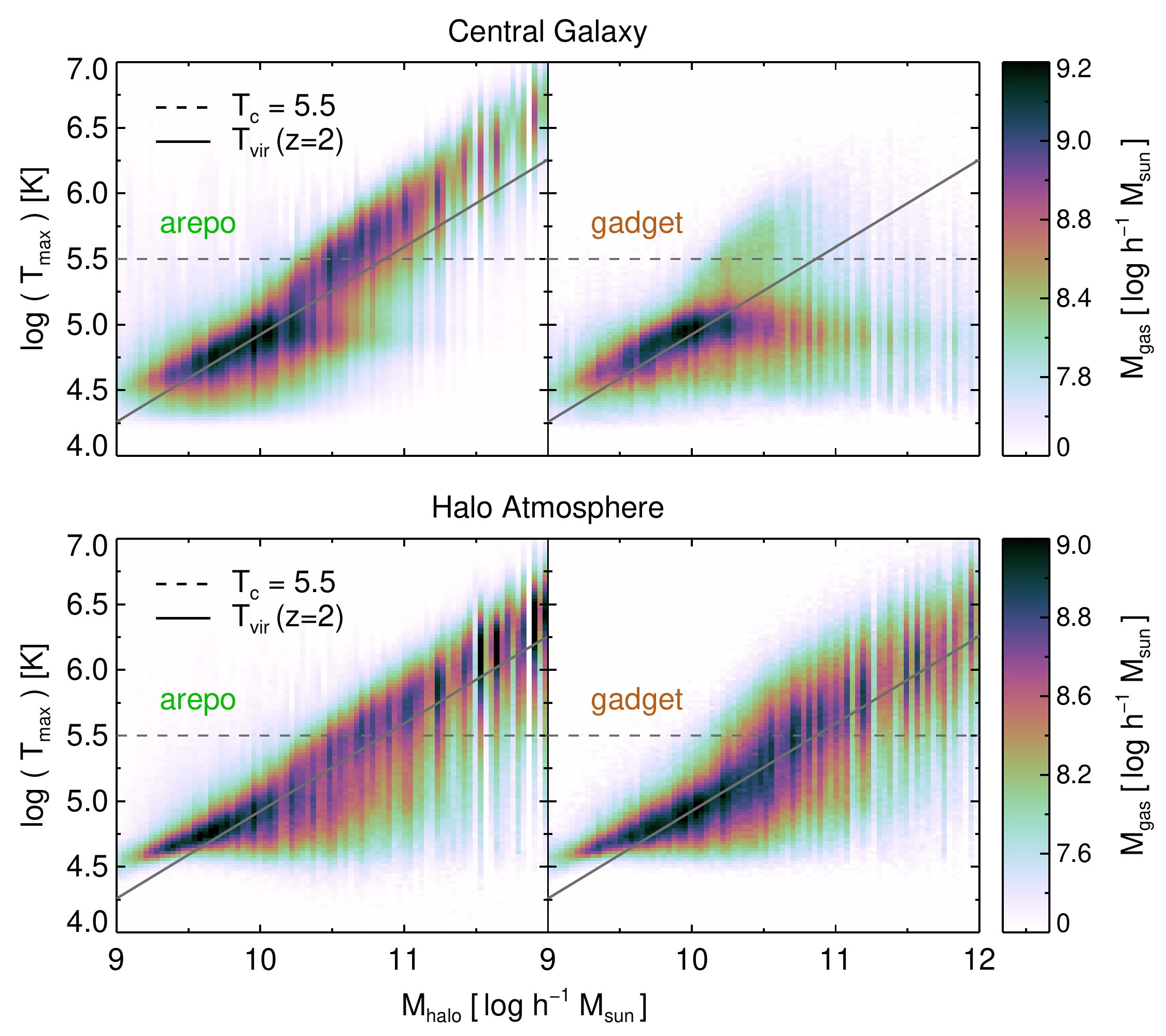, height=9cm}
\caption{The mass-weighted histogram of past maximum temperature $T_{max}$ for gas accreted onto central galaxies (top) and halo atmospheres (bottom) by z=2, as a function of the parent halo mass at z=2. AREPO and GADGET-3 are shown separately in the left and right panels, respectively. It is clear how the two codes deviate strongly for galaxies in haloes with masses above $10^{10.5}\Msun$, where the strong contribution of hot gas in AREPO, which scales approximately with $T_{vir}$, is mostly absent in GADGET-3. The cold $10^5K$ gas which dominates the accretion in GADGET-3 galaxies is largely absent in the AREPO simulation. Figure from \citep{2013MNRAS.429.3353N}.}
\label{fig:arepo}
\end{figure}

Highlighting the differences produced by the above mentioned techniques has  been a topic of  considerable interest, and simulators are still in the process of  converging on such key questions as how much hot versus cold gas is accreted at differing scales and times.
A summary of recent results on the accretion modes of gas in galaxies within hydrodynamical cosmological simulation of different types, specifically particle-based or grid-based codes, includes the following:
\begin{itemize}
\item \citep{2013MNRAS.429.3353N} investigated the nature of gas accretion within galaxies using cosmological hydrodynamic simulations run with the moving mesh code AREPO \citep{2010MNRAS.401..791S}
\item and with the smoothed particle hydrodynamics code GADGET-3 (an improved version of the publicly available GADGET-2  \citep{2005MNRAS.364.1105S})
\item and found that the fraction of gas supplied to galaxies in massive halos through cold flows is less in the AREPO simulation than in the GADGET-3 simulations, because with traditional SPH, much of the cold gas reaches the central galaxies in purely numerical "blobs". Their results are shown in fig. \ref{fig:arepo}.
The authors explain the discrepancies as due to numerical inaccuracies in the standard formulation of SPH. Such inaccuracies, however, can be overwhelmed within more refined SPH schemes.
\end{itemize}

Another indicator that all is not well in the galaxy formation simulation community is that there is still a serious overcooling problem. This has been with us for a very long time. Essentially, gas clouds collapse and star formation occurs too early. The prevalent solution is supernova feedback, but this fails unless unto $\sim 300\%$ of the supernova energy is tapped as kinetic energy to heat the gas. Alternatively, cooling is arbitrarily turned off within a few million years after the supernova explodes. New formulations of feedback are being explored that, for example, include cosmic-ray pressure in addition to thermal pressure in order to drive gas outflows \cite{2013arXiv1307.6215S, 2013ApJ...777L..16B} or refine feedback on GMC scales by resolving the effects of grain photoheating and radiation pressure \cite{2013arXiv1311.2073H}. \cite{2013arXiv1310.7572S} studied the impact of radiation pressure and photoionization feedback from young stars on surrounding gas and found that the latter is the dominant effect: in this sense the thermal pressure implementation of such "early stellar feedback" as in \citet{stinson13} seems promising.

\section{A global star formation law}
\begin{figure}[h!]
\centering
\epsfig{file=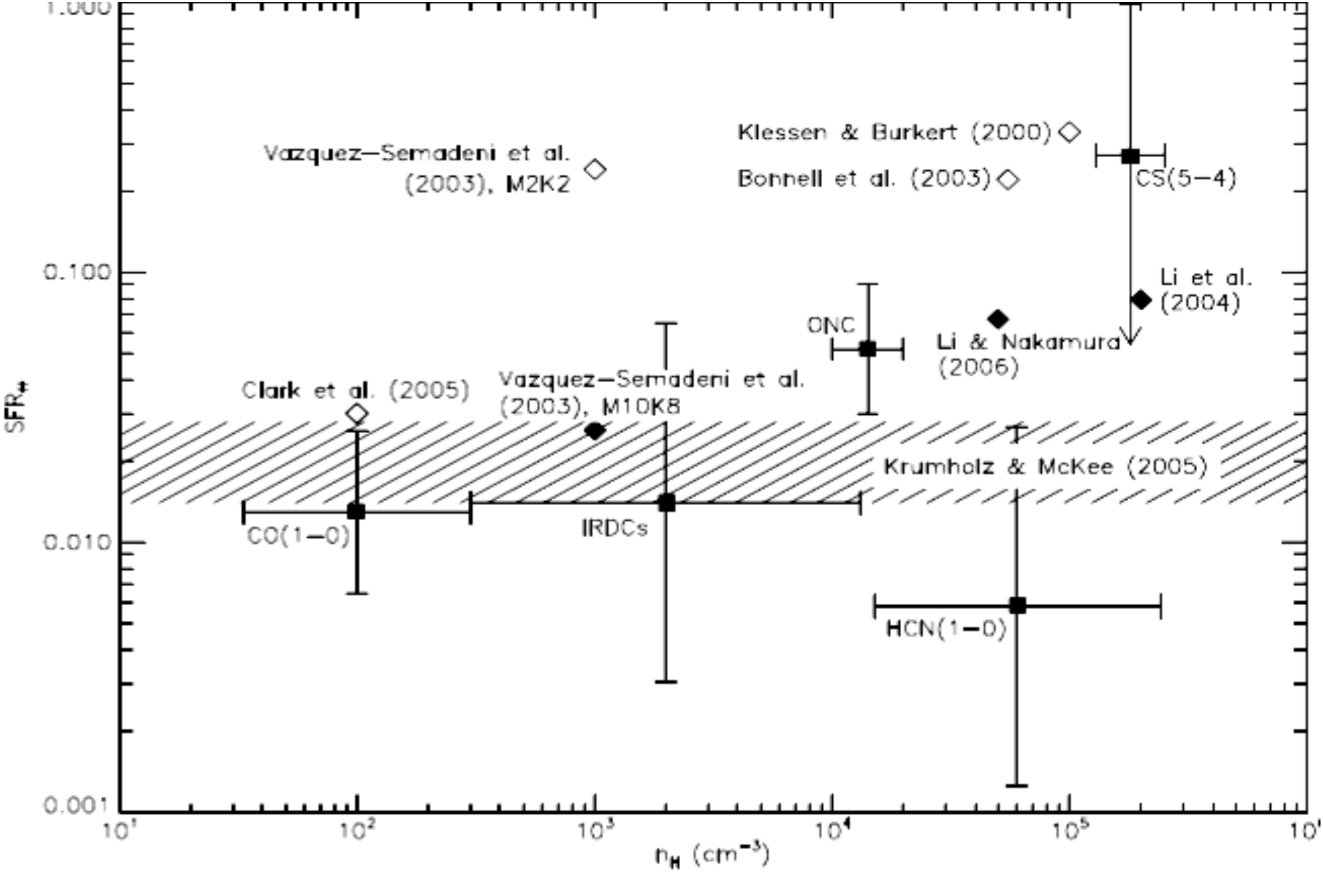, height=7cm}
\caption{Observed star formation efficiency per dynamical time as a function of mean gas density. Each data point represents a different method of measuring the gas, which is sensitive to different densities. GMC indicates giant molecular clouds, IRDCs indicates infrared dark clouds, ONC is the Orion Nebula cluster, HCN represents extragalactic measurements. Figure from \cite{2007ApJ...654..304K}.}
\label{fig:SFE}
\end{figure}

Stars form in giant molecular clouds, and feedback from massive stars within
these clouds results in a star formation efficiency (SFE) of about 2\%: only in
the dense molecular cores does the SFE rise to about  $\sim 30$\%, as  confirmed
with the prestellar core mass function.
In other words, there is a constant fraction of molecular gas turned into stars per free-fall timescale, as in fig.\ref{fig:SFE}.  A similar global  SFE is observed in star-forming disk galaxies, in which the star formation rate SFR can be described by:
\begin{equation}
SFR=SFE\rho_{gas}t_{dyn}^{-1}
\label{eq:global}
\end{equation}
where $\rho_{gas}$ is the gas density and $t_{dyn}$ is the dynamical time of the rotating disk.

\begin{figure}[h!]
\centering
\epsfig{file=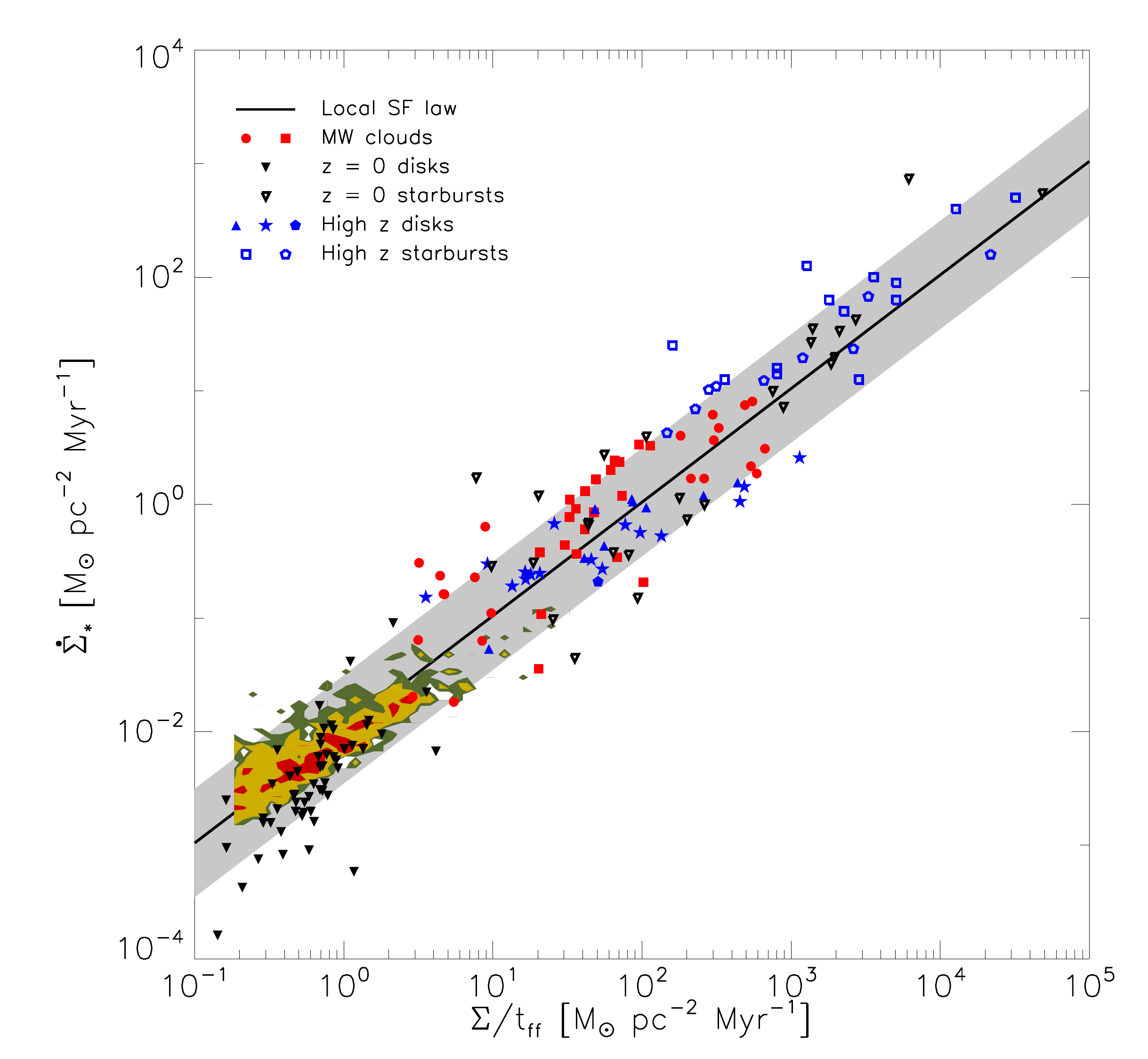,
width=9cm}
\caption{Star formation surface density versus gas surface density per dynamical time. The slope of the solid line represents the star formation efficiency SFE. Figure from \citep{2012ApJ...745...69K}}
\label{fig:krumholz}
\end{figure}

The reason for these relations resides in the gravitational instability of cold disks, while feedback physics, such as supernovae-driven turbulence, provides the observed efficiency of $\sim2\%$ which reproduces the normalization, SFE, of the global star formation law in eq.\ref{eq:global}.
The star formation rate per unit area, $\Sigma_{SFR}$, obeys the Kennicutt-Schmidt (KS) law, which can be expressed as:
\begin{equation}
\Sigma_{SFR}=SFE\frac{\Sigma_{gas}}{t_{dyn}}
\end{equation}
where $\Sigma_{gas}$ is the surface density of gas. Such a global law applies also to starburst galaxies, as in fig.\ref{fig:krumholz}

It is a remarkable coincidence that the SFE observed in giant molecular clouds is similar to that seen globally in nearby (as well as in distant) disk galaxies.
Massive OB stars provide a common link, but grain photoheating, winds and photo-ionization  dominate in the former case, and SNe in the latter case.

\begin{figure}[hb!]
\centering
\epsfig{file=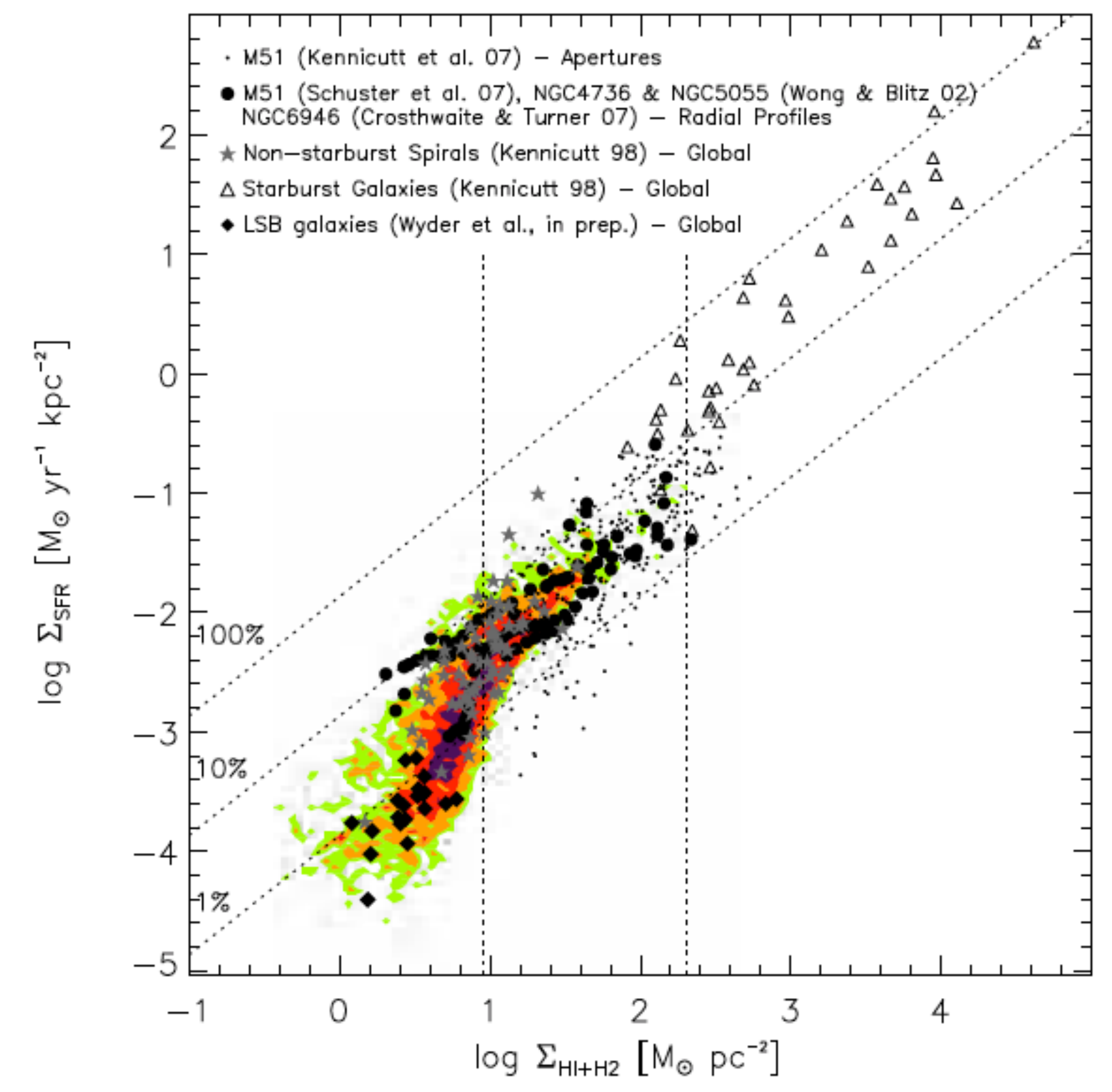, height=7cm,width=11cm}
\caption{$\Sigma_{SFR}$ vs $\Sigma_{HI+H_2}$. Diagonal dotted lines show lines of constant star formation efficiency SFE. Figure from \citep{2009eimw.confE..12B}.}
\label{fig:bigiel}
\end{figure}

An investigation of the KS law reveals that the key ingredient that regulates star formation is molecular gas, H$_2$, with an evident "knee"
in the $\Sigma_{SFR}-\Sigma_{HI+H_2}$ distribution at the transition point from a HI to an H$_2$-dominated interstellar medium, as in fig.\ref{fig:bigiel}.
Because of the saturation of $\Sigma_{HI}$ at $\sim 9\rm \Msun pc^{-2}$, this quantity as well as the total $\Sigma_{HI+H_2}$ cannot be used to predict either $\Sigma_{SFR}$ or the SFE in spiral galaxies \citep{2009eimw.confE..12B}.
Indeed, in the outer parts of galaxies, where the molecular gas H$_2$ is reduced due to the UV radiation field and lower surface density, the star formation rate per unit gas mass also declines.

Disk instabilities result in cloud formation and subsequent star formation, and one needs to supply cold gas in order to maintain such a cold disk. There is evidence for spiral galaxies to have reservoirs of HI in their outer regions, for example in NGC 6946 \citep{2008A&A...490..555B} and UGC 2082 \citep{2011A&A...526A.118H}, pointing to recent gas accretion.
In particular, the deep neutral hydrogen survey HALOGAS with WSRT,  presented in \citep{2011A&A...526A.118H},  has the goal of revealing the global characteristics of cold gas accretion onto spiral galaxies in the local Universe. Recent examples of extraplanar and HI gas reservoirs are \cite{2013A&A...554A.125G, 2012ApJ...760...37Z, 2011ApJ...740...35Z}.

\section{Disk galaxies}
\subsection{Formation of disk galaxies and spin alignment}

The disk galaxy formation process originates from the conservation of angular momentum from collapsing gas in extended dark matter haloes: the collapse stops once the system becomes rotationally supported.

The dimensionless spin parameter of a protogalactic system is
\begin{equation}
\lambda=J|E|^{1/2}G^{-1}M^{-5/2}
\end{equation}
where $J$, $E$ and $M$ are the total angular momentum, energy and mass of the system. The angular momentum of the gas arises in the same way as that of dark matter, from tidal torques from the surrounding large scale structure.
Hoyle \citep{Hoyle49} was the first to propose that protogalaxies acquired their angular momentum via tidal torques from neighbouring perturbations during a period of gravitational instability.

Peebles \citep{1969ApJ...155..393P} analyzed the tidal torque theory using linear perturbation theory and showed that it could account, roughly, for the angular momentum of the Milky Way.
Tidal torque theory was first applied to the angular momentum of disks in \citep{1980MNRAS.193..189F} and reviewed in \cite{2012ApJS..203...17R}.

\begin{figure}[hb!]
\centering
\epsfig{file=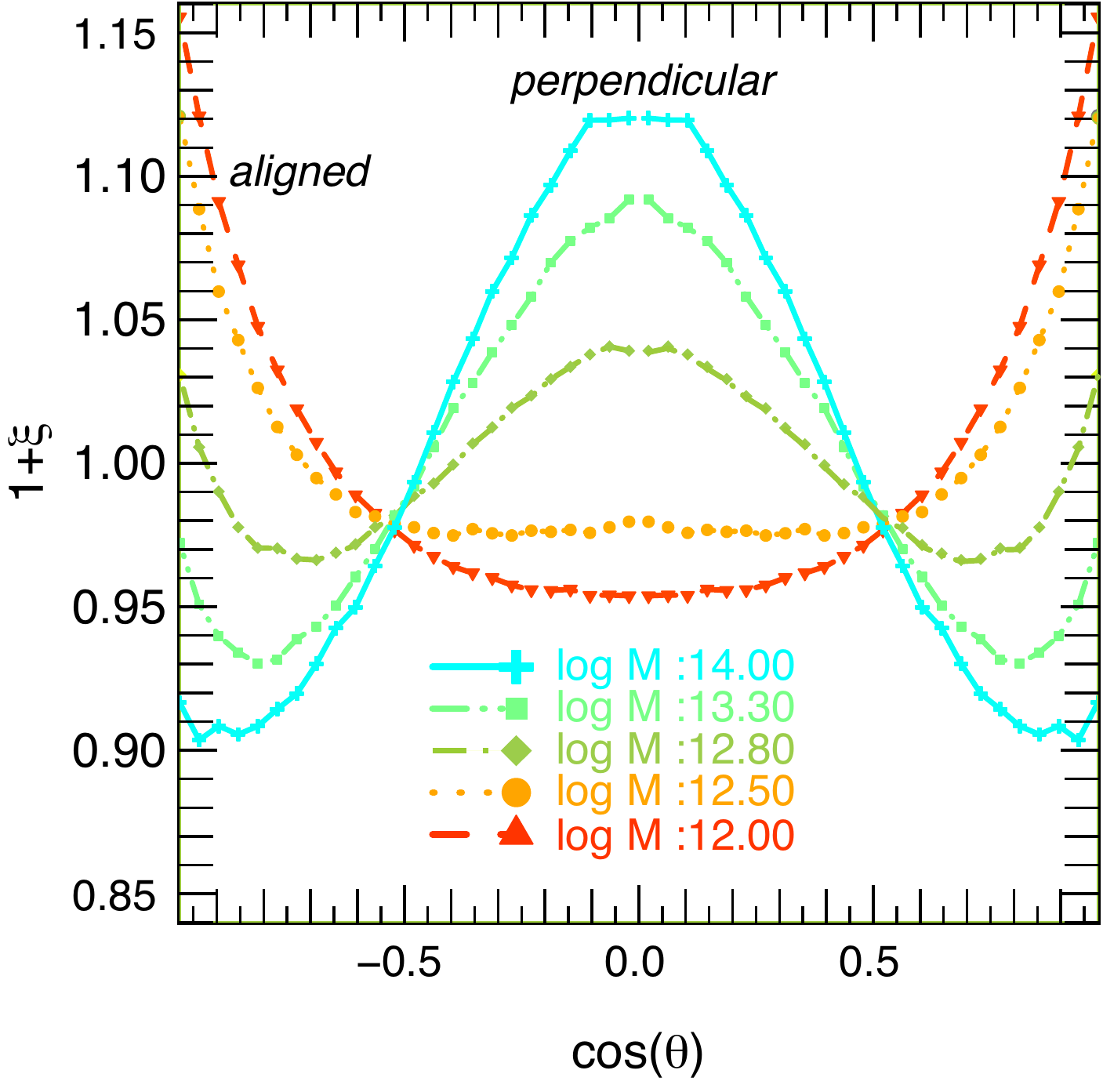, height=8cm}
\caption{Excess probability of alignment between the spin and the direction of the closest filament. Different colors correspond to different mass bins as labeled. Figure from \citep{Codis12}.}
\label{fig:codis}
\end{figure}

A topic that has recently seen a revival in interest is  the study of correlations between the spin of dark matter
halos and their large-scale environment, specifically the large-scale filamentary structure of the cosmic web.
\citep{Codis12} investigated the alignment of the spin of dark matter halos relative to the surrounding  filamentary structure, using a dark matter-only simulation which resolves over 43 million dark matter halos at redshift zero.
She detected a clear mass transition: the spin of dark matter halos above the critical mass $M_{crit}\sim5\cdot10^{12}\Msun$ tends to be perpendicular to
the closest large scale host filament, whereas the spin of lower mass halos is more likely to be aligned with the closest filament, see fig.\ref{fig:codis}.

The proposed explanation is that low mass halos mostly form
at high redshift within the filaments generated by colliding-collapsing walls, a process that naturally produces a net halo spin parallel to the filaments.
In contrast, high-mass
halos mainly form by merging with other halos along the
filaments at a later time when the filaments are themselves
colliding and/or collapsing. Therefore they acquire a spin which
is preferentially perpendicular to these filaments.
Hence, the correlations measured in \citep{Codis12} can
be understood as a consequence of the dynamics of large-scale
cosmic flows.

In the context of high redshift galaxy formation, this paper argued that galaxies form preferentially along filaments, and that the main nodes of the cosmic web are where galaxies migrate, not where they form.
Consequently, these galaxies would inherit the anisotropy of their birth place as spin orientation.

\subsection{Bulgeless disk galaxies}
\begin{figure}[h!]
\includegraphics[width=7.0cm,height=5.0cm]{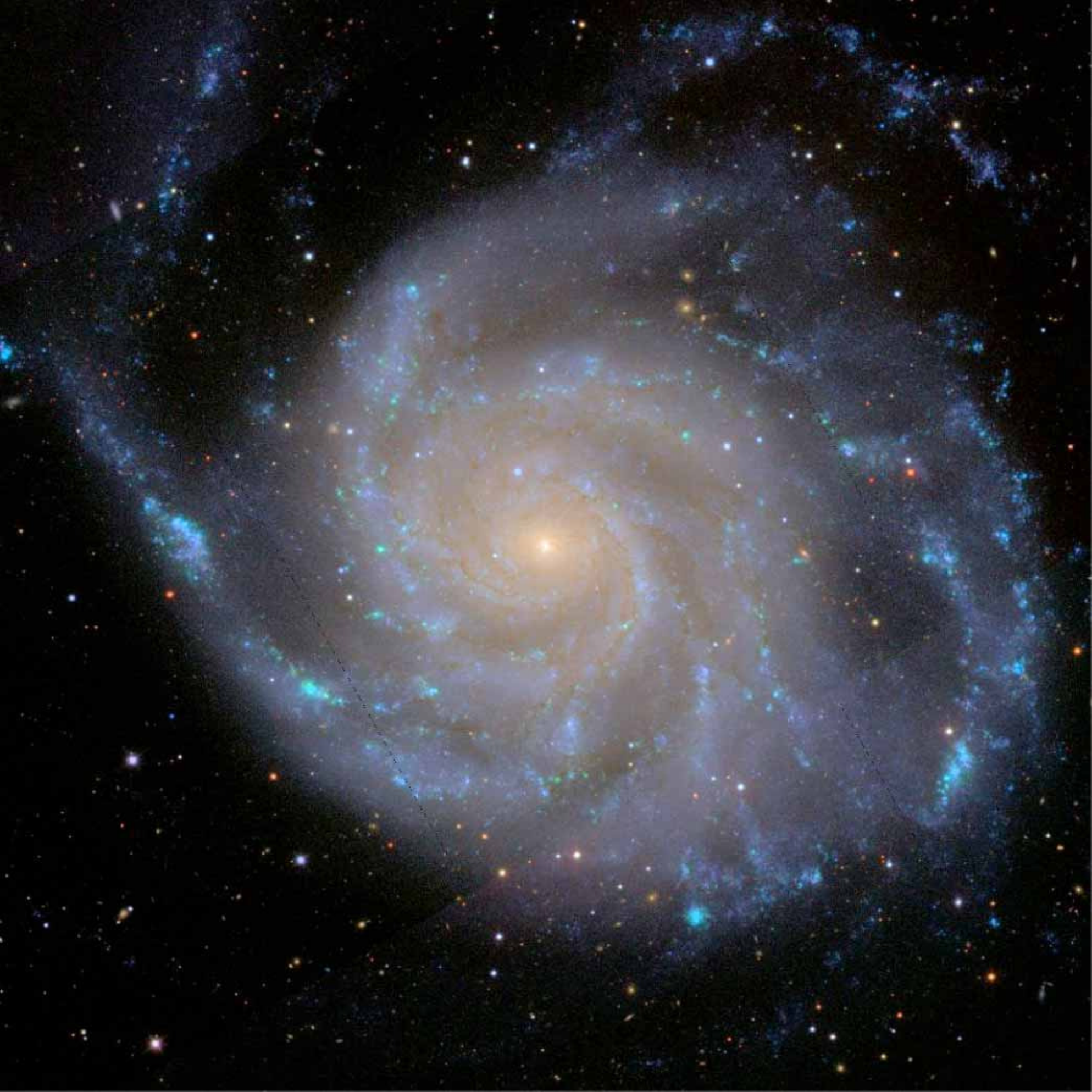}\includegraphics[
width=7.0cm,height=5.0cm]{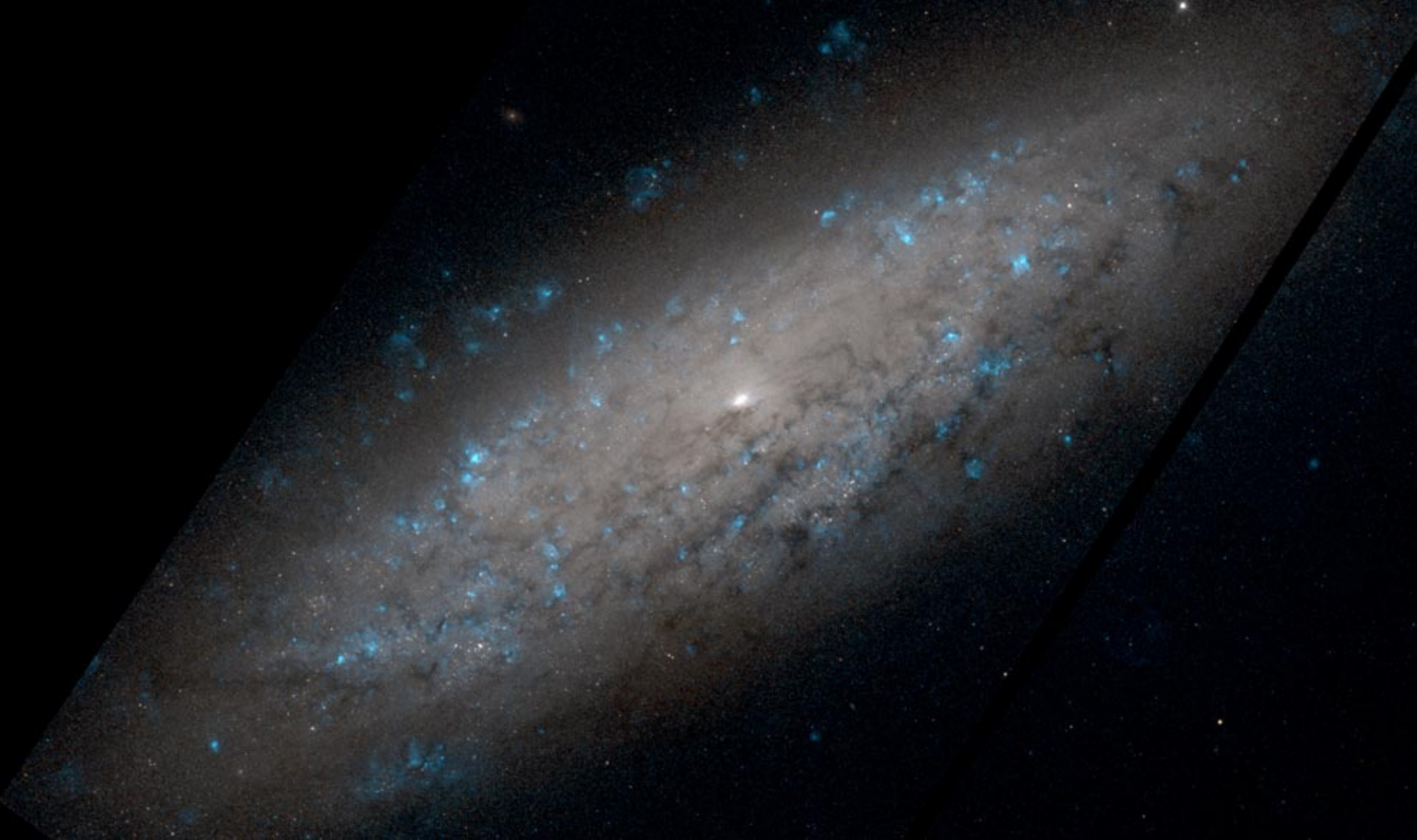}
\caption{Left: SDSS color image of NGC 5457,emphasizing how much this giant galaxy is dominated by its disk; the tiny, bright center is the pseudo-bulge. Right: Color image of NGC 6503, a slightly smaller galaxy, taken with the Hubble Space Telescope Advanced Camera for Surveys. The central tiny pseudo-bulge makes up $0.11\%$ of the I-band light of the galaxy. Figure from \citep{Kormendy10}.}
\label{fig:kormendy}
\end{figure}

Giant pure-disk galaxies are a challenge to our understanding of galaxy formation.
Reference \citep{Kormendy10} studied several pure disk galaxies, two of which are shown in fig.\ref{fig:kormendy}, concluding that, in the nearby field, much of the stellar mass is in pure disks.
The problem is that is difficult to understand how, within a hierarchical halo growth model, such galaxies could have formed, without converting any preexisting stellar disk into a classical bulge. Angular momentum loss and spheroid formation is inevitably found in galaxy formation simulations. This problem may, with extreme feedback, be avoided  in dwarf galaxies \cite{2011MNRAS.415.1051B}.
In fact, these massive galaxies only show a pseudo-bulge, which is presumably created by secular evolution of isolated galaxy disks.
Massive pure disk galaxies are not rare enough to be explained as mergerless galaxies.
It does not seems likely that physical processes such as energy feedback, are able to sufficiently  delay
star formation and thereby allow the halo to grow without forming a classical bulge, as might provide a solution to the problem \citep{Kormendy10}.

In the Virgo cluster, more than  $60\%$ of the stellar mass
is in elliptical galaxies and some additional mass is in classical bulges, whereas in Local-Group-like environments apparently the majority of galaxies with halo $v_{circ}>150\rm \,kms^{-1}$ form with no sign of the major merger that would have formed a classical bulge: \citep{Kormendy10} emphasizes that the problem of the formation of bulgeless giant galaxies appears to be a strong function of environment.

\section{The role of AGN}
\subsection{Feedback from supermassive black holes}

In order to solve the "overcooling" problem in massive galaxies (given the high rate at which gas can cool within galaxies, they should be, at the present epoch, much more massive and luminous than observed) some source of heating of the cold gas is needed.
While the evolution of low mass galaxies is mainly driven by feedback from SNe explosions, this type of feedback has little impact on the formation of massive galaxies.
An energy budget analysis suggests that a possible source of feedback in such massive galaxies is AGN feedback from supermassive black holes (SMBHs), which release an amount of energy up to a factor 20-50 higher than from SNe.

\begin{figure}[h!]
\centering
\epsfig{file=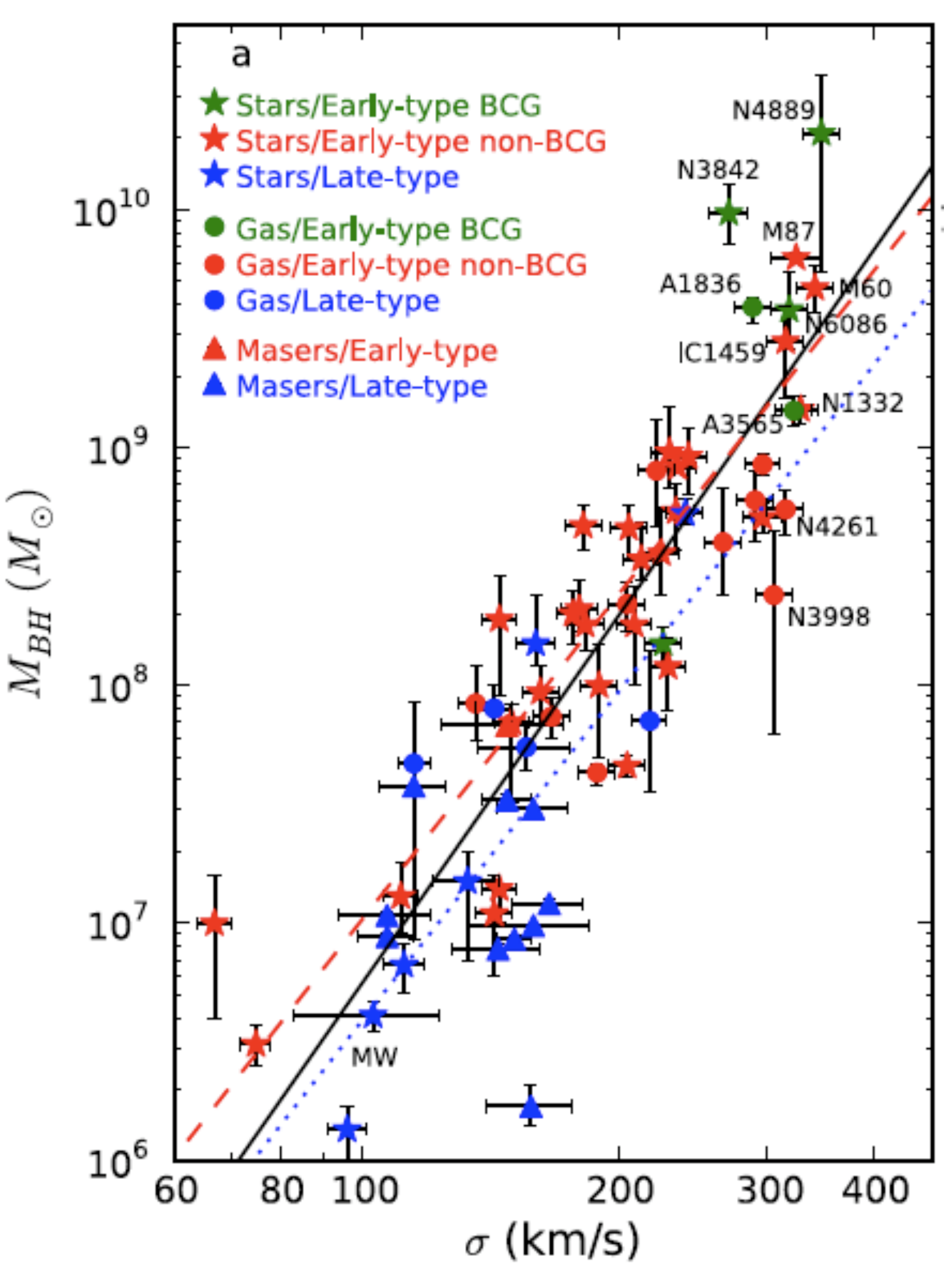,height=7cm,width=10cm}
\caption{Black hole mass vs spheroid velocity dispersion, figure from \citep{2011Natur.480..215M}.}
\label{fig:BH}
\end{figure}

Feedback from AGN will naturally result in a connection between the properties of a SMBH and its host galaxy. Therefore, this process  will also explain the observed correlation between the SMBH mass and galaxy mass or velocity dispersion, fig.\ref{fig:BH}. Indeed, coupling the energy released by the formation of the SMBH to the surrounding gas content of the forming galaxy will inevitably  lead to a relationship like the observed $M_{BH}-\sigma$ relation. If the BH is massive enough, outflows from its center will result in residual gas ejection, regulating star formation. This condition requires the Eddington luminosity of the BH to be $L_{Edd}/c=GM_{BH}M_{gas}/r^2$.
The MW is in the lower left corner of fig.\ref{fig:BH}, and represents a typical example of a late-type galaxy whose central BH is less massive than those typically observed in early-type galaxies.

The slope of $M_{BH}-\sigma$ depends on whether the quasar feedback is predominantly energy-conserving $M_{BH}\propto\sigma^5$  \cite {1998A&A...331L...1S} or momentum-conserving $M_{BH}\propto\sigma^4$ \cite{1999MNRAS.308L..39F,
2003ApJ...596L..27K}: both data \cite {2011MNRAS.412.2211G} and 3-d simulations \cite{Wagner13} favor the steeper slope.

AGN feedback can explain the exponential break in the galaxy luminosity function
(fig.\ref{fig:silk}) and, by quenching star formation, can reproduce the bimodal
distribution of galaxy colors: massive, early-type galaxies will be "red and
dead", with star formation quenched by the SMBH feedback. Reference
\citep{2007MNRAS.382.1415S} found observational evidence for AGN feedback in early-type galaxies, with star-forming early types inhabiting the blue cloud, while early types with AGN being located considerably closer to the red sequence.

\begin{figure}[h!]
\centering
\epsfig{file=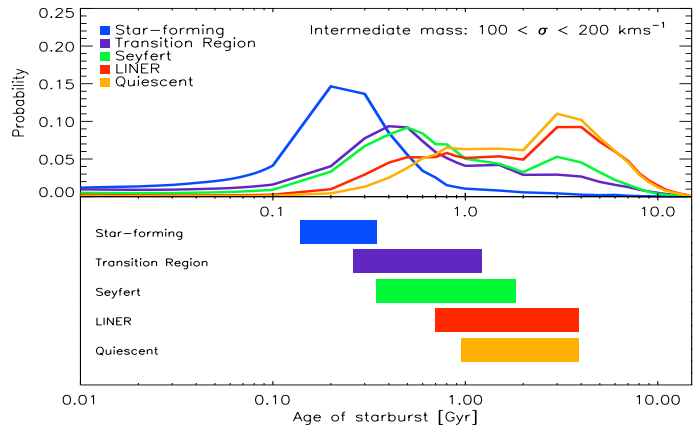,height=7cm,width=11cm}
\caption{Top panel, normalised probability distribution functions for the time elapsed since the start of the starburst $t_y$. Bottom panel, $50\%$ of such probability as a function of $t_y$. Figure from \citep{2007MNRAS.382.1415S}.}
\label{fig:shaw}
\end{figure}

Fig.\ref{fig:shaw} shows how star-forming objects have a starburst age around 150-300 Myr, while the most common starburst age $t_y$ for the transition region objects is around 300 to 500 Myr. The peak AGN phase occurs roughly half a Gyr after the starburst. The most likely interpretation is that star formation is suppressed by nuclear activity in these objects.

\subsection{Modes of AGN feedback}

There are two modes of AGN feedback: the quasar mode, occurring when large amounts of gas flow inwards, during the dominant accretion of BH mass, and the radio mode, during which the BH  accretes at a lower rate. During the radio mode, the AGN drives powerful jets and cocoons that heat the circumgalactic and halo gas, effectively shutting down cooling in massive haloes and resulting in agreement with the bright end of the observed luminosity function.

\begin{figure}[h!]
\centering
\epsfig{file=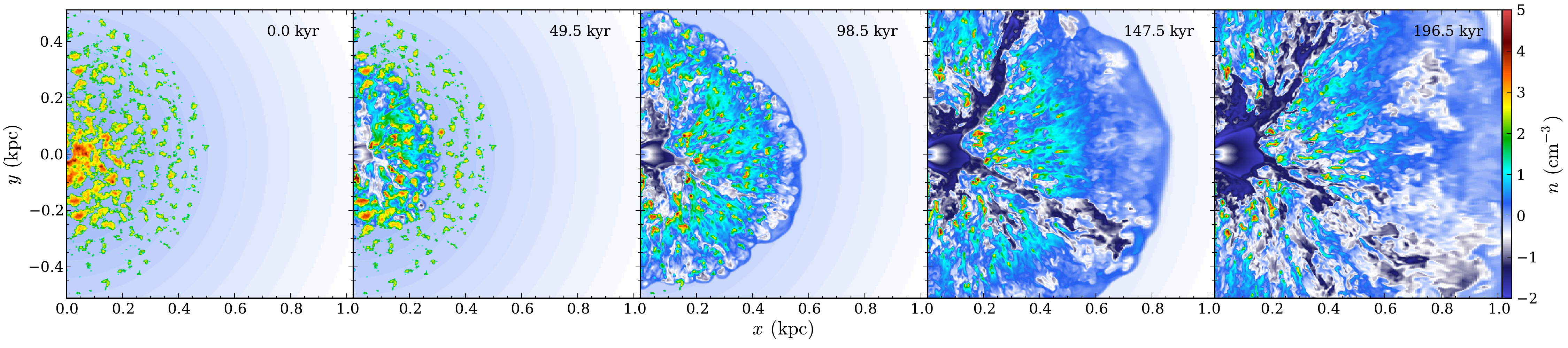,height=4cm,width=14cm}\\
\centering
\epsfig{file=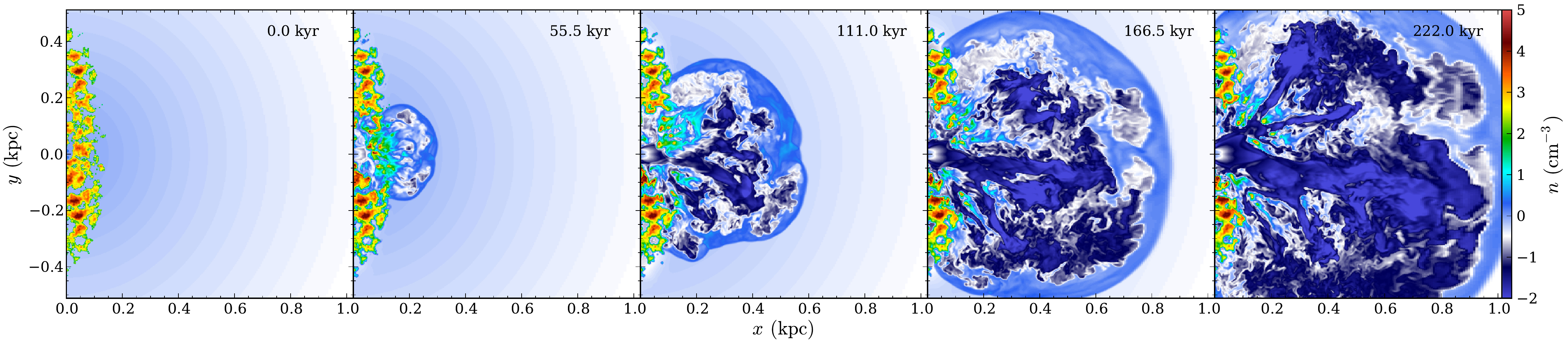,height=4cm,width=14cm}
\caption{Top panel, midplane density slices of the evolution of a $10^{44}ergs^{-1}$ ultra-fast outflow for a two-phase ISM with spherically distributed clouds. Lower panel, same as top but for a ISM with disc-distributed clouds. Figure from \citep{Wagner13}.}
\label{fig:wag}
\end{figure}

AGN activity is important in the cosmological feedback cycles of galaxy formation.
Reference \citep{2010MNRAS.405.1303A} studied the internal circulation within the cocoon arising from such a relativistic jet emanating from an AGN, performing 2D simulations, and found that backflows could feed the AGN and provide a self-regulatory mechanism of its activity. The study 
\citep{Wagner13} used 3D grid-based hydrodynamical simulations to show that ultra-fast outflows (UFO) from AGN result in considerable feedback of energy and momentum into the interstellar medium of the host galaxy. They performed simulations of the UFO interacting with a two-phase ISM in which the clouds
are distributed spherically or in a disc. Differences are show in fig.\ref{fig:wag}. Within 10 kyr after the start of the UFO, the evolution starts to differ between the cases of bulge-like and disc-like cloud distributions. In the former case, the UFO streams continue to channel and branch out quasi-isotropically and inflate a quasi-spherical energy bubble.

\subsection{Positive feedback from AGN}
We have seen that negative feedback from AGN helps account for the BH mass-$\sigma$ correlation and for the luminosity function of massive galaxies. At the same time, AGN activity could result in positive feedback on the star formation rate \citep{2009ApJ...700..262S, 2013ApJ...772..112S}.
A phase of positive feedback is motivated by evidence for AGN triggering of star formation \citep{Feain07, 2013ApJ...774...66Z}, discussed further below.

AGN outflows can trigger star formation by compressing dense clouds.
Propagation of jets into a clumpy interstellar medium will lead to the formation of an expanding, over-pressurized cocoon at $v_{co}$, which is much larger than the velocity field associated with the
gravitational potential well. Therefore, protogalactic clouds that are above the Jeans, or the more appropriate Bonnor-Ebert, mass may be induced to collapse.

 The region where AGN feedback can be positive is determined by the condition that the AGN-induced pressure exceeds the dynamical pressure that controls the ambient interstellar medium. A key ingredient  in star formation is molecular hydrogen. The molecular hydrogen fraction correlates with interstellar pressure in nearby star-forming galaxies
\cite{2006ApJ...650..933B}.
Enhanced pressure from AGN is likely to accelerate molecular cloud formation and thereby star formation.

\begin{figure}[h!]
\centering
\epsfig{file=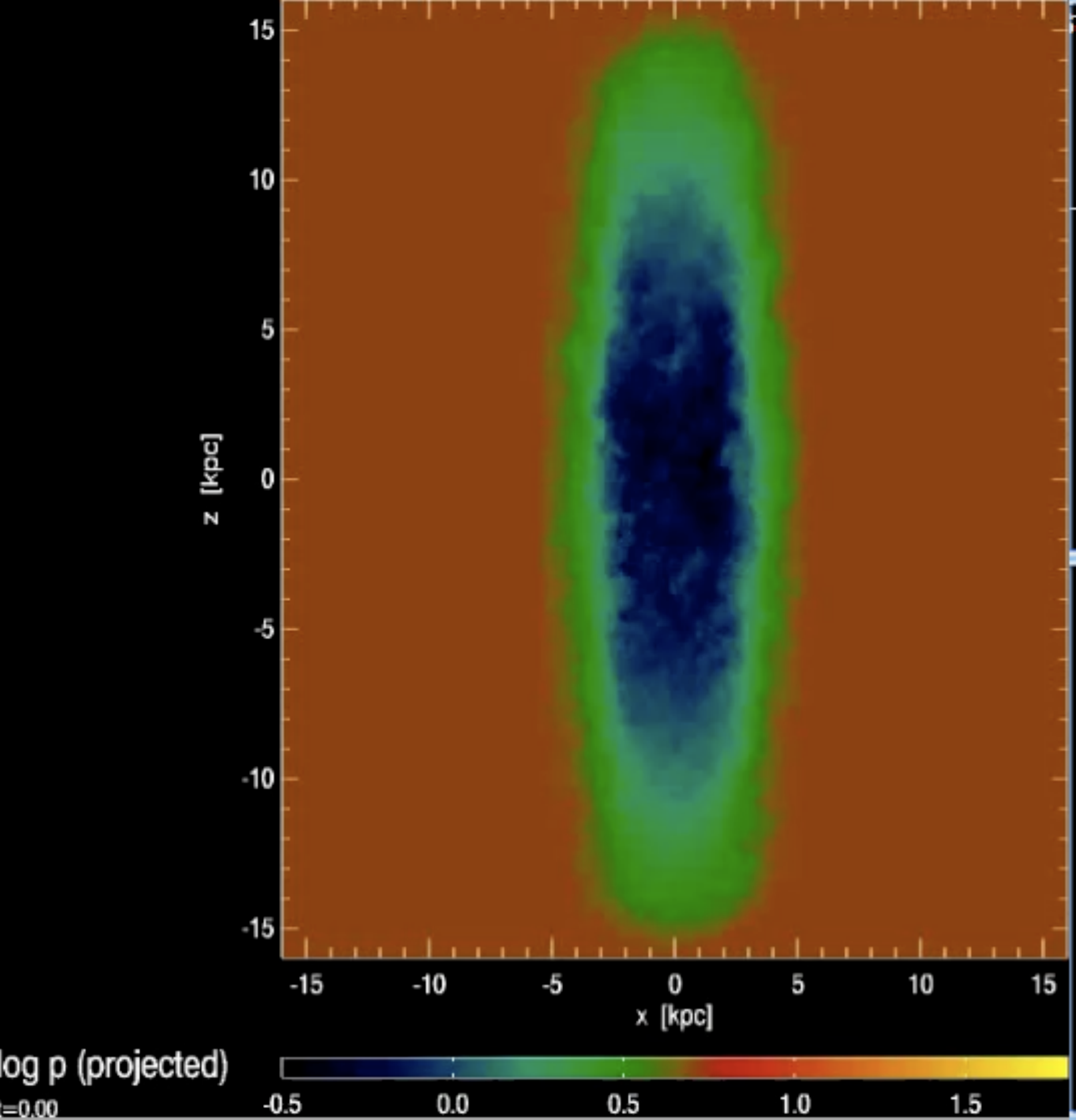,height=6cm}\epsfig{file=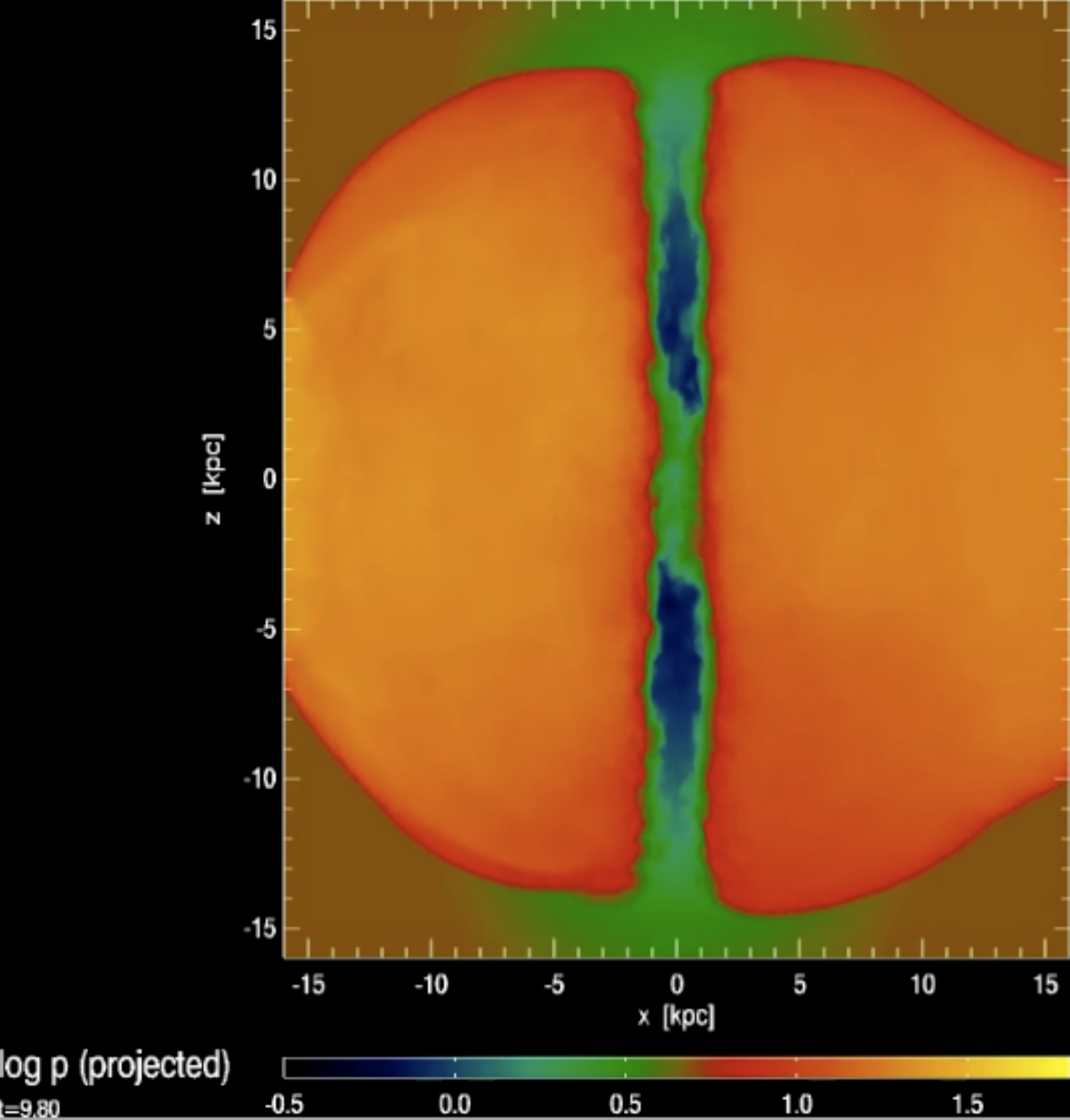,height=6cm}
\caption{The gaseous disc gets compressed by the expanding over-pressurized cocoon driven by AGN outflows. Left panel, initial state;right panel, final stage of evolution. The figures are 
color-coded according to increasing pressure, from blue to yellow.
Simulation from \citep{2012MNRAS.425..438G}.}
\label{fig:coocon}
\end{figure}

If one replaces the gas pressure, $\rho_g$, by the AGN-driven pressure, $\rho_{AGN}$, then the AGN-driven-star-formation-enhancement factor is $(\rho_{AGN}/\rho_g)^{1/2}\approx (v_{co}/\sigma)\tau^{1/2}$ where $\tau$ is the optical depth. Since $\epsilon_{SN}\sim\sigma$ the fraction of stars formed per
dynamical time is boosted for spheroids relative to disks.
Numerical simulations \citep{2012MNRAS.425..438G} of the interaction of a powerful AGN jet with the massive gaseous disc of a high-redshift galaxy demonstrate that such enhanced AGN-driven pressure from jets is effectively able to compress the disk gas and to enhance star formation, as shown in fig.\ref{fig:coocon}.

\subsection{SMBH formation}
The epoch of first quasars, first galaxies and first stars is at about z=6-10.
But how does a SMBH form? Before supermassive black holes can grow via accretion or merging, there must be some pre-existing seed black holes.
There are two models for the creation of SMBH seed: via remnants of PopIII stars \citep{2001ApJ...552..459H} or by direct halo gas collapse \citep{2006MNRAS.370..289B}.
In the first scenario, the seed is created at high redshift by the remnants of the earliest generation of Population III stars, which have reached the end of their stellar lifetimes. Gas collapse in a $10^6\Msun$ halo leads to an BH seed that will grow into an AGN by gas accretion.
In the second scenario, the halo gas collapse directly to form a massive IMBH seed.
Since the halo virial temperature is:
\begin{equation}
T_{vir}=10^4\left(\frac{M}{10^8\Msun}\right)^{2/3}\left(\frac{1+z}{11}\right)K
\end{equation}
at $10^8\Msun,$ the Lyman-alpha cooling operates at $10^4$K, so direct collapse via atomic cooling is possible.
In this latter case, there are two dynamical problems: the angular momentum barrier prohibits the gas from collapsing and the fragmentation depletes the accreting gas. However, while the gas collapses and becomes
turbulent, the fragmentation is suppressed and, once a gaseous bar is formed, it redistributes $J$, overcoming the angular momentum barrier by a sequence of bar formation and dissolution to progressively smaller scales \cite{2013ApJ...774..149C}.

\section{Gas accretion in galaxies}\label{sec:gas_acc}

There are two modes of gas accretion, cold flow accretion and hot gas accretion via major mergers.
In the first scenario, cold gas is provided along filaments, while in the major merger case, a source of hot gas, which will eventually feed star formation by cooling, is supplied to the galaxy.

\begin{figure}[h!]
\centering
\epsfig{file=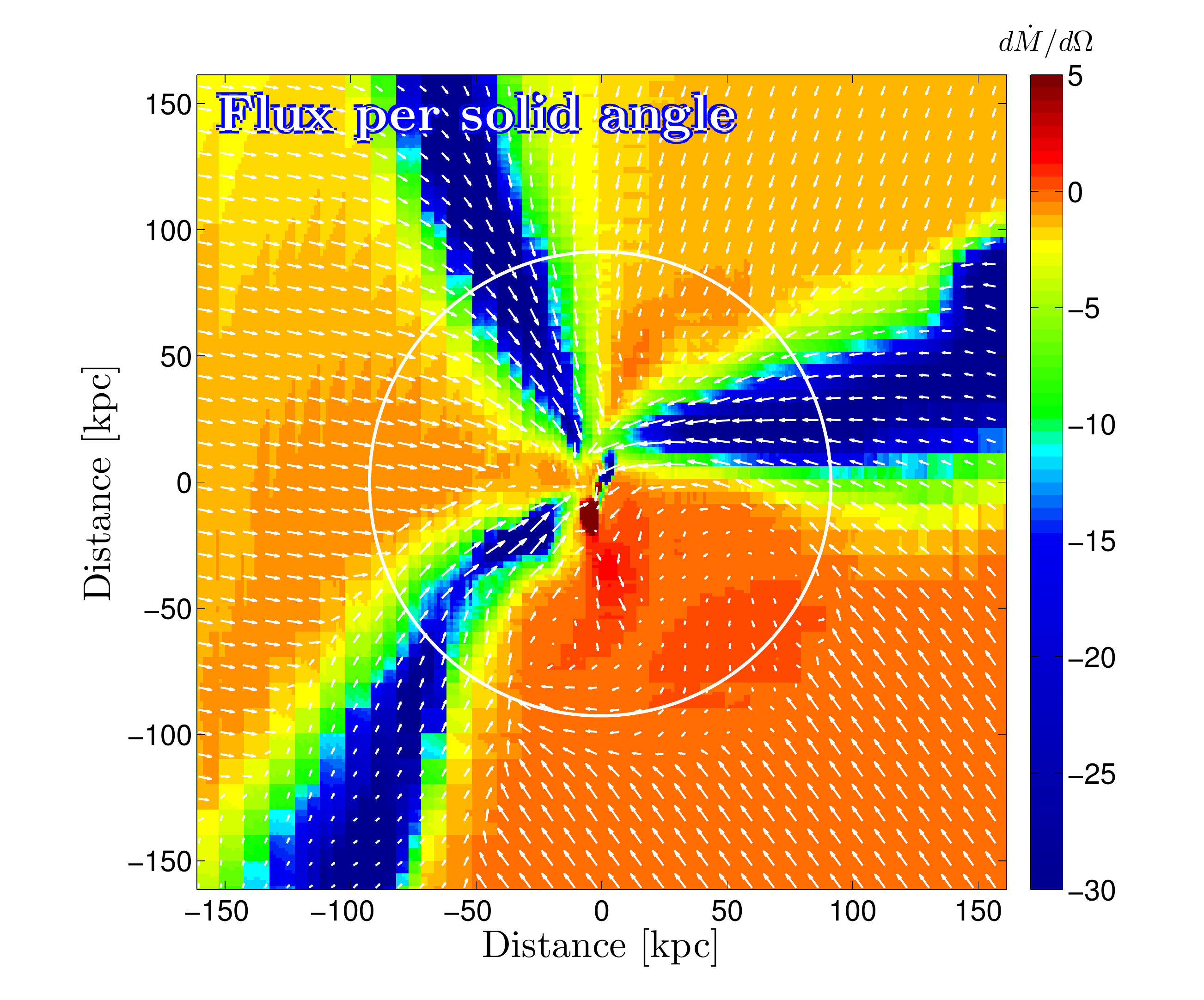, height=8cm}
\caption{Flux of cold streams through a hot halo at z=2.5. The circle marks the halo virial radius. The hydrodynamical simulated galaxy has $M=10^{12}\Msun$. Figure from \citep{2009Natur.457..451D}.}
\label{fig:dekel}
\end{figure}

 The cold flows occur in filamentary streams following the cosmic web of large-scale structure, fig.\ref{fig:dekel}. However, cold flows are rarely observed, but this can be due to the small covering factor of the filaments: the best indirect evidence in favor of cold flows is perhaps the study of star formation in dwarf galaxies, where several distinct episodes of star formation are detected.

Major mergers are subdominant in star-forming galaxies at $z\sim 2$ where the star formation rate density peaks, as measured by Sersic profile fitting. A sample of submillimeter galaxies at $z=1-3, $ considered to be the most extreme star-forming galaxies, shows that their morphologies with WFC3/IR imaging are predominantly disk-like \cite{2013MNRAS.432.2012T}. However the most luminous starbursts  almost  invariably show signs of major mergers \cite{2012ApJ...757...23K}. Moreover, AGN host galaxies are mostly disks as measured by morphology \cite{2012ApJ...744..148K}.

\begin{figure}[h!]
\centering
\epsfig{file=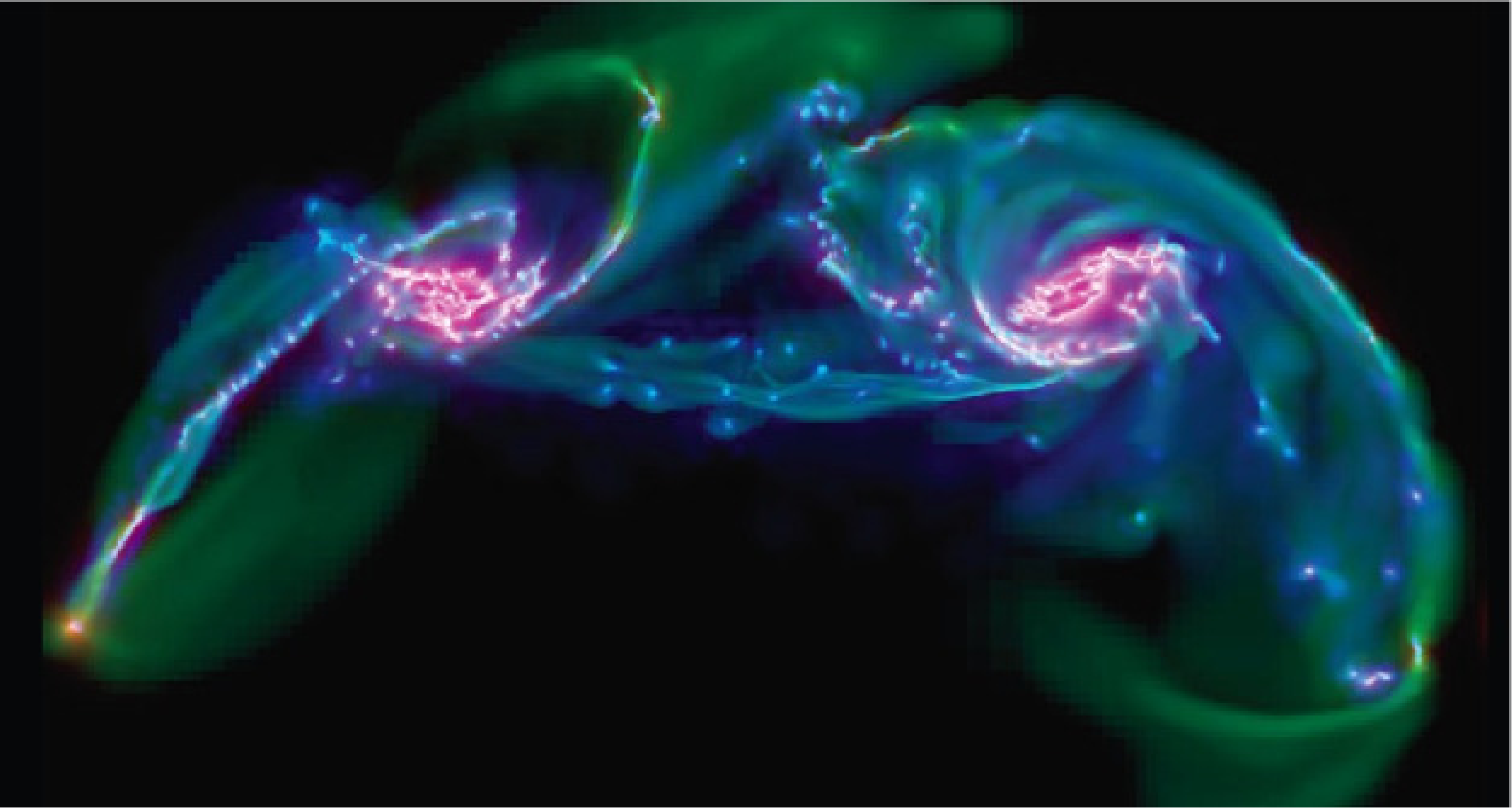, height=3.7cm}\epsfig{file=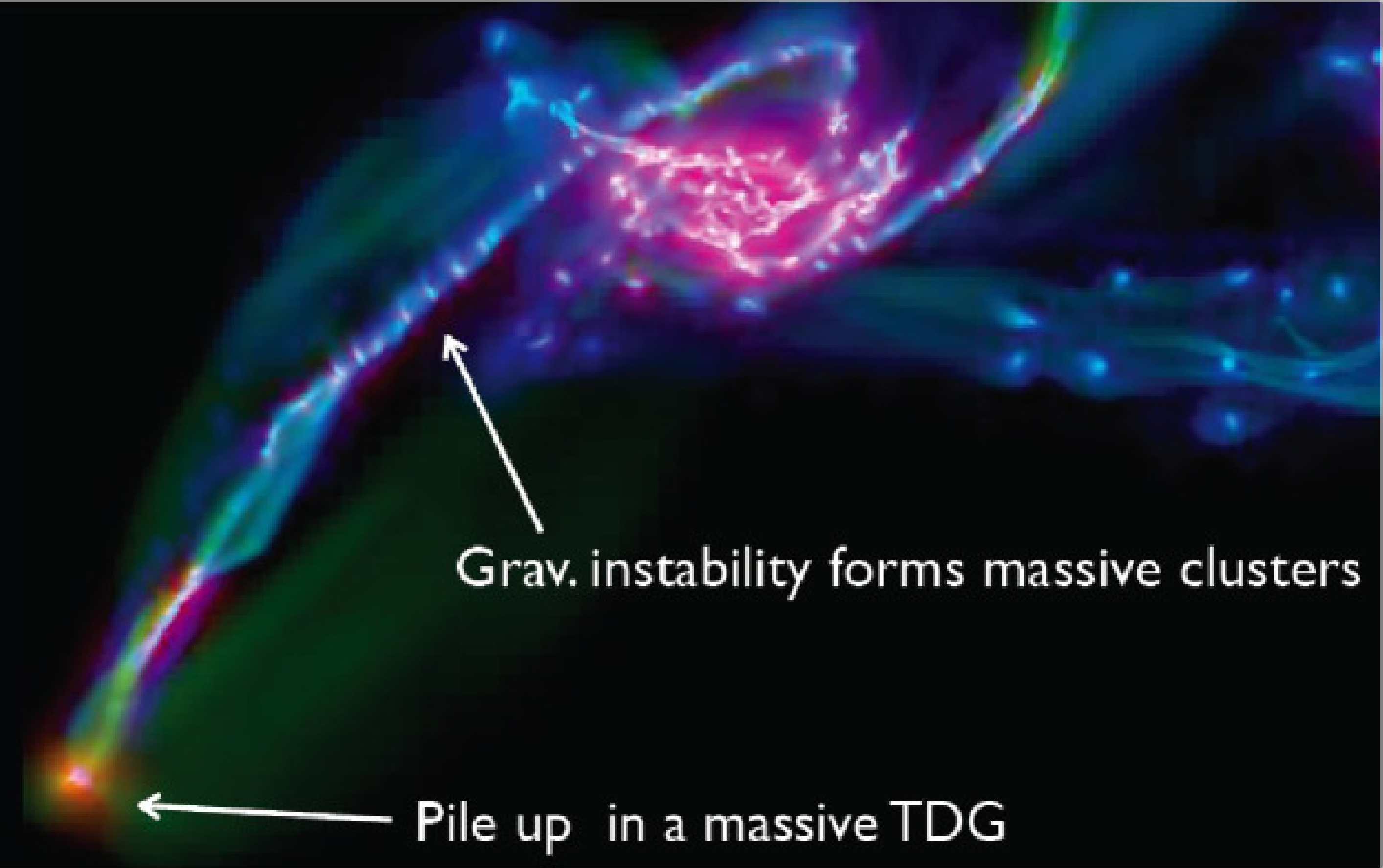,height=3.7cm}
\caption{AMR grid hydrodynamic simulation of a major merger from Chapon et al. Pile-up of material in a massive tidal dwarf galaxy at the tip of a tidal tail is visible. Figure from \citep{2009arXiv0909.1812B}.}
\label{fig:chapon}
\end{figure}

At high redshift, major mergers especially between the most massive galaxies are more common. Usually, the mechanism invoked in merger-induced starbursts is a global inflow of gas towards the central kpc, resulting in a nuclear starburst.
Major galaxy mergers lead to cloud agglomeration, angular momentum loss and cooling flows that feed star formation \citep{2009arXiv0909.1812B, 2011IAUS..271..160B}.
Galaxy interactions and mergers drive star formation, and a variety of stellar structures can then be formed: for example, it has been proposed that the pile-up mechanism forms massive tidal dwarf galaxies which survive as dwarf satellites around the merger remnant, fig.\ref{fig:chapon}.

\section{The two modes of star formation}

\subsection{The SFR main sequence and starburst galaxies}
\begin{figure}[h!]
\centering
\epsfig{file=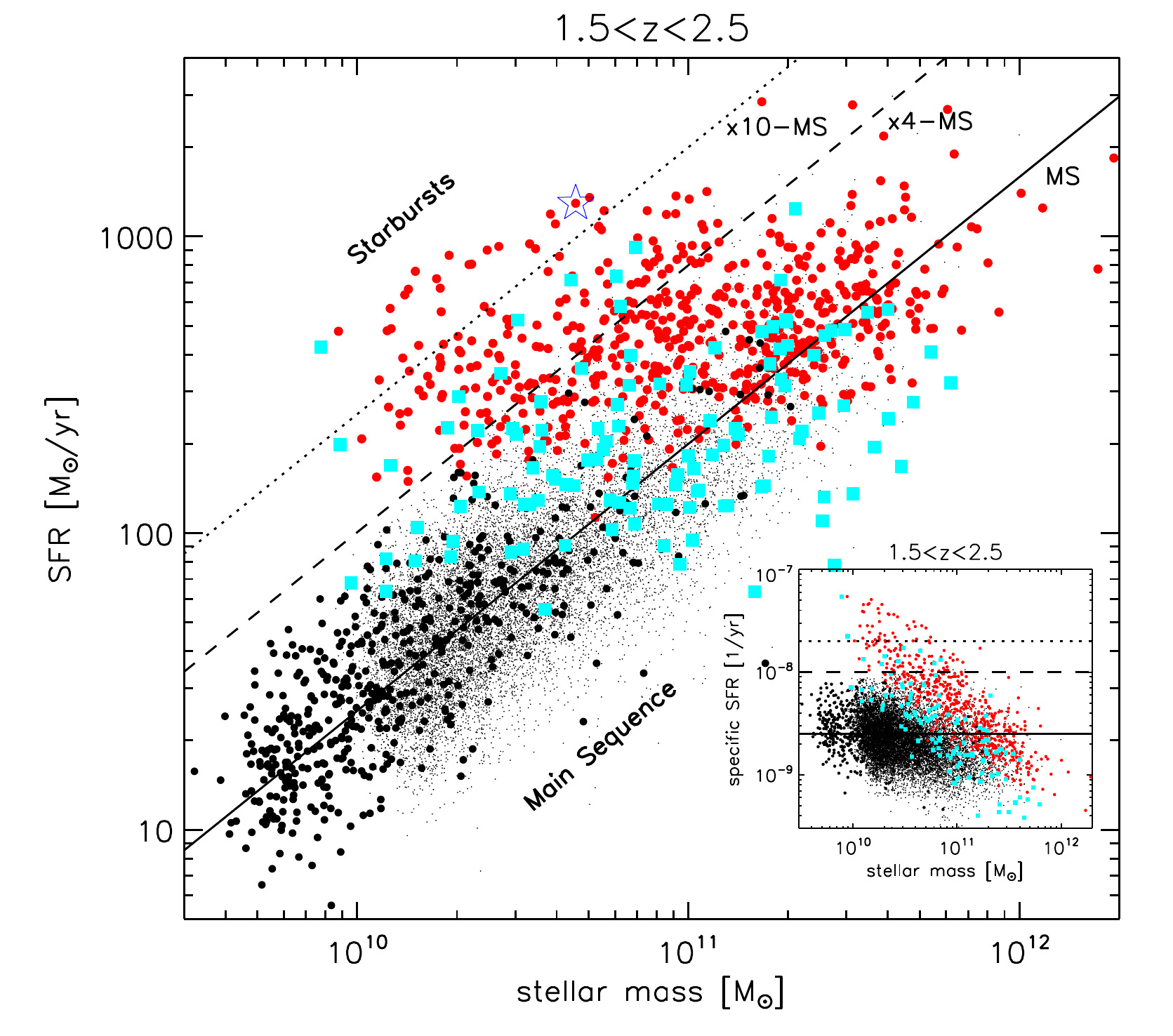, height=8cm}
\caption{Star formation rate vs stellar mass relation at $1.5<z<2.5$, for different samples of galaxies (various symbols). The solid black line indicates the Main Sequence of star forming galaxies, and a population of starbursts is evident in the top left panel. In the inset, the same relation is shown but as a function of specific SFR.
Figure from \citep{2011ApJ...739L..40R}.}
\label{fig:rodigh}
\end{figure}

Alongside with the two modes of gas accretion, two main modes of star formation are known to control the growth of galaxies: a relatively steady rate  in disk galaxies, although intermittent in dwarfs, which defines the main star formation rate-stellar mass sequence, and a starburst mode in outliers of such a sequence, fig.\ref{fig:rodigh}.
Such starburst galaxies, which are generally interpreted as driven by mergers, are relatively rare 
at $z\sim 2$ and have considerably higher SFRs.
In the aim of establishing the relative importance of these two modes, reference \citep{2011ApJ...739L..40R} analyzed several sample of galaxies.
They conclude that merger-driven starbursts play a minor role for the formation of stars in galaxies, whereas they may represent a critical phase towards the quenching of star formation and morphological transformation in galaxies.

Morphologies confirm this picture for luminous star-forming galaxies, both locally and at $z\sim 1$ \cite{2013arXiv1309.4459H}. Most galaxies with IR luminosities above
$3\times 10^{11}\rm L_\odot$ are found to show merger-induced morphological disturbances and the fraction of such deviations from normality increases systematically with distance above the galaxy main sequence.
%Hung2013

Star-forming galaxies at $z\sim 2$  often contain  a small number of extremely massive star-forming clumps. Resolution using adaptive optics reveals that
the clumps are undergoing  extreme rates of star formation \cite{2013ApJ...767..151M}. Such rates may be difficult to achieve with purely gravitational triggering as found in local star-forming galaxies.
%Menendez 2013.
Simulations suggest that such clumps, presumably formed by violent disk gravitational instabilities,  may generate outflows but survive for $\gtsim10^8\rm yr,$ long enough to fall into the central regions and contribute to bulge formation and black hole growth \cite{2013arXiv1307.7136B}.

\subsection{Evolution of early-type galaxies (ETG)}
\begin{figure}[h!]
\centering
\epsfig{file=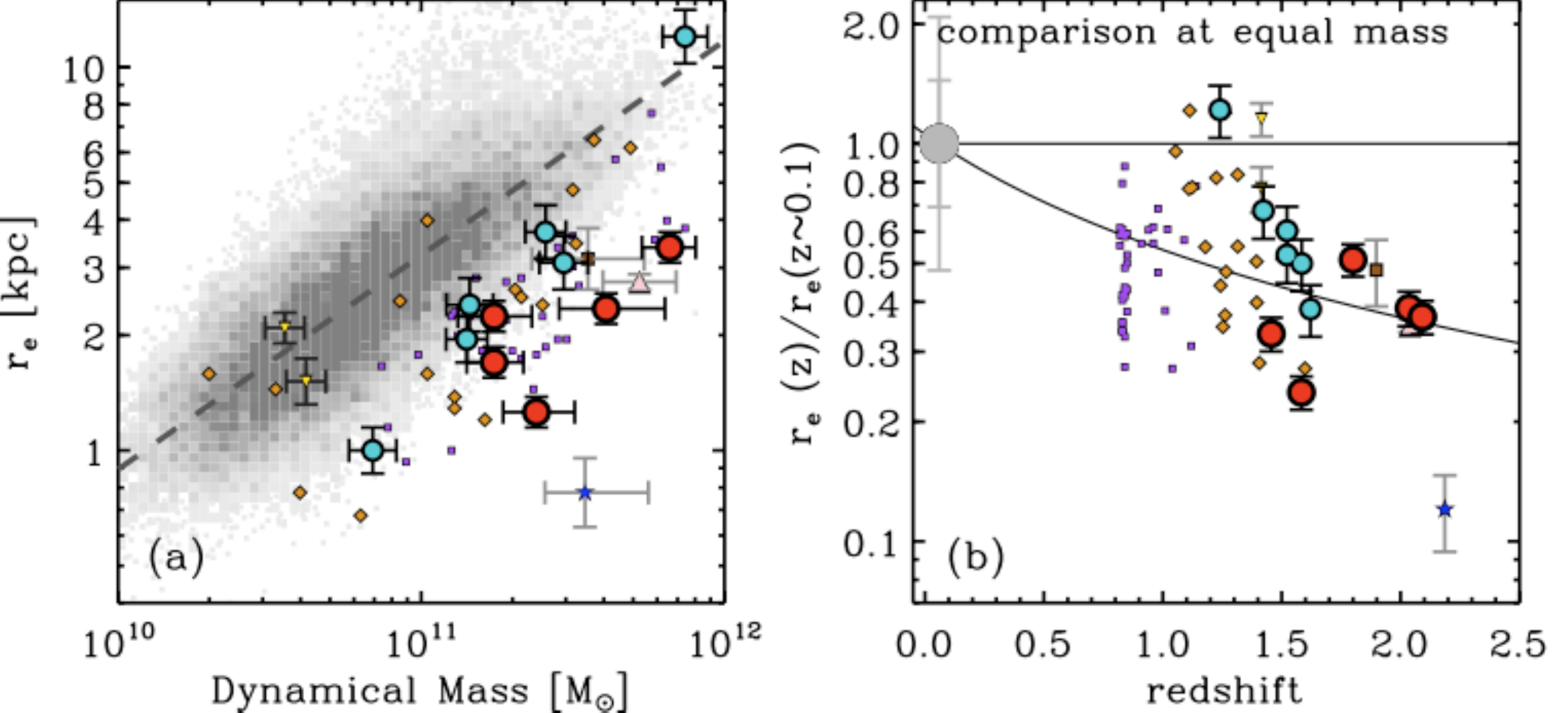, height=6cm}
\caption{Evolution of the effective radius of passive galaxies with redshift. Left panel, size vs. dynamical mass: $z\sim2$ galaxies (red circles) are smaller by a factor 3 compared to low-redshift galaxies. Right panel, evolution of the effective radius at fixed dynamical mass vs. redshift. The solid line is the best-fit $r_e \approx (1+z)^{-0.97\pm0.10}$. Figure from \citep{2013ApJ...771...85V}.}
\label{fig:passive}
\end{figure}

Isolated, early-type galaxies such as ellipticals and S0s, usually evolve in a passive way, with few signs of any on-going star formation.
Such passive galaxies, however, are observed to grow in size. The mass-size relation of ETGs has been largely studied to probe their mass assembly history.
The general picture is that the inner parts of ETGs form in situ after gas accretion and gas-rich mergers at $z>2$, while the outer parts form via minor mergers at $z<2$. Reference 
\citep{2013ApJ...771...85V} used stellar kinematics measurements to investigate the growth of massive, quiescent galaxies from $z\sim2$ to today, fig.\ref{fig:passive}. They found an inside-out growth of quiescent galaxies, consistent with expectations from minor mergers.

\begin{figure}[h!]
\centering
\epsfig{file=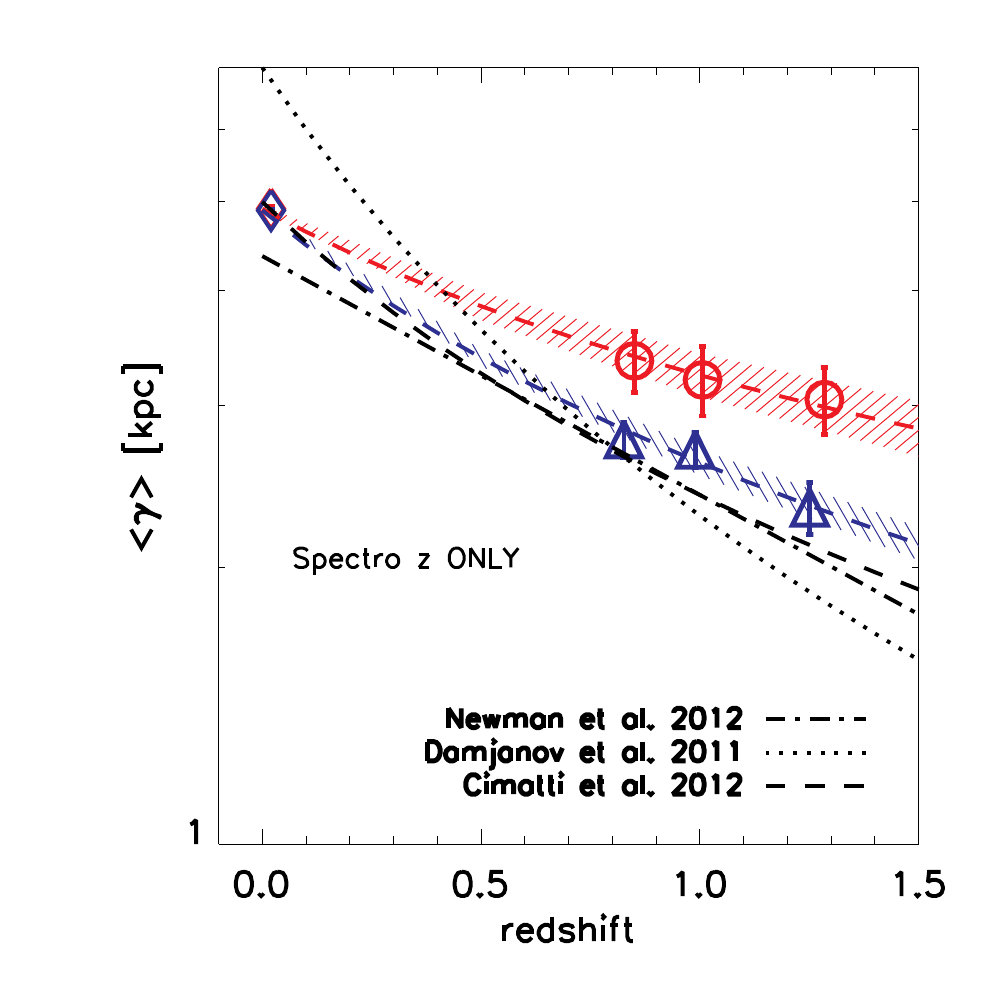, height=8cm}
\caption{Evolution of $\gamma$ for passive ETGs with $log(M/\Msun)\geqslant10.5$ in clusters (red circles) and in the field (blue triangles). Blue and red dashed lines show the best-fit model for field and
cluster galaxies respectively. Cluster passive ETGs are on average larger at $z\sim1$ and present a less steep evolution than field galaxies at fixed stellar mass.Figure from \citep{2013arXiv1307.0003D}.}
\label{fig:delaye}
\end{figure}

Moreover, passive galaxies have been found to be larger in clusters than in the field.
\citep{2013arXiv1307.0003D} studied the mass-size relation of quiescent massive ETGs living in massive clusters at $0.8<z<1.5$, as compared to those living in the field at the same epoch. The authors find that ETGs in clusters are $\sim30-50\%$ larger than galaxies with the same stellar mass residing in the field. They parametrize the size using the mass-normalized size, $\gamma=R_e/M_{star}^{0.57}$, fig.\ref{fig:delaye}. The size difference seems to be essentially driven by the galaxies residing in the cluster cores. They conclude that part of the size evolution is due to mergers: the observed differences between cluster and field galaxies could be due to higher merger rates in clusters at higher redshift.

\subsection{Multiple stellar populations and the AGN connection}

\begin{figure}[h!]
\centering
\epsfig{file=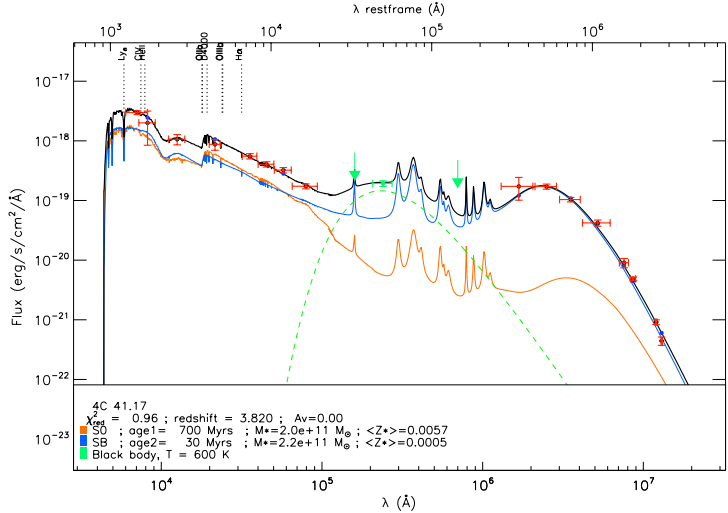, height=8cm}
\caption{The spectral energy distribution of the radio galaxy 4C 41.17 (dots with red error bars) and its best fit (black line) obtained by the sum of the S0-type population evolution scenario at age of 0.7Gyr (orange line), an AGN model (green),  and a starburst with an age of 30 Myrs after the initial 1 Myr duration burst (blue line) from a dense medium. Figure from \citep{2013arXiv1301.1983R}. The AGN model SED is from reference\cite{1993ApJ...418..673P}.}
\label{fig:rocca}
\end{figure}

Recently, \citep{2013arXiv1301.1983R} studied several distant radio galaxies, and conclude that a single stellar population is insufficient to fit their spectral energy distributions. Instead, we have to simultaneously follow the passive evolution of the galaxy as well as that of an on-going starburst to explain the overall spectral energy distribution. Their best fits are a sum of two evolving stellar populations, a recent starburst plus an old population, fig.\ref{fig:rocca}.
The two stellar components are a $\sim10^{11}\Msun$ starburst  some $\sim30$ Myrs after formation and an old, massive $\sim10^{11-12}\Msun$ early-type galaxy population, formed $\sim1.0$ Gyr previously. Comparable masses may be in these two populations. 
This finding suggests that most of the stellar population in high redshift radio galaxies may be  formed by massive starbursts in the early universe, and the fact that similar characteristics have been found in distant radio galaxies suggests that multiple stellar populations, one old and one young, may be a generic feature of the luminous infrared radio galaxy population.

AGN generally have enhanced star formation \cite{2013arXiv1310.1922R}. This is usually interpreted as negative feedback, that is, the AGN are in the process of quenching star formation. Of course one might then expect that in non-AGN galaxies of similar stellar mass at high redshift, there should be even higher star formation rates. This is very difficult to establish if one uses AGN-selected samples.
 % Rosario 2013.
%Zinn 2013

 The case for radio-jet induced positive feedback on star formation, presumably via pressurized cocoon formation, has been made in a study of the
correlation of radio-selected AGN with host galaxy star formation rates.
 The presence of radio jets is found to correlate with enhanced star formation far more strongly  than with x-rays from AGN \cite{2013ApJ...774...66Z}.

It is possible that when positive feedback occurs, this is at such an early stage that the sources are Compton-thick to x-rays. Indeed  \citep{2013arXiv1306.5235R}  finds a high z ($\sim  4$) example of a luminous AGN obscured in x-rays but detectable via  both its AGN signature  (rest- frame NIR emission by warm dust) and highly enhanced star formation (detectable by strong polycyclic aromatic hydrocarbon emission).

\subsection{Specific star formation rate}
{Stellar masses and spectral energy distributions can be used to probe the evolution of the specific star formation rate  ($\dot M_\ast/M_\ast $). The observed sSFR
was originally thought to plateau at $z\gtsim 4$, but  after correction for dust and nebula emission is now found to rise to $z\sim 6 $ or beyond \cite{2013ApJ...763..129S}.

A rising  or declining star formation history seems to be preferred by the data \cite{2012arXiv1207.3663D}. Whether this is due to a process such as merging or quenching or to a more exotic means of triggering star formation cannot currently be ascertained.
%de Barros 2012
}

\section{Efficiency of galaxy formation and downsizing of galaxies}
Galaxy formation theory generically favors formation of small galaxies before larger ones, and tends generically to overproduce early star formation. Observations unambiguously favor downsizing for both galaxies and for SMBH.

\begin{figure}[h!]
\centering
\epsfig{file=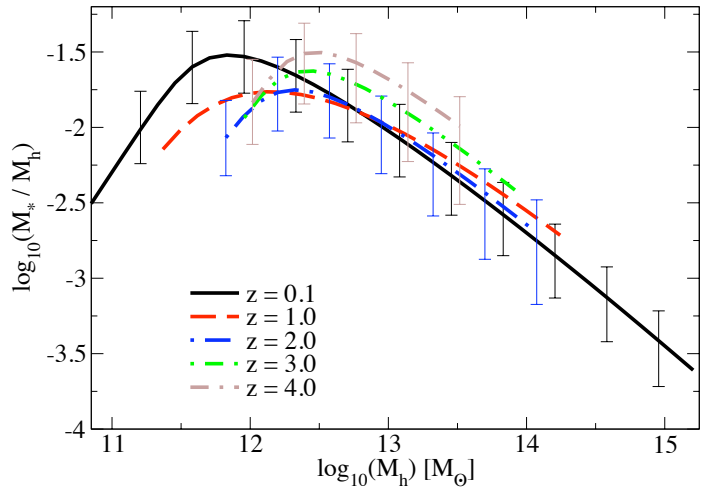, height=8cm}
\caption{Evolution of the derived stellar mass fractions for central galaxies,
from z = 4 to the present. Figure from \citep{2010ApJ...717..379B}.}
\label{fig:behroozi}
\end{figure}

In order to reconcile theory and observations, what is needed is some process which preferentially suppresses the formation of stars in lower mass dark matter halos.
The efficiency of galaxy formation must thus depend strongly upon halo mass.
Using the abundance matching technique, \citep{2010ApJ...717..379B} constrained the relationship between stellar mass and halo mass, and therefore the galaxy formation efficiency, and how this relationship evolves with redshift. The idea is to match the cumulative number of observed galaxies with the cumulative number of dark matter haloes, the latter being derived either from theory or from cosmological N-body simulations.
Such a technique reproduces the galaxy correlation function from SDSS galaxies; however, it does not tell us what the underlying physics responsible for such a correlation may be.
Fig.\ref{fig:behroozi} shows how the mass corresponding to the peak efficiency for star formation evolves slowly, being a factor of five larger at z = 4. The peak is at a halo mass of $7\cdot10^{11}\Msun$ at z = 0 and increases toward higher redshift. At lower and higher masses, the star formation is substantially less efficient.
The integrated star formation at a given halo mass peaks at $10-20\%$ of available baryons for all redshifts from 0 to 4.
Of course, at high redshift, the statistical uncertainties do not allow one to derive strong conclusions about the evolution of the abundance matching relation.

\begin{figure}[h!]
\centering
\epsfig{file=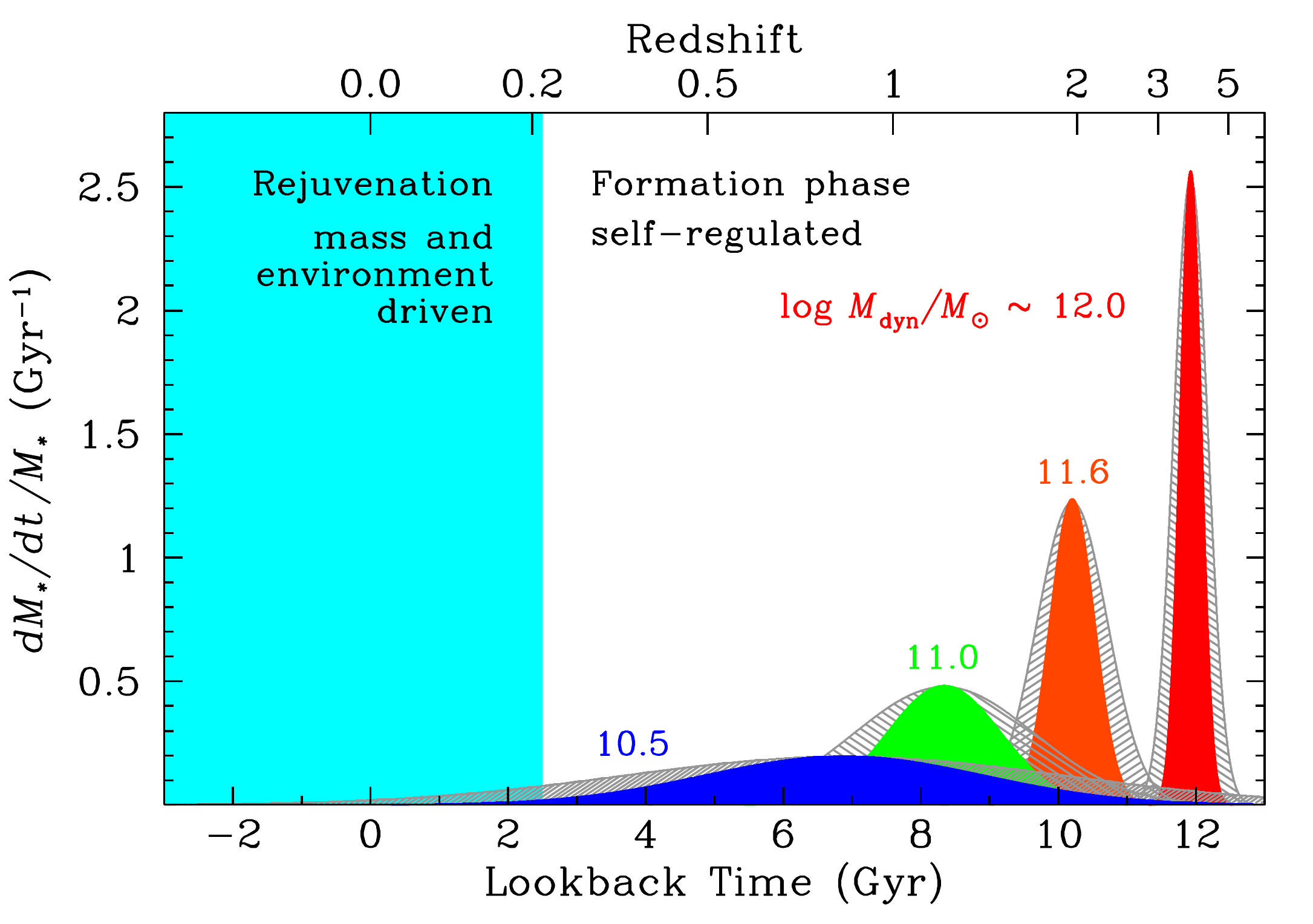, height=7cm}
\caption{Specific star formation rate as function of look-back time for early-type galaxies of various masses. The grey hatched curves indicate the range of possible variation in the formation time-scales that are allowed. The upper x-axis represents the  redshift. These star formation histories are meant to sketch the typical formation history averaged over the entire galaxy population. Figure from \citep{2010MNRAS.404.1775T}.}
\label{fig:thomas}
\end{figure}

One of the most intriguing aspects of galaxy formation is the so-called \it downsizing \rm of galaxies. While in a hierarchical galaxy formation scenario the first haloes to form are the smallest ones, there is observational evidence in favor of more massive, early-type galaxies being in place before smaller galaxies.
Downsizing is observed in stellar mass from IR data to $z\sim 4$ \cite{2008ApJ...675..234P}. It is also observed in the star formation time-scale by use of $[\alpha/Fe]$ as a chronometer. The $\alpha$-rich elements form in SNII supernovae, formed by core-collapse of short-lived massive stars, whereas $[Fe]$ is produced in SNIa, associated with an old stellar population and attributed to white dwarf mergers.
It has been shown \citep{2010MNRAS.404.1775T} that stars in most massive galaxies tend to have formed earlier and on a shorter time span, fig.\ref{fig:thomas}

Downsizing is  observed in SMBH masses and in black hole accretion rates (via Eddington ratios) \cite{2013ApJ...764...45K}.  Evidence for downsizing  of  chemical abundances in galaxies  is also found to $z\sim 3$ \cite{2008A&A...488..463M}.

A possible explanation \citep{2008MNRAS.389..567C} for the downsizing phenomenon is that galaxies cannot accrete or retain cold gas in massive halos, either because of AGN feedback or because of virial shocks that prevent gas supply to the disk via the cold filaments predicted in the simulations.

Of course downsizing of galaxies is not necessarily anti-hierarchical \citep{2006MNRAS.372..933N}: while the main progenitor of a galaxy shows the usual hierarchical  behavior, the integrated mass over all the progenitors down to a given minimum mass shows downsizing that is similar to what has been observed. In this sense, downsizing of galaxies can be partly environmental, a natural outcome of the bottom-up clustering process of dark matter haloes. This applies to star formation time-scale downsizing but not to global stellar mass downsizing: here baryonic gas physics must be invoked as discussed above.

\section{Current issues}
Let us now summarize  the most important unsolved questions in galaxy formation, at small and high mass scales, respectively:
\begin{itemize}

\item{Dwarf galaxies}

Alongside with the excessive predicted numbers of dwarf galaxies, there is a problem
with their inner dark matter density profiles: observations show that they have a core \citep{oh11b}, whereas simulations generally predict a cusp \citep{1996ApJ...462..563N}.
Moreover, another aspect of the "missing satellite problem" \citep{1999ApJ...522...82K} is the so-called too-big-to-fail problem \citep{2011MNRAS.415L..40B}: the dwarf spheroidal galaxies of the Milky Way live in dark matter haloes which are less concentrated than expected from N-body simulations. Several solutions have been proposed, amongst them modifications of dark matter via self-interactions \citep{2012MNRAS.423.3740V} or an Einasto-like density profile for satellite galaxies combined with a mass of $8.10^{11}\rm M_\odot$  for the total mass of the MWG  \citep{2013MNRAS.431.1220D, Vera-Ciro12}.
Probably both of these problems could be solved by simply invoking appropriate feedback mechanisms. Supernovae feedback is able to lower the dwarf central densities \citep{2012ApJ...761...71Z} \citep{2012MNRAS.422.1231G} but cannot apparently resolve the too-big-to-fail problem \cite{2013MNRAS.433.3539G}. One could perhaps invoke feedback by IMBH to resolve the latter problem: these most likely form  ubiquitously  in subhalos if IMBH are the building blocks of SMBH, as is commonly assumed in some scenarios of SMBH formation.

\item{Massive galaxies}

How are bulgeless thin disk galaxies formed? This type of galaxy has been observed to be common \citep{Kormendy10}, while numerical simulations produce galaxies with thick disks and bulges. One appealing solution involves SN feedback, which can drive a galactic fountain that feeds the bulge: this mechanism of redistribution of angular momentum can solve the bulgeless problem\citep{2012MNRAS.419..771B}.  Another proposal includes energy from massive stars as well as SNe \citep{stinson13}. An inevitable consequence of supernova or radiation feedback on dust is likely to be a  thick disk. Whether these ideas can be reconciled with observations of spiral galaxies at low $z$ is not clear.

Why is star formation so inefficient with regard to the total baryon reservoir? There is a serious shortfall of baryons in typical galaxies by $50\%$ or more \cite{2010ApJ...708L..14M}. Baryons are lacking on  larger scales too, such as groups  and even clusters \cite{2013ApJ...778...14G}. The circumgalactic environment may provide the reservoir where the bulk of the baryons reside although this is far from clear, in part because ejected baryons are enriched and cool rapidly in this environment, hence begging the question of what keeps them out. Recirculation mechanisms such as galactic fountains actually bring gas into the disk, indeed these supernova-driven phenomena help explain the current star formation rate in galaxies like the MWG
\cite{2012MNRAS.419.1107M}.

Finally, the massive galaxies predicted today are generally too many and too blue.
The simulated evolution of the galaxy luminosity function contradicts the data, either at high or at low redshift. For example, in order to match the rest-frame K-band luminosity function of galaxies, reference \citep{2011MNRAS.415.3571H} used semi-analytic models, including asymptotic giant branch stars in the stellar populations, which result in a more rapid reddening time-scale: in this case , however, there are too many blue galaxies predicted at  $z \sim 0.5.$

\end{itemize}

\section*{Acknowledgements}
ADC and ID thank the Italian INFN for the financial support during the "New
Horizons for Observational Cosmology" courses.
ADC thanks the MICINN (Spain) for the financial support through the grant
AYA2009-13875-C03-02 and the MINECO grant AYA2012-31101 and the MultiDark
project, grant CSD2009-00064.  The research of JS has been supported at IAP by
the ERC project  267117 (DARK) hosted by Universit\'e Pierre et Marie Curie -
Paris 6   and at JHU by NSF grant OIA-1124403.

\bibliographystyle{varenna}
\bibliography{gf}

%
% \begin{thebibliography}{0}
% \bibitem{ref:apo} \BY{Boccaccio~G. \atque de~Cam\~oes~L.}
%   \IN{Phys. Rev. A}{13}{1999}{12};
%   \SAME{69}{999}{1666}.
% \bibitem{ref:pul} \BY{Pulci~L.}
%   preprint INFN 8181.
% \bibitem{ref:bra} \BY{Bragg~B.}
%   \TITLE{Tender comrade},
%   in \TITLE{Workers Playtime},
%                   edited by \NAME{Tizio A. \atque Caio B.}
%                   (Unexeditor, Bologna) 1997, pp.~1-10.
% \end{thebibliography}

\end{document}